\documentclass[11pt]{article}
\usepackage{geometry} 
\usepackage{amsmath}        % See geometry.pdf to learn the layout options. There are lots.
\usepackage{graphicx}

\usepackage{amsthm}
\usepackage{amssymb}
\usepackage{amsmath}
\usepackage[mathscr]{euscript}
\usepackage{bm}
\usepackage{mathrsfs}

\usepackage{epstopdf}
\usepackage{amsfonts}
\usepackage{hyperref}
\theoremstyle{definition}

\theoremstyle{lemma}

\usepackage{url}
\usepackage{caption}
\usepackage{stmaryrd}
\usepackage{booktabs}
\usepackage{placeins}
\usepackage{xcolor}
\usepackage{float}
\usepackage[caption = false]{subfig}

\usepackage[inline]{enumitem}

 \usepackage{algpseudocode}
 \usepackage{algorithm,eqparbox,array}

\usepackage{natbib}
\newcommand{\powerset}{\raisebox{.15\baselineskip}{\Large\ensuremath{\wp}}}

\usepackage{epstopdf}
\AppendGraphicsExtensions{.tiff}

\newcommand{\Var}{\mathbb{V}\text{ar}}
\newcommand{\Cov}{\mathbb{C}\text{ov}}

\DeclareMathOperator{\Sum}{sum}

\DeclareMathOperator*{\argmin}{arg\,min}

\def\D{\mathrm{d}}
\providecommand{\norm}[1]{\lVert#1\rVert}

\newcommand{\mc}[2]{\multicolumn{#1}{c}{#2}}

\newtheorem{theorem}{Theorem}

\begin{document}

\title{High Dimensional Model Representation as a Glass Box in Supervised Machine Learning}

\author{Caleb Deen Bastian, Herschel Rabitz}

\maketitle

\begin{abstract}%   <- trailing '%' for backward compatibility of .sty file
Prediction and explanation are key objects in supervised machine learning, where predictive models are known as black boxes and explanatory models are known as glass boxes. Explanation provides the necessary and sufficient information to interpret the model output in terms of the model input. It includes assessments of model output dependence on important input variables and measures of input variable importance to model output. High dimensional model representation (HDMR), also known as the generalized functional ANOVA expansion, provides useful insight into the input-output behavior of supervised machine learning models. This article gives applications of HDMR in supervised machine learning. The first application is characterizing information leakage in ``big-data'' settings. The second application is reduced-order representation of elementary symmetric polynomials. The third application is analysis of variance with correlated variables. The last application is estimation of HDMR from kernel machine and decision tree black box representations. These results suggest HDMR to have broad utility within machine learning as a glass box representation. 
\end{abstract}

\tableofcontents

\section{Introduction}

In many areas of science and engineering one is typically interested in using empirical data $\textbf{D}$ to understand an input-output map $f$ from some class of systems $F$, often having a large number of input and output variables. We indicate the input space by $(X,\mathscr{X})$, the output space by $(Y,\mathscr{Y})$, and the system (model) space as $F=Y^X$, and put $n\equiv|X|\in\mathbb{N}$ and $p\equiv|Y|\in\mathbb{N}$ respectively. Suppose $\textbf{D}$ is a collection of $iid$ random variables taking values in $X\times Y$, denoted by $\textbf{D}=\lbrace (x_i,y_i)\rbrace$, and regard each $(x_i,y_i)$ as a realization from the probability measure $\nu$.

This setting can be identified to \emph{supervised machine learning} \citep{erm}, where $f$ is a prediction function, $F$ is some class of candidate functions, and $\nu$ is the joint probability measure on $X\times Y$ whose independent random realizations form $\textbf{D}$. The task for supervised machine learning is solving \[f^* = \argmin_{f\in F} R(f),\] where $R$ is a risk functional \[R: f\mapsto \frac{1}{N}\sum_{(x,y)\in\textbf{D}}L(y,f(x))\] with sample size $N\in\mathbb{N}$ and where $L$ is a loss function, $L: X\times Y\mapsto\mathbb{R}_+$. The learned $f^*$ is used to assign an element $f^*(x)$ of $Y$ to each $x$ in $X$. In this article we consider quadratic loss function $L(y,f(x))=(y-f(x))^2$. 

\subsection{Motivations of representation}

The input-output relation $f$ can be represented many ways. \emph{Black box} representation is solely concerned with the task of assigning an element $f(x)$ of $Y$ to each $x$ in $X$ (\emph{prediction}) and includes all such functions from $X$ into $Y$. In other words, the particular structure of $f$ is irrelevant so long as it is contained in $F$, and notions of input variable importance and dependence of $f$ upon the important variables are abstract. General-purpose black box learning algorithms in machine learning include neural networks, kernel machines, and decision tree models. \emph{Glass box} representation restricts $F$ to include only those functions which provide information on variable importance and dependence (\emph{explanation}). Glass box representation means that $F$ is equipped with necessary and sufficient information to interpret the model output in terms of the model input. This information is referred to as \emph{interpretative diagnostics} and is defined by the variable importance and variable dependence sets $\text{I}(f)$ and $\text{P}(f)$ respectively, each indexed on $\powerset(\lbrace1,\dotsc,n\rbrace)$. \cite{hooker} discusses diagnostics for high-dimensional functions based upon projection operators, and functionals can be further defined to measure variable importance from projections. These ideas are adopted in this article to construct $\text{I}(f)$ and $\text{P}(f)$. %A key notion for representation is \emph{information leakage}, which we use to describe models or interpretative diagnostics that either lack information on the true system $f$ (bias) or that contain information irrelevant to the intended input-output behavior (variance). Both black box and glass box models can experience information leakage.

%Whenever a black box is distinct from the true system of interest or exhibits defects in regard to its asymptotic behavior on $\textbf{D}$ it is said to experience \emph{information leakage}. 
%The construction of a full space representation of $f\in F$ is computationally or observationally NP-complete, with complexity growing exponentially in $n$ \citep{stone}. 

A consideration is managing the sizes of $\text{I}(f)$ and $\text{P}(f)$ as the number of input variables increases, each of size $2^n$. The exponential growth is known as the \emph{curse of dimensionality}. For large enough $n$, construction cost outstrips available resources, and feasibility is achieved through simplifying assumptions on $F$ or $\mu$, e.g., stipulating restrictions on model form or distribution. A large body of work supports the \emph{ansatz} that many high-dimensional systems especially those of a physical (or real-world) nature are accurately, if not exactly, described by representations whose sizes grow polynomially in $n$. %This is found in mathematical investigations of dynamical systems, wherein the embedding dimension is often very low, infrequent statistical reporting of tri and higher-order variances, ephemeral many-body collisions in physiochemical systems, highly effective use of reduced-order modeling techniques in computational simulations, etc. For example, experimental or observational designs on $(X,\mathscr{X})$ are specified to be fractionated for many systems, and Gaussian distributions are frequently assumed. 

The ansatz is elevated to theorem in a result of Kolmogorov \citep{kol}: his superposition theorem, an existence result, establishes that every multivariate continuous function on the unit cube $X=[0,1]^n$ can be exactly represented using a finite number of univariate continuous functions and the binary operation of addition, \[f(x) = \sum_{i=1}^{2n+1}\phi_i\circ\left(\sum_{j=1}^n\psi_{ij}\circ x_j\right),\] where $\phi_i$ are $\psi_{ij}$ are continuous functions. These functions are highly non-smooth, limiting their utility in approximation. %The trouble arises in seeking a finite representation strictly in terms of univariate functions, which, while attaining concision of form, destroys learnability. 

Instead of seeking a representation in terms of univariate functions, suppose a hierarchy of projections into subspaces of increasing dimensions, with the expectation that subspace contributions to output(s) rapidly diminish with increasing dimension. This is the idea behind a mathematical model called \emph{high dimensional model representation} (HDMR), a finite multivariate representation that efficiently manages the curse of dimensionality and that provides structured information for input-output relationships. When applied to supervised machine learning models, HDMR is a glass box that interprets the model output in terms of the model input. %Because HDMR enjoys a number of useful mathematical properties and over its long history has been formulated for and applied to seemingly distinct domains, we discuss the decomposition and provide illustrative applications in forthcoming sections to motivate its relevance to supervised machine learning.

\subsection{Related work} HDMR is discussed as early as \cite{fisher} in ANOVA analysis and has enjoyed extensive application to statistics. When $F$ is the collection of symmetric functionals of \emph{iid} variables HDMR is known as the Hoeffding decomposition, a fundamental object in U-statistics \citep{hoeffding}. The univariate terms of HDMR are sometimes known as Hajek projections and are extensively used to establish asymptotic normality of various statistics \citep{hajek1968}. Global sensitivity analysis (GSA) measures the importance of variables to $F$ \citep{Sobol01,Sobol93,Sobol2004}. In GSA applications functional representation of $F$ is avoided where instead sensitivity indices are directly estimated. \citep{alis99} discussed HDMR as a general decomposition of $F$ where the input space resides in $\mathbb{R}^n$ or in the $n$-fold product space of arbitrary linear topological function spaces. These ideas have been further developed to $F$ for general (non-degenerate) distributions in \cite{hooker}, wherein HDMR is known as generalized functional ANOVA, which in turn provisions GSA in terms of structural and correlative sensitivity indices \citep{li12}. Sometimes HDMR is known as the Hoeffding-Sobol decomposition \citep{hoeff} or Sobol decomposition \citep{Arwade20101}. See \cite{takemura} for references on HDMR's earlier history.

\subsection{Contributions}  %We demonstrate applications of HDMR to goodness-of-fit, reduced-order modeling, and analysis of variance. 
%HDMR enjoys a number of useful properties and over its long history has been formulated for and applied to seemingly distinct domains. 

We discuss HDMR and provide illustrative applications to motivate its utility to supervised machine learning. First, we illustrate that HDMR can diagnose information leakage. This is demonstrated for Pearson goodness-of-fit settings for ``big-data'' wherein HDMR characterizes estimator efficiency loss and for popular machine learning black boxes wherein HDMR identifies biases. Second, we illustrate that HDMR characterizes the information leakage experienced by partial dependence, another interpretative diagnostic, whenever input variables are correlated. In such settings, HDMR reveals that the interpretative diagnostics functionally depend upon the distribution correlation. Third, we demonstrate that HDMR admits efficient reduced-order representations of high-dimensional models, managing the curse of dimensionality. In particular, we illustrate that the input distribution regulates the efficiency of reduced-order representation in polynomials and demonstrate this effect in a learning setting. Fourth, we demonstrate that HDMR can be applied as a wrapper method for black boxes to provision glass boxes. Fifth, we estimate HDMR of trained kernel machines or ensembles of decision trees.

\subsection{Organization} 
In section~\ref{sec:setup} we formulate interpretative diagnostics using projection operators and functionals and related quantities. In section~\ref{sec:hdmr} we discuss HDMR and provide three examples of applications: (i) to goodness-of-fit in ``big-data'' settings (subsection~\ref{sec:pgof}), (ii) high-dimensional modeling (subsection~\ref{sec:rom}), and (iii) analysis of variance (subsection~\ref{simple}). In section~\ref{sec:id} we consider alternative interpretative diagnostics and compare to HDMR for a simple mathematical model having an analytic solution. In section~\ref{sec:gb} we apply HDMR as a wrapper method for black boxes to convey information on variable importance and dependence, and we consider two machine learning black boxes---kernel machines (subsection~\ref{sec:km}) and decision trees (subsection~\ref{sec:dt})---on analytic and empirical (test dataset) problems. We conclude with a discussion.% and areas of future interest. %We make the software available. Details of calculations and supplemental results are contained in the appendices.

%Notwithstanding the large literature to date, HDMR is an active area of research and development. Oftentimes, HDMR is identified to the particular function class and distribution, such as ANOVA HDMR (ordinary Lebesgue measure on unit hypercube), functional HDMR (Wiener measure on $n$-fold product space of continuous functions), cut HDMR (Dirac measure on the space of grids) \citep{alis99}, and so on. Various computational strategies have been employed in estimating HDMR, each of which either makes assumptions on the interpretation or structure of the underlying measure (and implicitly conferring properties on $F$): Monte Carlo \citep{rsli,rs2,rs3,rs,rs4}, analytic models under assumed distributions \citep{li14}, fast Fourier transform \citep{outside16}, piecewise continuous functions \citep{outside17}, stochastic partial differential equations \citep{outside23}, and so on. HDMR has seen extensive application to a variety of application domains: semi-conductor optimization \citep{semi,semi2}, material and molecular discovery \citep{ma1,reorder},  design of experiments \citep{outside35,doe},  biological network analysis \citep{feng,bio2,bio},  optimization of spectral inversion \citep{opt1,opt2,opt3,opt4},  safety and reliability estimation in engineering \citep{outside9,outside12,safe1,safe2},  combustion models and atmospheric physics and chemistry \citep{transport2,transport,transport3,io2,io1}, aircraft and vehicle design \citep{veh3,veh2,veh1}, global sensitivity analysis \citep{sa2,sa3,sa1,outside18,outside20,outside21,outside40,Arwade20101}, and so on.

\section{Preliminaries}\label{sec:setup} In this article we consider the model space to be square-integrable functions $F=L^2(X,\mathscr{X},\mu)$. We formulate interpretative diagnostics for the subspaces $u\subseteq\lbrace1,\dotsc,n\rbrace$ (\emph{notationally we use $\mathbb{N}_n=\lbrace1,\dotsc,n\rbrace$ throughout this article}) using projection operators and functionals. For every $u\subseteq\mathbb{N}_n$ let $\textbf{P}^u$ be a projection operator that profiles the dependence of $f$ on $x_u$ and put $f_u\equiv\textbf{P}^u f$; $f_u$ is said to be the \emph{variable dependence} or projection of $f$ on $x_u$. Let $\textbf{I}^u: F\mapsto\mathbb{R}_+$ be a functional that measures the importance of $f\in F$ on $x_u$ and put $S_u\equiv\textbf{I}^u f$; $S_u$ is said to be a measure of \emph{variable importance} of $f$ on $x_u$.  Using $\lbrace\textbf{I}^u\rbrace$ and $\lbrace\textbf{P}^u\rbrace$, we define respective variable importance and variable dependence sets $\text{I}(f) \equiv \lbrace\textbf{I}^uf\rbrace$ and $\text{P}(f) \equiv \lbrace\textbf{P}^uf\rbrace$ on $\powerset(\mathbb{N}_n)$. 

Sometimes total and relative notions of variable importance are necessary. We define the \emph{total variable importance} of $f$ on $x_i$ as $T_i\equiv\textbf{T}^i f$ where $\textbf{T}^i \equiv \sum_{u\supset i}\textbf{I}^u$. We define the \emph{relative variable importance} of $f$ on $x_i$ as $R_i\equiv\textbf{R}^i f$ where $\textbf{R}^i f\equiv \textbf{T}^i f/ \sum_j \textbf{T}^j f$. Using $\lbrace\textbf{T}^i\rbrace$ and $\lbrace\textbf{R}^i\rbrace$ we define total variable importance and relative variable importance sets $\text{T}(f)=\lbrace \textbf{T}^i f\rbrace$ and $\text{R}(f)=\lbrace\textbf{R}^i f\rbrace$. 

Using $\text{I}(f)$, $\text{P}(f)$, $\text{T}(f)$, and $\text{R}(f)$, we can pose various questions for $f$ as super level-sets:
\begin{enumerate}[label=(\roman*)]
\item which variables are important overall? \[\text{T}_\epsilon(f)=\lbrace T_i\in\text{T}(f) : T_i \ge \epsilon\rbrace\] 
\item what are the relative importances of the variables? \[\text{R}_\epsilon(f)=\lbrace R_i\in\text{R}(f) : R_i\ge \epsilon\rbrace.\] 
\item which variables are important individually? \[\text{I}_\epsilon(f)=\lbrace S_u\in\text{I}(f) : S_u \ge \epsilon, |u|=1\rbrace\] 
\item which variables participate in interactions?  \[\text{X}_\epsilon(f)=\lbrace S_u\in\text{I}(f) : S_u \ge \epsilon, |u|>1\rbrace\] 
\item how does the system depend upon the important variables and interactions?  \[\text{P}_\epsilon(f)=\lbrace f_u\in\text{P}(f) : I(f)\ni S_u \ge \epsilon \rbrace\]
\item does the system admit a reduced order representation? \[\exists\, T\in\lbrace 1,\dotsc,n\rbrace: \,\,\,f\simeq f^T=\sum_{u: |u|\le T}f_u\]
\end{enumerate} 
%and is defined as a hierarchy of projections into subspaces of increasing dimensions, with the expectation that subspace contributions to output(s) rapidly diminish with increasing dimension. 

%High dimensional model representation (HDMR) is a family of tools useful for efficiently capturing high-dimensional input-output behavior using low-dimensional projections \citep{alis99,hooker}. Mathematically, it is a non-orthogonal representation of $F=L^2(X,\mathscr{X},\mu)$ in terms of $2^n$ component function subspaces $\lbrace\mathcal{V}_u\rbrace$, \[F = \mathcal{V}_0\oplus\sum_i\mathcal{V}_i\oplus\sum_{i<j}\mathcal{V}_{ij}\oplus\dotsc\oplus\mathcal{V}_{1\dotsb n},\] so that \[f(x) = f_0 + \sum_i f_i(x_i) + \sum_{i<j}f_{ij}(x_i,x_j)+\dotsc+f_{1\dotsb n}(x_1,\dotsc,n_n).\] The collection $\lbrace f_u(x_u)\rbrace$ are known as \emph{component functions}. They are attained using the projection operators $\lbrace\mathscr{P}_u\rbrace$ where $f_u(x_u)\equiv\mathscr{P}_uf(x)$. 

\section{High dimensional model representation}\label{sec:hdmr} Writing $F=L^2(X,\mathscr{X},\nu)$, the HDMR of $f(x)\in F$ is the solution to the variational problem \[\min_{u}\norm{f(x) - u},\,\,\,\,\,u\in\mathcal{V}_0\oplus\sum_i\mathcal{V}_i\oplus\sum_{i_1<i_2}\mathcal{V}_{i_1i_2}\oplus\dotsc\oplus\sum_{i_1<\dotsb<i_l}\mathcal{V}_{i_1\dotsc i_l},\] where the norm $\norm{\cdot}$ is induced by the inner product as $\norm{\cdot}=\langle\cdot,\cdot\rangle^{1/2}$ and the $\lbrace\mathcal{V}_u\subset F: |u|\le l\rbrace$ are subspaces having certain null integral properties. It is uniquely minimized by \[u = \left(\mathscr{P}_0 + \sum_i\mathscr{P}_i + \sum_{i_1<i_2}\mathscr{P}_{i_1i_2} + \dotsb + \sum_{i_1<\dotsb<i_l}\mathscr{P}_{i_1\dotsc i_l}\right) f(x)\] using the collection of non-orthogonal projection operators $\lbrace\mathscr{P}_u\rbrace$. Putting \[\varepsilon_l(x) \equiv f(x) - u = f(x) - f_0 - \sum_i f_i(x_i) - \sum_{i_1<i_2}f_{i_1i_2}(x_{i_1},x_{i_2}) - \dotsb - \sum_{i_1<\dotsb<i_l}f_{i_1\dotsb i_l}(x_{i_1},\dotsc, x_{i_l}),\] we express \[\norm{f(x)-u} = \langle \varepsilon_l,\varepsilon_l\rangle^{1/2},\] which for scalar valued functions is \[\norm{f(x)-u}=\int_X\varepsilon_l^2(x)\D\nu(x).\]

% Putting $\mathcal{V}_u\equiv L^2(X_u,\mathscr{X}_u,\nu_u)$, we the unique solution to the variational problem, or equivalently with $f_0\in\mathcal{V}_0$, $f_i(x_i)\in\mathcal{V}_i$

The HDMR of $f\in F$ is written as \[f(x) = f_0 + \sum_i f_i(x_i) + \sum_{i_1<i_2}f_{i_1i_2}(x_{i_1},x_{i_2}) + \dotsc + f_{1\dotsc n}(x_1,\dotsc,x_n),\], a sum of $2^n$ \emph{component functions} \citep{alis99,hooker}. Although many additional component functions appear when compared to Kolmogorov's result, most of these are identically zero or insignificant for many $F$ of practical interest: it is often observed \[f(x) \backsimeq f^T(x) = f_0 + \sum_i f_i(x_i) + \sum_{i_1<i_2}f_{i_1i_2}(x_{i_1},x_{i_2}) + \dotsc + \sum_{i_1<\dotsc<i_T}f_{i_1\dotsc i_T}(x_{i_1},\dotsc,x_{i_T})\] for $T\ll n$, i.e. \emph{$f^T$ is a reduced-order representation of $f$}. Whenever $T<n$ and $f(x)\simeq f^T(x)$, we achieve a polynomial-scaling representation in $n$.  In most applications $T\le3$ is sufficient. In general, HDMR is expected to exhibit favorable convergence in $T$ whenever the input variables are meaningfully defined in relation to the output variable(s), typically the case with physical systems where input and output variables correspond to observable physical states. Some properties of HDMR:
\begin{enumerate}[label=(\roman*)]
\item it is unique given $\mu$
\item $\lbrace\mathcal{V}_u\rbrace$ are hierarchically orthogonal and partition variance, $\Var(f)=\sum_{u,v}\Cov\left(f_u,f_v\right)$
\item $\lbrace\mathcal{V}_u\rbrace$ maximize explained variance, i.e. no other hierarchically-orthogonal functions achieve higher explanatory variance than those belonging to $\lbrace\mathcal{V}_u\rbrace$
\item the elements of $\lbrace\mathcal{V}_u\rbrace$ can be attained through the action of projection operators $\lbrace\mathscr{P}_u\rbrace$ such that $f_u(x_u)\equiv\mathscr{P}_uf(x)$, i.e. the component functions are  $L^2$ projections of $f(x)$
\item all HDMR's converge at the same order: if $f^T(x)$ converges for $\mu$ with error $\mathcal{O}(\epsilon)$, then $f^T(x)$ converges for $\mu'$ with error $\mathcal{O}(\epsilon)$, although the constant can be substantially different
\item for a set of functions $\lbrace f^i(x)\rbrace$ obeying linear-superposition conservation laws, the corresponding HDMR's obey such for every $T$.
\end{enumerate}
The notion of hierarchical-orthogonality guarantees the existence and uniqueness of the decomposition such that $\Var(f)=\sum_{u,v}\Cov(f_u,f_v)$ \citep{hooker}. Note that hierarchical-orthogonality is a generalization of mutual orthogonality where functions are orthogonal only to functions on nested subspaces. For example, for $\mu_{123}\ne \mu_1\mu_2\mu_3$, hierarchical-orthogonality implies $\langle f_{12},f_1\rangle=\langle f_{12},f_2\rangle=0$ but neither implies $\langle f_1,f_2\rangle=0$ nor $\langle f_{12},f_3\rangle=0$.\\

Given the variance decomposition, sensitivity indices can be defined in terms of normalized variances and covariances. This is known as \emph{structural and correlative sensitivity analysis} (SCSA) \citep{li12}, where SCSA defines structural, correlative, and overall sensitivity indices $(\mathbb{S}_u^\text{a},\mathbb{S}_u^\text{b},\mathbb{S}_u)$ for each component function $u\subseteq\mathbb{N}_n$, \begin{align*}
\mathbb{S}_u^\text{a} &\equiv \frac{1}{\Var(f)}\Var(f_u)\\
\mathbb{S}_u^\text{b}&\equiv\frac{1}{\Var(f)}\sum_{v: v\ne u}\Cov(f_u,f_v)\\
\mathbb{S}_u&\equiv\mathbb{S}_u^\text{a}+\mathbb{S}_u^\text{b}, \\
\end{align*}
and these satisfy \[\sum_u\mathbb{S}_u = 1.\] Note that when $\mu\ne\prod_i\mu_i$, the projections $f_u=\mathscr{P}_u f$ are hierarchically-orthogonal with $\sum_u\mathbb{S}_u^\text{a}<1$ and $\sum_u(\mathbb{S}_u^\text{a}+\mathbb{S}^\text{b}_u)=1$. \\

When the input variables are independent, $\mu=\prod_i\mu_i$, the component functions are mutually orthogonal and can be recursively constructed, 
\begin{align}\label{eq:rec}
f_{i_1\dotsc i_l}(x_{i_1},\dotsc,x_{i_l}) \equiv&\,\textbf{M}^{i_1\dotsc i_l}f(x) - \sum_{j_1<\dotsb<j_{l-1}\subset\lbrace i_1,\dotsc,i_l\rbrace}f_{j_1\dotsc j_{l-1}}(x_{j_1},\dotsc,x_{j_{l-1}})\nonumber\\
& - \sum_{j_1<\dotsb<j_{l-2}\subset\lbrace i_1,\dotsc,i_l\rbrace}f_{j_1\dotsc j_{l-2}}(x_{j_1},\dotsc,x_{j_{l-2}}) - \dotsb\nonumber\\
&-\sum_{j\subset\lbrace i_1,\dotsc,i_l\rbrace}f_{j}(x_{j}) - f_0
\end{align}
where $\textbf{M}^{i_1\dotsc i_l}f(x) \equiv \int_{X_{-i_1\dotsc i_l}}f(x)\prod_{j\notin\lbrace i_1,\dotsc,i_l\rbrace}\D\mu_j(x_j)$ (note that \eqref{eq:rec} does not hold for correlated $\mu$). Because of the mutual orthogonality of the component functions, we have $\Cov(f_u,f_v)=0$ for $u\ne v$ such that $\sum_u\mathbb{S}^\text{a}_u=1$. This is known as global sensitivity analysis (GSA) \cite{Sobol93}.\\ %Presaging the ideas of this article, we see that the choice of $\mu$ regulates the properties of the HDMR of $f$ and its decomposition of variance. \\%Some formulations include ANOVA HDMR (ordinary Lebesgue $\mu=\leb$ on $X=[0,1]^n$), functional HDMR (Wiener $\mu$ for $F=\mathcal{L}(X)$ with $X=C^n[0,1]$), cut HDMR (Dirac $\mu=\delta_a$) \citep{alis99,anchored}, multi-cut HDMR (Dirac mixture) \citep{io1,outside1}, and so on. \\
%T, where 
%The representation is unique using the definitions of $\lbrace\mathcal{V}_u\rbrace$ for any non-degenerate $\mu$. \\

In regard to the projection operators $\lbrace\textbf{P}^u\rbrace$ and functionals $\lbrace\textbf{I}^u\rbrace$ of \emph{interpretative diagnostics}, we put for every $u$ \[{\color{black}\textbf{P}^u_\text{HDMR}} f(x) \equiv \mathscr{P}_u f(x) = {\color{black}f^\text{HDMR}_u(x_u)}\] and \[{\color{black}\textbf{I}^u_\text{HDMR}} f(x) \equiv (\mathbb{S}^\text{a}_u, \mathbb{S}^\text{b}_u, \mathbb{S}_u).\]

We provide three examples of HDMR in machine learning. 

\subsection{Eliminating information leakage in ``big-data'' settings}\label{sec:pgof} Consider an independency of iid random variables $\textbf{D}=\lbrace X_i\rbrace$ with distribution $\mu$ on $X$. Suppose $(X,\mathscr{X})$ is discrete and that $\textbf{D}$ is represented as an infinite double-array $[X_{nj}]$ of real-valued random variables (that is, for each $n\in\mathbb{N}$ there is a $k_n\in\mathbb{N}$ such that $X_{nj}=0$ for all $j>k_n$). The probability law of $X_n$ is given by $\mu_n\lbrace k\rbrace\equiv \mathbb{P}(X_n = k)$ for $k\in\lbrace1,\dotsc, k_n\rbrace$. We consider ``big-data'' to be the case of $k_n\rightarrow\infty$ as $n\rightarrow\infty$, i.e., the dimension of $X_n$ (denoted by $k_n$) increases to infinity as the sample size (denoted by $n$) increases so. Consider the goodness-of-fit Pearson chi-square (PGOF) statistic $\chi_n^2$. For fixed $k_n=k$ the statistic $\chi_n^2$ asymptotically follows the $\chi^2$-distribution with $(k-1)$ degrees of freedom and whenever $k$ is large the standardized statistic $(\chi^2_n-(k-1))/\sqrt{2(k-1)}$ is approximated by the standard Gaussian distribution. However, $k_n=k$ is not the case in ``big-data'' applications where $k_n\rightarrow\infty$ is observed as $n\rightarrow\infty$. As shown in \cite{rempala} the Gaussian approximation may or may not be valid for the doubly infinite case; furthermore it turns out that the asymptotic behavior of $\chi_n^2$ is characterized by a subset of its HDMR component functions. Employing the material in the appendices, we define a new statistic using this subset of component functions, denoted by $\chi_{n\text{HDMR}}^2$,  \begin{align*}\chi_{n\text{HDMR}}^2(x_1,\dotsc,x_n)&\equiv f_0+\sum_{i<j}f_{ij}(x_i,x_j) \\&\equiv \chi_n^2(x)-\sum_if_i(x_i),\end{align*} and compare its relative efficiency to $\chi_n^2$ for the power law distribution, \[\alpha\in[0,1),\,\,\,\mu_n\lbrace k\rbrace = (C_\alpha k^\alpha)^{-1},\] where \[C_\alpha=\sum_{k\in[k_n]}k^{-\alpha}\simeq k_n^{1-\alpha}/(1-\alpha).\] The HDMR statistic dominates the PGOF statistic.%This setting is common, as a wide variety of phenomena empirically follow a power law distribution. For example, it is observed in biomedical ``big data'' settings in the analysis of biodiversity in next-generation sequencing (NGS) data, including comparing T-cell receptor populations' diversities in transgenic mice or identifying transcriptomic profiles of hepatocellular carcinomica in humans \citep{rempala2}. 

\begin{figure}
\centering
\begingroup
\captionsetup[subfigure]{width=2in,font=normalsize}
\color{black}
\subfloat[$n=5000$ ($\lambda = 100$)\label{fig:est5000}]{\includegraphics[width = 4in]{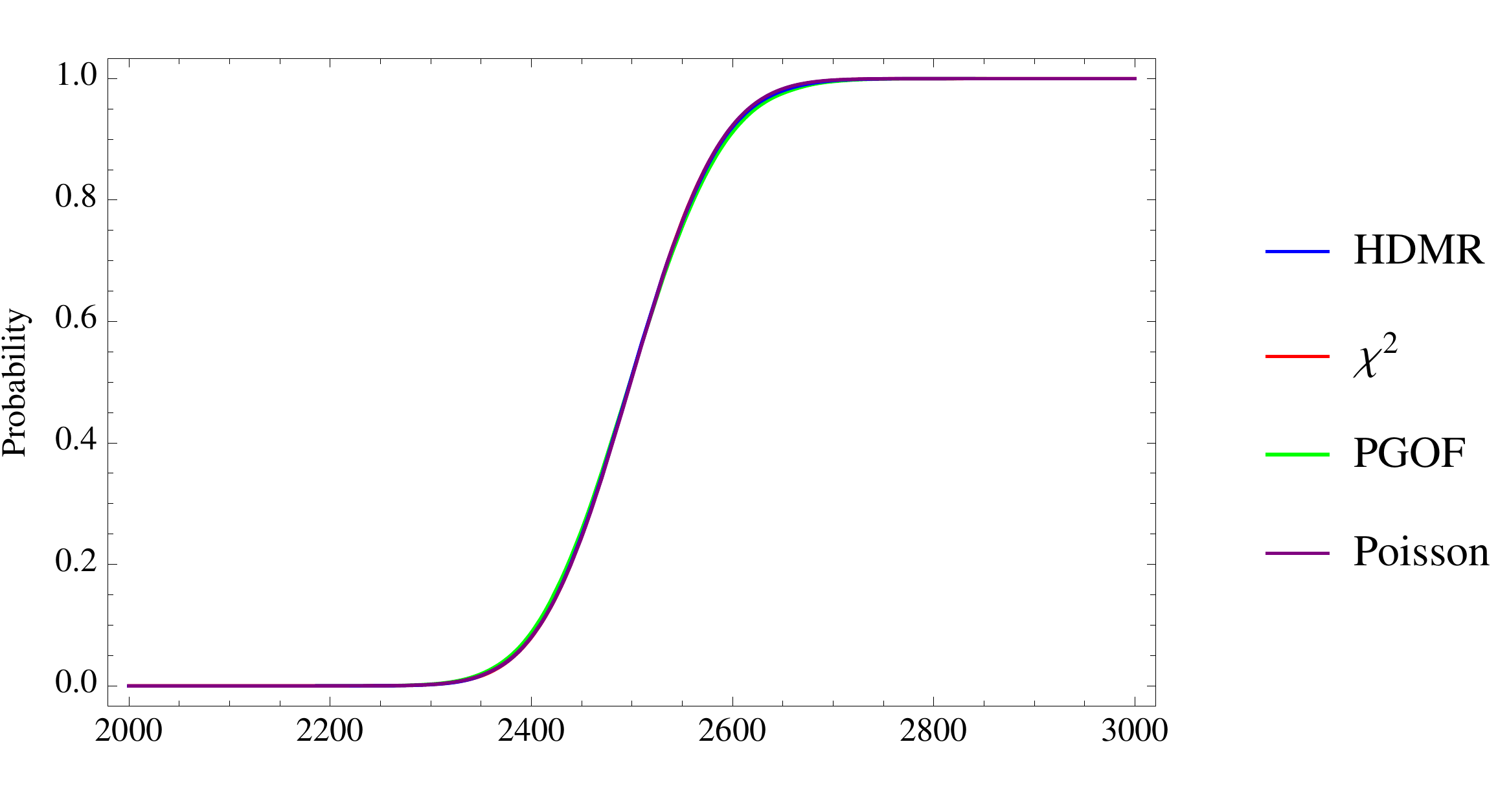}}\\
\subfloat[$n=500$ ($\lambda = 10$)\label{fig:est500}]{\includegraphics[width = 4in]{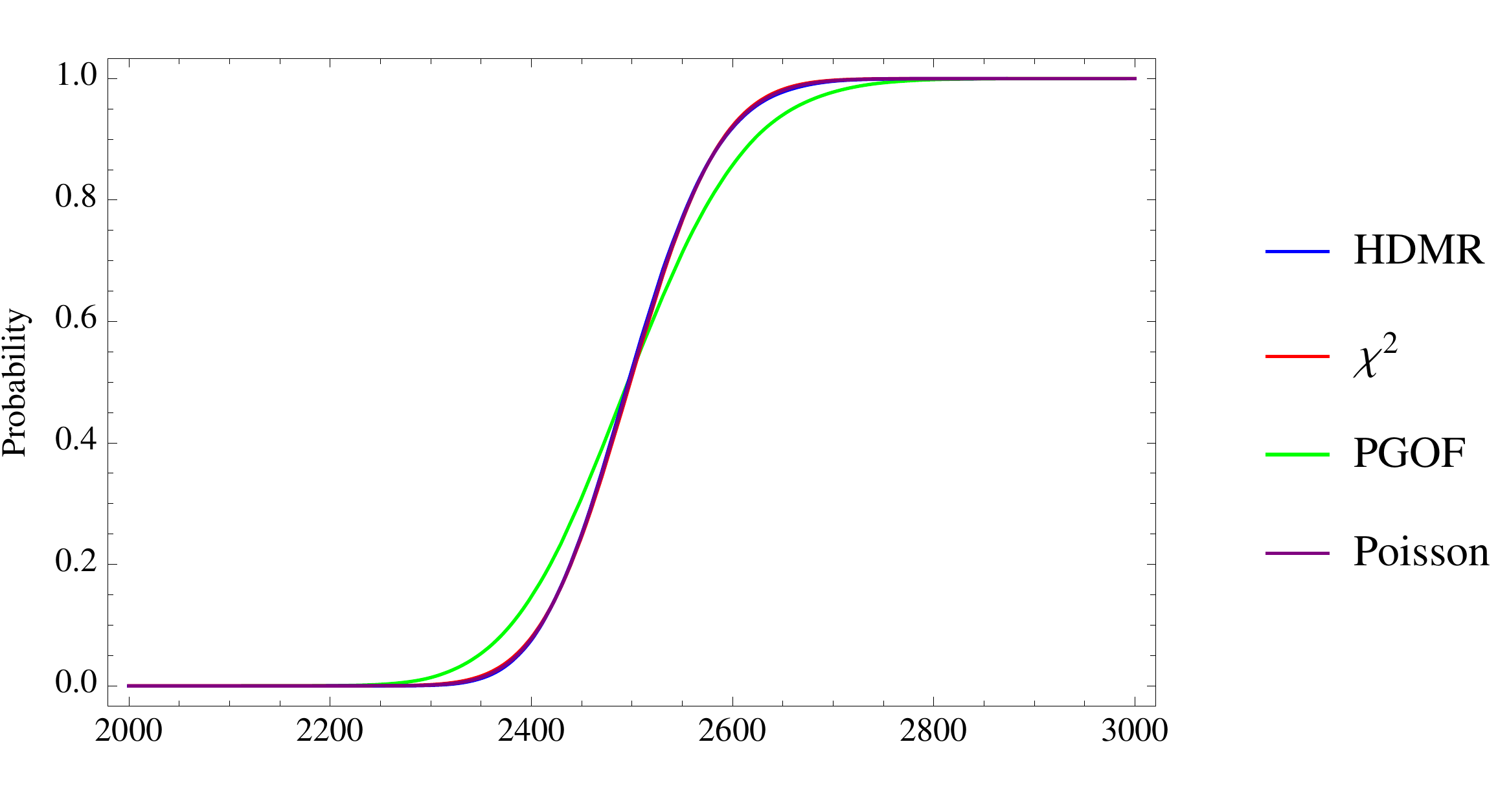}}\\
\subfloat[$n=50$ ($\lambda = 1$)\label{fig:est50}]{\includegraphics[width = 4in]{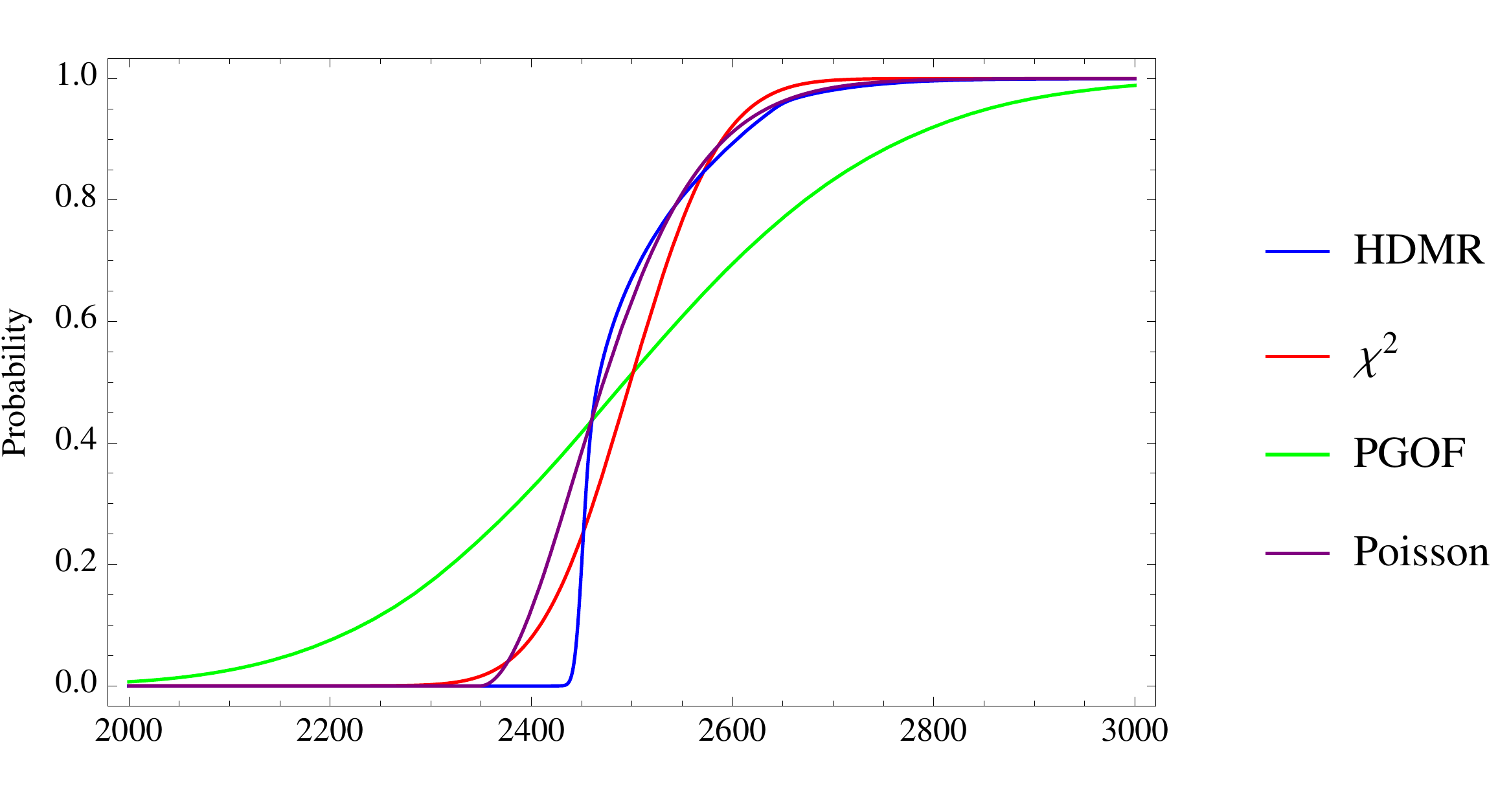}}\\
\endgroup
\caption{Empirical distributions for $k_n=2500$ and power law distribution with $\alpha=1/2$}\label{chi}
\end{figure}

In Figure~\ref{chi}, we exhibit empirical distributions for PGOF, HDMR chi-square, and chi-square for $k_n=2500$ and $n\in\lbrace50,500,5000\rbrace$, having corresponding $\lambda\in\lbrace1,10,100\rbrace$, for the power law distribution with $\alpha=1/2$. We show statistics for the estimators in Table~\ref{tab:est}. For $\lambda=100$, Figure~\ref{fig:est5000} shows both $\chi_n^2$ and $\chi_{n\text{HDMR}}^2$ are asymptotically $\chi^2$. For $\lambda = 10$, Figure~\ref{fig:est500} shows that $\chi_n^2$ is not asymptotically $\chi^2$ but $\chi_{n\text{HDMR}}^2$ is. For $\lambda = 1$, Figure~\ref{fig:est50} shows that neither $\chi_n^2$ nor $\chi^2_{n\text{HDMR}}$ is asymptotically $\chi^2$. In particular, $\chi^2_{n\text{HDMR}}$ is expressed in terms of a Poisson law (see appendices for a precise characterization), although observe that the empirical distribution is truncated due to undersampling of the power law tails. \textbf{The HDMR statistic $\chi_{n\text{HDMR}}^2$} dominates the PGOF statistic $\chi_n^2$, i.e. PGOF experiences \emph{information leakage} in comparison to the HDMR. For $\lambda = 1$ we observe PGOF to have a relative efficiency of $\sim10\%$ to that of HDMR. %For these and other settings we observe $\chi_{n}^2\ne\chi_{n\text{HDMR}}^2$ with the efficiency of $\chi_{n\text{HDMR}}^2$ greater than $\chi_{n}^2$. As a result, \textbf{the PGOF $\chi_n^2$ is said to exhibit \emph{information leakage}}. \\ 

\begin{table}[h]
\caption{Estimator statistics from $5\times 10^4$ simulations  for $k_n=2500$ and power law distribution with $\alpha=1/2$}
\label{tab:est}
\begin{center}
\begin{tabular}{ccccccc}
\toprule
&&&\mc{2}{$\mathbb{E}$}&\mc{2}{$\Var$}\\
\cmidrule(lr){4-5}\cmidrule(r){6-7}$n$&$k_n$&$\lambda\equiv\frac{n}{\sqrt{k_n}}$&$\chi_{n}^2$&$\chi_{n\text{HDMR}}^2$&$\chi_{n}^2$&$\chi_{n\text{HDMR}}^2$\\\midrule
5000 &2\,500&100 & 2499.20 &2499.18&5464.45&5054.05\\
500 &2\,500&10 & 2498.84&2498.83&9023.85&5032.56\\
50&2\,500&1 & 2499.30&2499.45&44188.40&5014.13\\
\bottomrule
\end{tabular}
\end{center}
\end{table}

\FloatBarrier

\subsection{Reduced-order representation of high-dimensional behavior}\label{sec:rom} \emph{Note that in this section we use $\nu$ to indicate the measure on $X$, and $\mu$ and $\sigma$ denote the mean and standard deviation common to the input variables.}\\

In this application we compute HDMR for polynomial models, in particular the function $\prod_{i=1}^n x_i$ and a sum of elementary symmetric polynomials in $n$ variables. We show that the coefficient of variation $\rho$ of the input distribution $\nu$ regulates the efficiency of low-dimensional HDMR approximations. We illustrate these properties for polynomial models. %\textbf{Reduced-order representations of high-dimensional polynomial models by HDMR constitute explanatory models for complex systems.} %These results demonstrate that high-degree polynomial systems can be effectively represented as low-degree polynomial systems for small enough $\rho$. %This has potential application as a trapdoor to public key schemes based on the problem of solving multivariate quadratic polynomials and the isomophism of polynomials problem, the trapdoor being masking of quadratic polynomials as high-degree polynomials using suitable $\rho$.

\begin{figure}[h]
\centering
\includegraphics[scale=0.55]{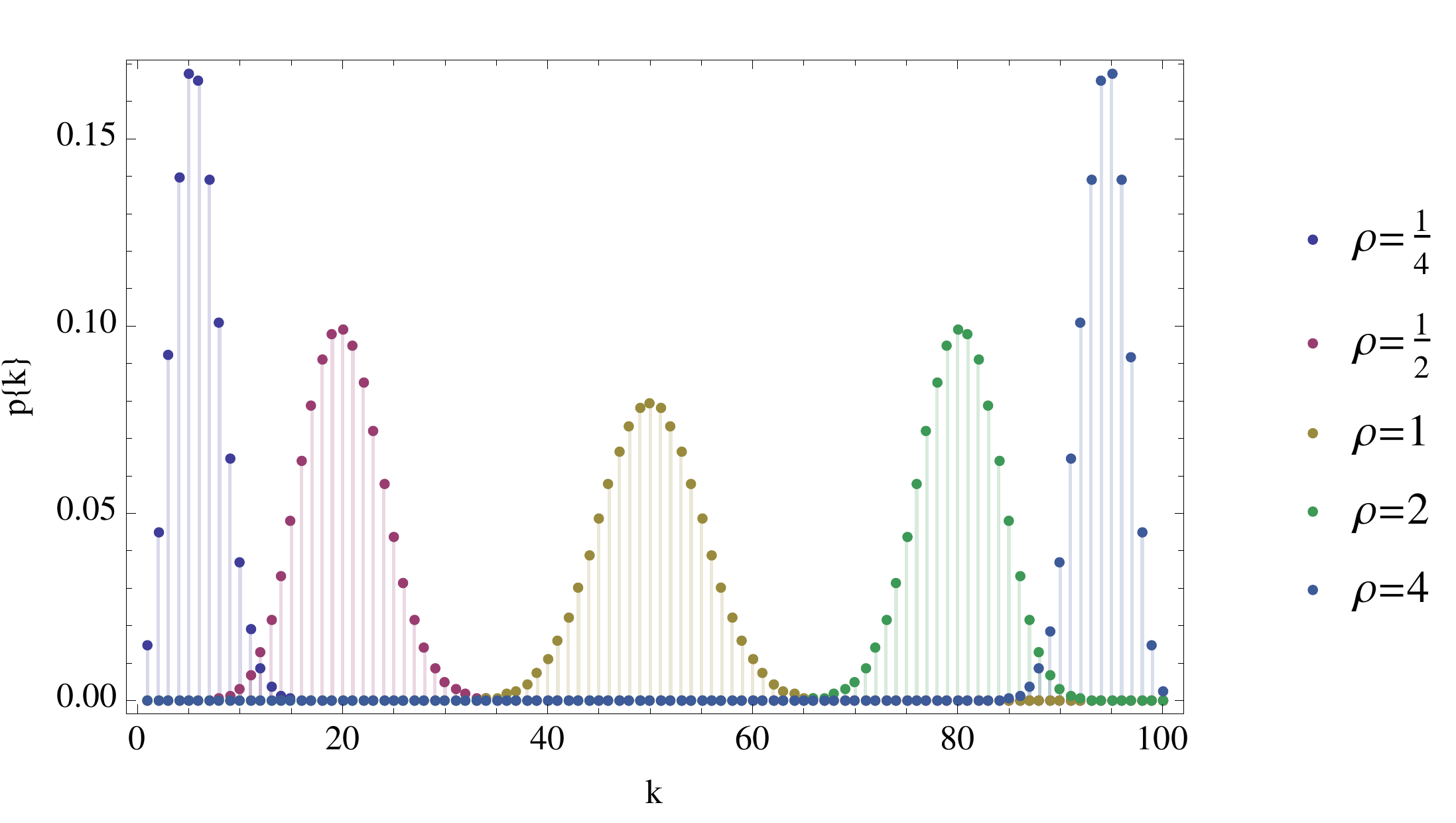}
\caption{$p\lbrace k\rbrace=\sum_{i_1<\dotsc<i_k}\mathbb{S}_{i_1\dotsc i_k}$}\label{fig:probrho}
\end{figure}

%\textcolor{black}{\[f(x) = f_0 + \sum_i f_i(x_i) + \sum_{i<j}f_{ij}(x_i,x_j) + \dotsc + f_{1\dotsc n}(x_1,\dotsc,x_n)\]}
%where
Consider \[f(x) = \prod_{i=1}^n x_i,\;\;\;\text{iid}\,x,\;\;\;\rho\equiv\sigma/\mu\ne 0\] with \[\Var\, f =\mu^{2n}\left(\left(1+\rho^2\right)^n-1\right). \]
Per \eqref{eq:rec} the component functions of $f(x)$ are 
\begin{align*}
f_0 &= \mathscr{P}_0f(x) = \mu^n\\%2^{-n}(a+b)^n\\
f_i(x_i) &= \mathscr{P}_if(x) =\mu^{n-1}x_i - f_0\\
f_{ij}(x_i,x_j)&=\mathscr{P}_{ij}f(x) = \mu^{n-2}x_ix_j - f_i(x_i) - f_j(x_j) - f_0\\
\vdots&.
\end{align*}
Employing the material in the appendices, the sensitivity indices satisfy \[\sum_{\substack{k\\i_1<\dotsb<i_k}}\mathbb{S}_{i_1\dotsb i_k} = 1\] and at each order follow
\begin{align*}
p\lbrace k\rbrace\equiv\sum_{i_1<\dotsb<i_k}\mathbb{S}_{i_1\dotsb i_k} &= \frac{ \binom{n}{k}\rho^{2 k}}{\left(\rho^2+1\right)^n-1}
\end{align*}
where $\sum_k p\lbrace k\rbrace = 1$. We have $\mathbb{E}\, p = \frac{n \rho^2 \left(\rho ^2+1\right)^{n-1}}{\left(\rho ^2+1\right)^n-1}$ and $\Var\, p = \frac{n \rho ^2 \left(\rho ^2+1\right)^{n-2} \left(\left(\rho ^2+1\right)^n-n \rho ^2-1\right)}{\left(\left(\rho ^2+1\right)^n-1\right)^2}$. \textbf{Observe that when \bm{$\rho<1$} the explained variance of low dimensional approximations increases.} For context, the uniform distribution on the unit interval (maximum entropy on $[a,b]$) has $\rho(\text{Unif}[0,1])=\frac{\sqrt{3}}{3}\approx 0.58$. In Figure~\ref{fig:probrho}, we plot the percent of explained variance by subspace order $\bm{p}\equiv(p\lbrace k\rbrace: k\in\mathbb{N}_n)$ for $n=100$ and $\rho\in\lbrace\frac{1}{4},\frac{1}{2},1,2,4\rbrace$. We observe that for $\rho<1$ the probability mass of the probability vector $\bm{p}$ is concentrated about small $k$. Similar results are observed for sums of elementary symmetric polynomials in $n$ variables \[E_n(x) = 1+\sum_{\substack{k\in\mathbb{N}_n\\i_1<\dotsb<i_k}}x_{i_1}\dotsb\,x_{i_k},\] where $\tau\equiv\frac{\sigma}{1+\mu}$ regulates efficiency, and are contained in the appendices. 

\FloatBarrier

\begin{figure}
\centering
\begingroup
\captionsetup[subfigure]{width=2in,font=normalsize}
\color{black}
\subfloat[$\rho=\frac{1}{4}$\label{fig:rho14}]{\includegraphics[width = 3.5in]{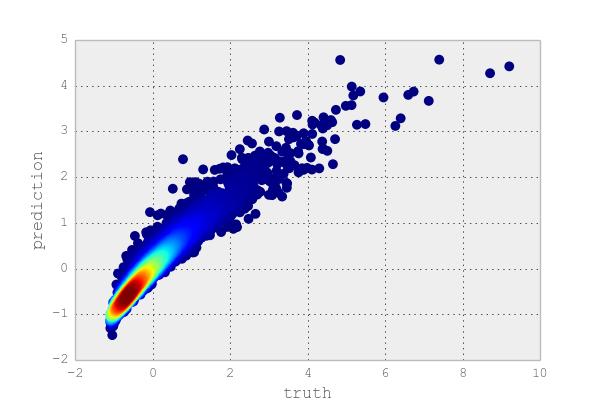}}\\
%\subfloat[Structural, $\mathbb{S}^\text{a}_{12}$\label{fig:poly:b}]{\includegraphics[width = 3in]{../../../Dissertation/corr12a.pdf}}\\
\subfloat[$\rho=\frac{1}{2}$\label{fig:rho14}]{\includegraphics[width = 3.5in]{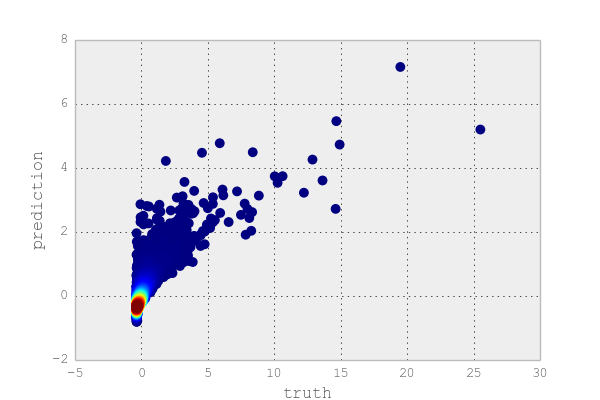}}\\
%\subfloat[Correlative, $\mathbb{S}^\text{b}_{12}$\label{fig:poly:b}]{\includegraphics[width = 3in]{../../../Dissertation/corr12b.pdf}}\\
\subfloat[$\rho=1$\label{fig:rho14}]{\includegraphics[width = 3.5in]{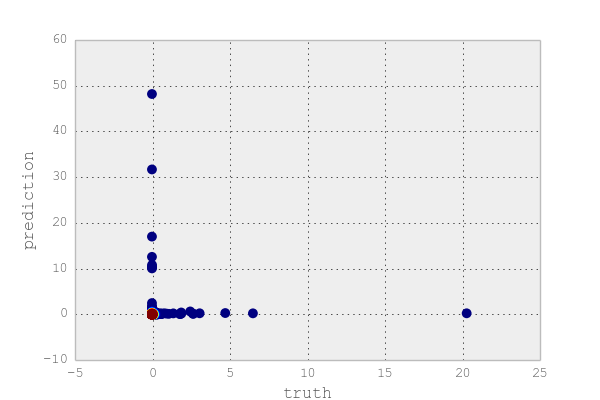}}\\
%\subfloat[Overall, $\mathbb{S}_{12}$\label{fig:poly:12}]{\includegraphics[width = 3in]{../../../Dissertation/corr12.pdf}}\\
\endgroup
\caption{Density truth plots for test data for $\rho\in\lbrace\frac{1}{4},\frac{1}{2},1\rbrace$}\label{fig:boot_tp}
\end{figure}

We demonstrate this effect through Monte Carlo estimation of mean-squared-error test performance using the gradient boosting regressor (GBR) machine (Monte Carlo cross-validation), i.e. the generalization ability of the learning algorithm. We take $n=10$ and form $\textbf{D}=\lbrace(x_i,f(x_i))\rbrace_{i=1}^{10^4}$ from \emph{iid} random variables from a $\text{Beta}(a,b)$ distribution on $X=[0,1]^n$. For the beta distribution, we take $\rho\in\lbrace\frac{1}{4},\frac{1}{2},1\rbrace$ using $(a,b)\in\lbrace(\frac{15}{2},\frac{15}{2}),(\frac{3}{2},\frac{3}{2}),(\frac{1}{2},\frac{3}{2})\rbrace$. We take a tree depth of six with $5\times10^3$ trees. We split the data into training and test data. This is performed for 50 independent estimates, where $(\textbf{D}_i)$ are independently standardized using training data. Mean and standard deviation Monte Carlo estimates are shown below in Table~\ref{tab:boot} and density truth plots in Figure~\ref{fig:boot_tp} (inverse-transformed): MSE markedly increases as $\rho$ increases.

%The choice of $\nu$ changes $T$ such that $f\simeq f^T$ for a given percent of explained variance. Note that this is not inconsistent with the claim that all HDMR's converge at the same order: the constant for a percent of explained variance in this example depends upon $\rho$. 

\begin{table}[h]
\caption{Monte Carlo MSE mean and standard deviation in $p$}
\label{tab:boot}
\begin{center}
\begin{tabular}{lcccccc}
\toprule
$\rho$ & Mean  & Std \\\midrule
1/4 & 0.1207 & 0.0145\\
1/2 & 0.3866 &0.1160 \\
1 & 10.1248& 45.55 \\\bottomrule
\end{tabular}
\end{center}
\end{table}

\FloatBarrier

\begin{figure}[h]
\centering
\includegraphics[scale=0.55]{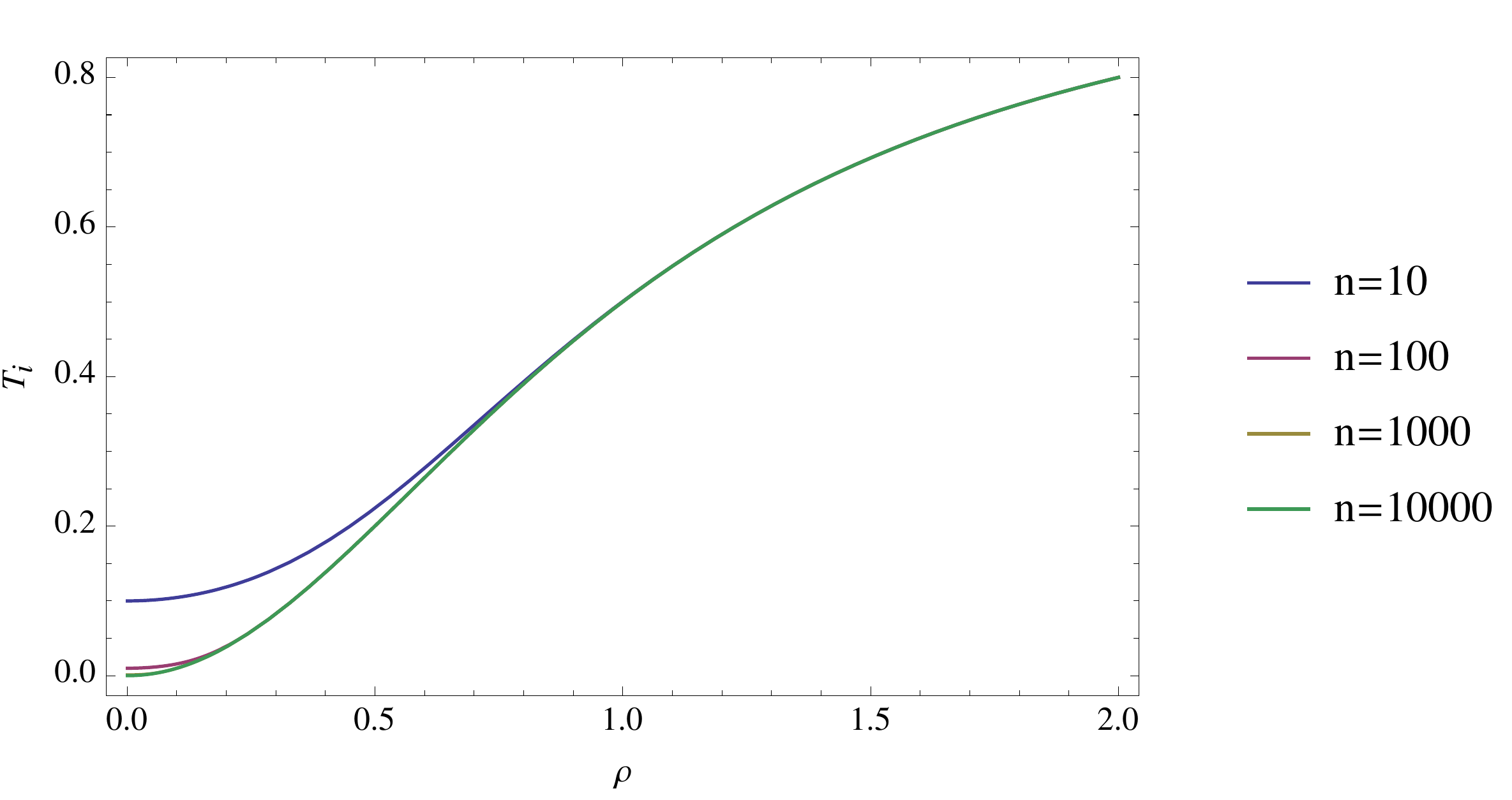}
\caption{$T_i$ for $(n,\rho)\in\lbrace10,100,1000,10000\rbrace\times[10^{-4},1]$}\label{fig:total_si}
\end{figure}

The total variable importance of $x_i$ to $f$ is computed as \[T_i=\sum_{k}\frac{\rho ^{2 i} \binom{n-1}{k-1}}{\left(\rho ^2+1\right)^n-1} = \frac{\rho^2}{\rho^2+1}\frac{\left(\rho ^2+1\right)^{n}}{\left(\rho ^2+1\right)^n-1}\approx\frac{\rho^2}{\rho^2+1}.\] Observe that \textbf{$T_i$ is independent of the value of $n$ whenever $n>k$ for sufficiently large $k$}. This is depicted in Figure~\ref{fig:total_si} and observed for $k\sim 100$. Relative variable importance is trivially $R_i=n^{-1}$ and is independent of $\rho$. 

%For comparison we bootstrap average variable importance as estimated from a gradient boosting regressor (GBR) machine for the uniform distribution on the unit hypercube ($\rho=\sqrt{3}/{3}$) and $n\in\lbrace10,100,1000\rbrace$. We take thirty bootstrap estimates.\\

\FloatBarrier

\subsection{Analysis of variance for correlated input variables}\label{simple} \emph{Note that in this section and subsequent sections, we use $\mu$ to indicate the probability measure on $(X,\mathscr{X})$.}\\

In this application we show how \textbf{variable importance $\text{I}(f)$ and variable dependence $\text{P}(f)$ by HDMR depend upon correlation $\rho$ in the input variables and this is necessary to preserve model variance}. This provisions an analysis of variance that is consistent for correlated or degenerate input variables: total variance is preserved for all values of $\rho$.\\

Let \[F=L^2(\mathbb{R}^n,\mathcal{B}(\mathbb{R}^n),\mu)=\Pi^n\equiv\text{Span}\lbrace x^\alpha: |\alpha|\le n, \,\alpha\in\mathbb{N}_0^{n}\rbrace\] where $\alpha = (\alpha_1,\dotsc)$, $|\alpha|=\sum_i\alpha_i$, and $\mu$ is Gaussian. In particular, take $n=2$ and let \begin{equation}\label{poly} f(x) = \beta_0 + \beta_1x_1 + \beta_2x_2 + \beta_{12}x_1x_2.\end{equation} Its projections $\lbrace f_u\gets\mathscr{P}_u f\rbrace$ decompose $f$, \[f(x)=f_0+f_1(x_1)+f_2(x_2)+f_{12}(x_1,x_2),\]
where

\resizebox*{\textwidth}{!}{
\begin{minipage}{\linewidth}
\begin{align*}
f_0 &= \beta_{0}+\beta_{1} \mu_1+\beta_{2} \mu_2+\beta_{12} \mu_1 \mu_2+\beta_{12} \rho \sigma_1 \sigma_2\\
f_1(x_1) &= (x_1-\mu_1) (\beta_{1}+\beta_{12} \mu_2)+\left(\frac{\rho}{\rho^2+1}\right)\frac{\beta_{12} \sigma_2}{\sigma_1}\left((x_1-\mu_1)^2-\sigma_1^2\right)\\
f_2(x_2) &=(x_2-\mu_2) (\beta_{2}+\beta_{12} \mu_1)+\left(\frac{\rho}{\rho^2+1}\right)\frac{\beta_{12} \sigma_1}{\sigma_2}\left((x_2-\mu_2)^2-\sigma_2^2\right)\\
f_{12}(x_1,x_2)&= \frac{\beta_{12} \left(-\rho \sigma_2^2 \left(\left(\rho^2-1\right) \sigma_1^2+(x_1-\mu_1)^2\right)+\left(\rho^2+1\right) \sigma_1 \sigma_2 (x_1-\mu_1) (x_2-\mu_2)+\rho \sigma_1^2 \left(-(x_2-\mu_2)^2\right)\right)}{\left(\rho^2+1\right) \sigma_1 \sigma_2}.\\
\end{align*}
\end{minipage}}
Computational details are contained in the appendices and these results are special cases of those in \cite{Li2014}.  Observe that HDMR projections contain elements not observed in $f$ and that these additional elements are functions on $X$ that depend upon the parameters of $\mu$. For example, \textbf{\bm{$f_1(x_1)$} contains a quadratic term in \bm{$x_1$}, whereas \bm{$f(x)$} does not.} The sensitivity indices of $f$ exhibit complex dependence on the model coefficients and distribution parameters,

\resizebox*{\textwidth}{!}{
\begin{minipage}{\linewidth}
\begin{align*}
\mathbb{S}^\text{a}_1 &= \frac{\sigma_1^2 \left((\beta_{1}+\beta_{12} \mu_2)^2+\frac{2 \beta_{12}^2 \rho^2 \sigma_2^2}{\left(\rho^2+1\right)^2}\right)}{2 \rho \sigma_1 \sigma_2 (\beta_{1}+\beta_{12} \mu_2) (\beta_{2}+\beta_{12} \mu_1)+\sigma_1^2 (\beta_{1}+\beta_{12} \mu_2)^2+\sigma_2^2 \left((\beta_{2}+\beta_{12} \mu_1)^2+\beta_{12}^2 \left(\rho^2+1\right) \sigma_1^2\right)}\\
\mathbb{S}^\text{b}_1 &= \frac{\rho \sigma_1 \sigma_2 \left(\left(\rho^2+1\right)^2 (\beta_{1}+\beta_{12} \mu_2) (\beta_{2}+\beta_{12} \mu_1)+2 \beta_{12}^2 \rho^3 \sigma_1 \sigma_2\right)}{\left(\rho^2+1\right)^2 \left(2 \rho \sigma_1 \sigma_2 (\beta_{1}+\beta_{12} \mu_2) (\beta_{2}+\beta_{12} \mu_1)+\sigma_1^2 (\beta_{1}+\beta_{12} \mu_2)^2+\sigma_2^2 \left((\beta_{2}+\beta_{12} \mu_1)^2+\beta_{12}^2 \left(\rho^2+1\right) \sigma_1^2\right)\right)}\\
\mathbb{S}^\text{a}_{12} &= \frac{\beta_{12}^2 \left(\rho^2-1\right)^2 \sigma_1^2 \sigma_2^2}{\left(\rho^2+1\right) \left(2 \rho \sigma_1 \sigma_2 (\beta_{1}+\beta_{12} \mu_2) (\beta_{2}+\beta_{12} \mu_1)+\sigma_1^2 (\beta_{1}+\beta_{12} \mu_2)^2+\sigma_2^2 \left((\beta_{2}+\beta_{12} \mu_1)^2+\beta_{12}^2 \left(\rho^2+1\right) \sigma_1^2\right)\right)}\\
\mathbb{S}^\text{b}_{12} &= 0,\\
\end{align*}
\end{minipage}
}
where \[\mathbb{S}_i = \mathbb{S}^\text{a}_i + \mathbb{S}^\text{b}_i,\,\,\,i\in\lbrace1,2,12\rbrace\] and satisfy for general parameters \[ \mathbb{S}_1 + \mathbb{S}_2 + \mathbb{S}_{12} = 1.\] Note that when $\rho=0$, then $\mathbb{S}^\text{b}_1=\mathbb{S}^\text{b}_2=\mathbb{S}^\text{b}_{12}=0$. \textbf{All non-trivial projections depend upon \bm{$\rho$} and \bm{$\beta$}}. In other words, the projections are functions of the measure $\mu$ and of the model $f$. Note that $\mathbb{S}^\text{b}_{12}=0$ for any $\rho$ due to the hierarchical-orthogonality of $f_{12}$, e.g., for every $\rho$ we have $\langle f_{12},f_1\rangle=\langle f_{12},f_2\rangle=0$ but $\langle f_1,f_2\rangle=0$ only for $\rho=0$.\\ 

\begin{figure}
\centering
\begingroup
\captionsetup[subfigure]{width=2in,font=normalsize}
\color{black}
\subfloat[Structural, $\mathbb{S}^\text{a}_1+\mathbb{S}^\text{a}_2$\label{fig:poly:a}]{\includegraphics[width = 3in]{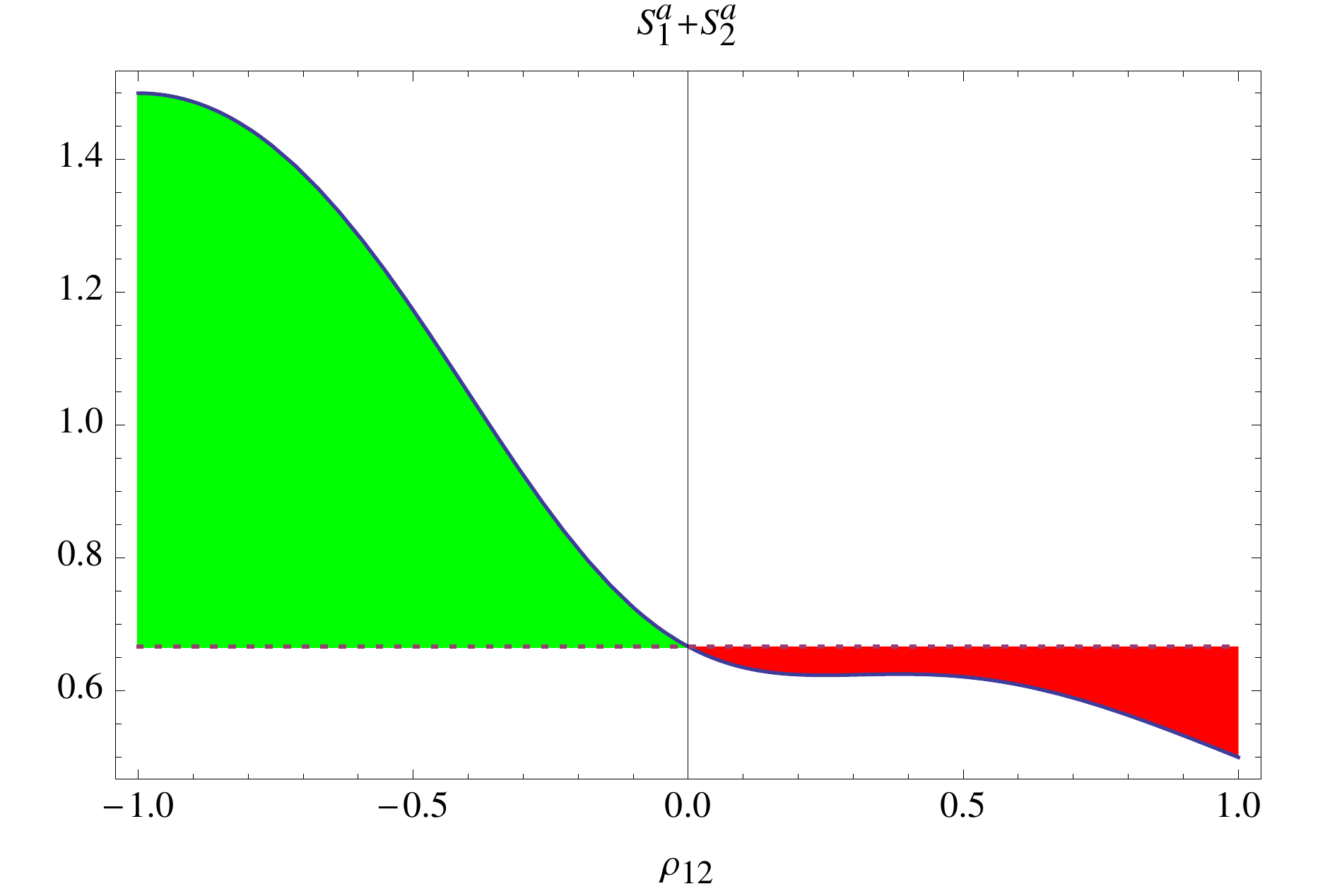}}
\subfloat[Structural, $\mathbb{S}^\text{a}_{12}$\label{fig:poly:b}]{\includegraphics[width = 3in]{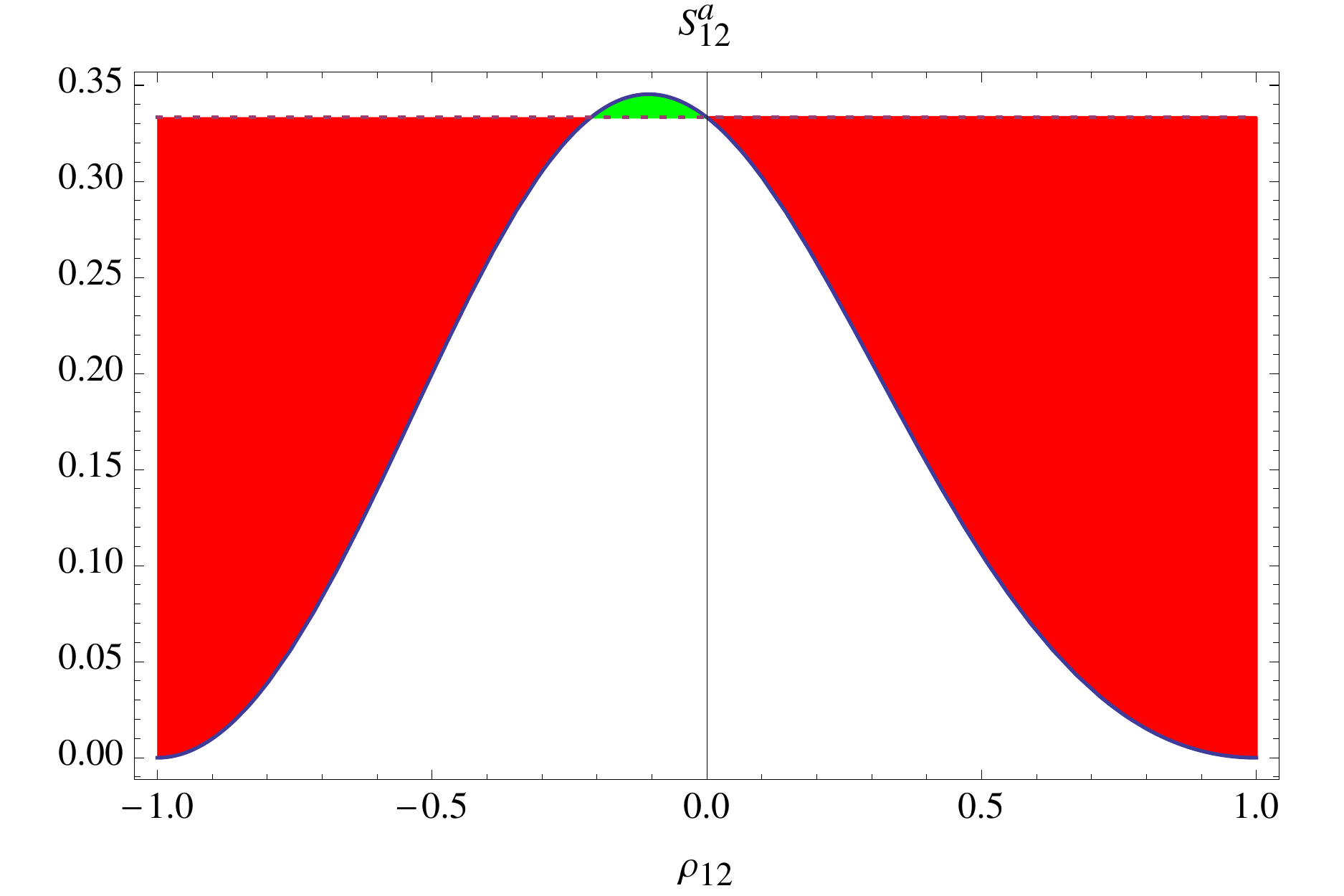}}\\
\subfloat[Correlative, $\mathbb{S}^\text{b}_1+\mathbb{S}^\text{b}_2$\label{fig:poly:a}]{\includegraphics[width = 3in]{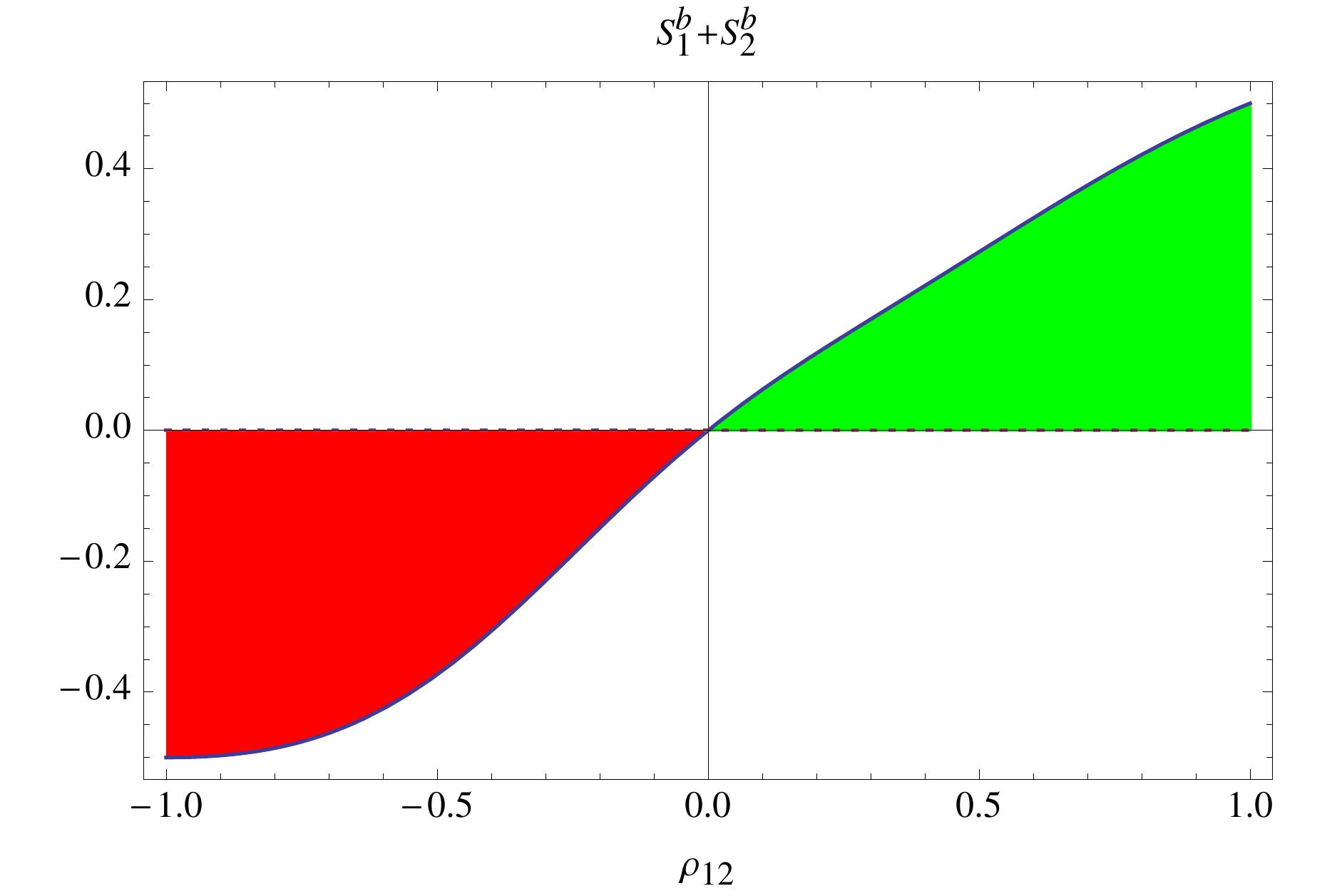}}
\subfloat[Correlative, $\mathbb{S}^\text{b}_{12}$\label{fig:poly:b}]{\includegraphics[width = 3in]{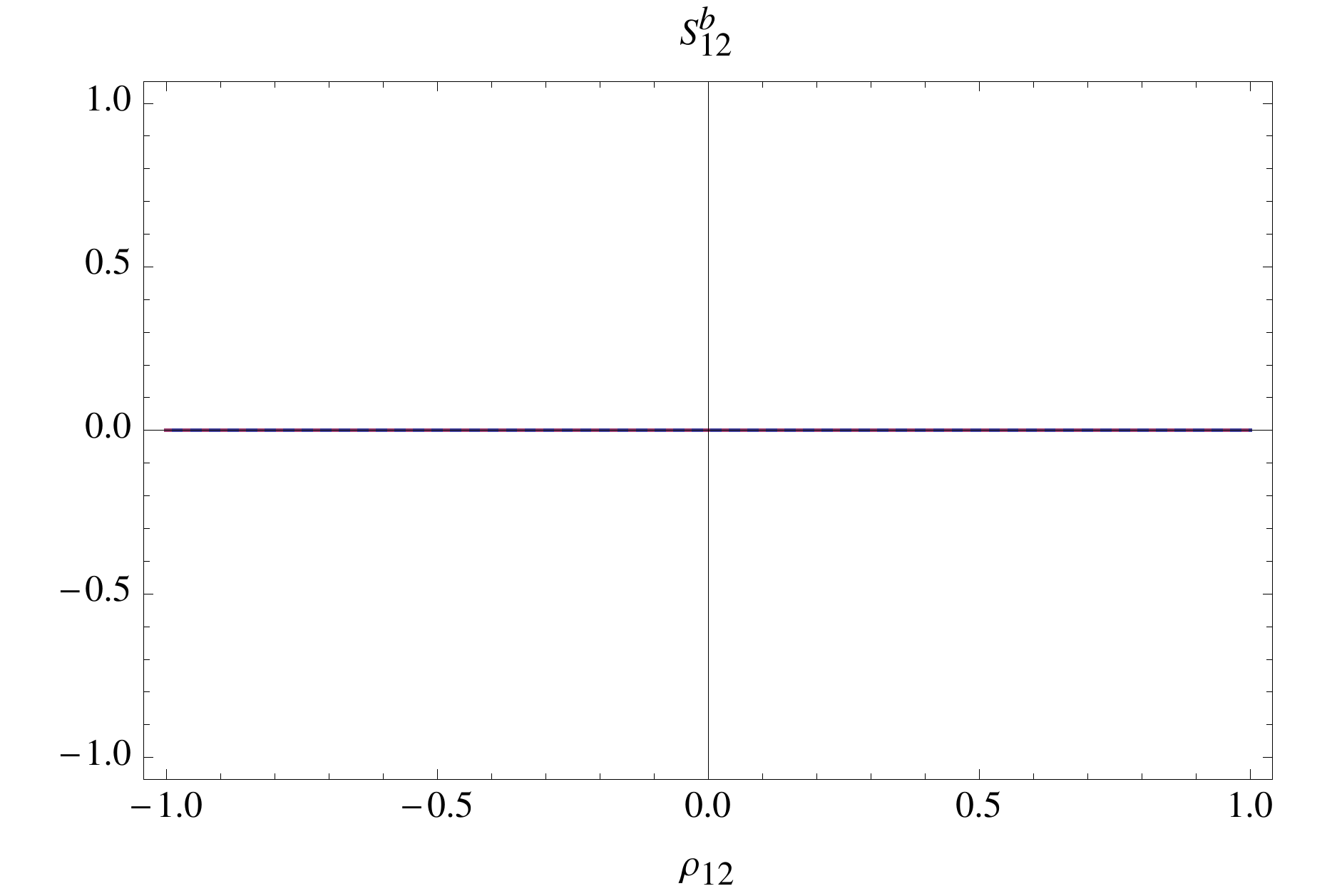}}\\
\subfloat[Overall, $\mathbb{S}_1+\mathbb{S}_2$\label{fig:poly:a}]{\includegraphics[width = 3in]{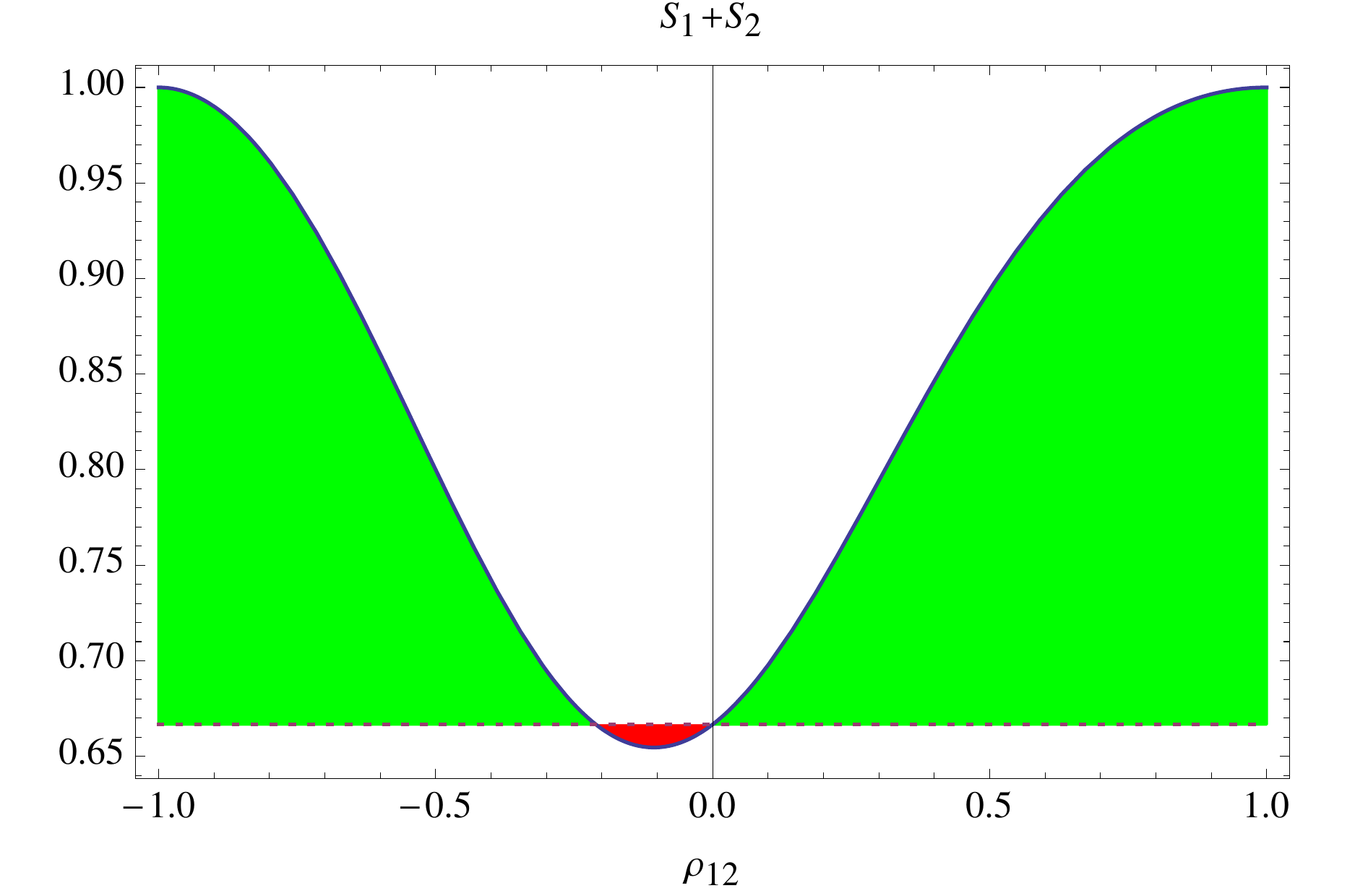}}
\subfloat[Overall, $\mathbb{S}_{12}$\label{fig:poly:12}]{\includegraphics[width = 3in]{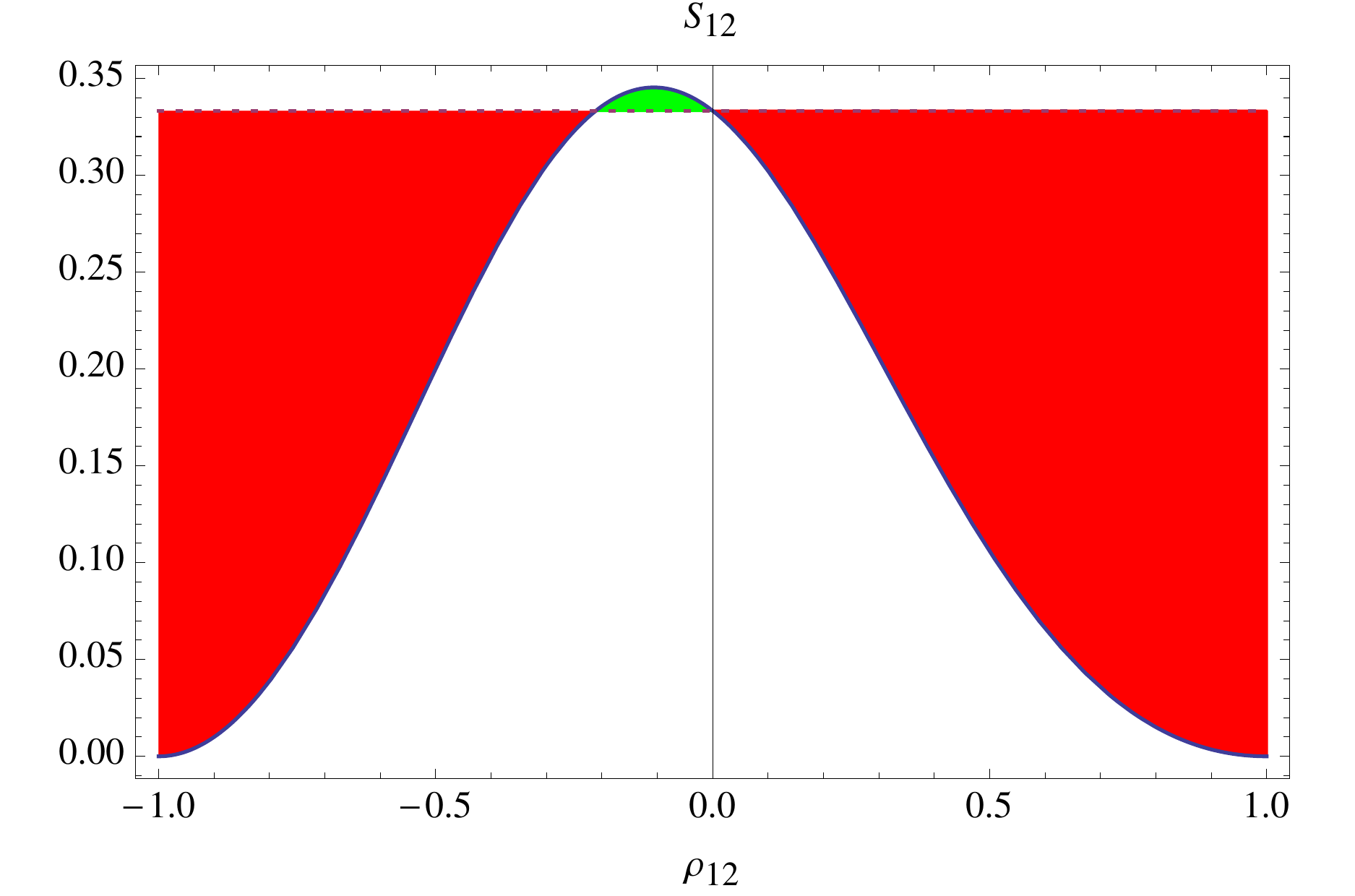}}\\
\endgroup
{\color{black}\caption{sensitivity indices for $\beta=\bm{1}$, $\mu=\bm{0}$ and $\sigma=\bm{1}$ as a function of $\rho$, colored to reflect the correlative contributions (\textcolor{green}{positive} and \textcolor{red}{negative})}\label{corr}}
\end{figure}

Consider $\rho\rightarrow 1$. We know from the direct form of $f(\beta,x)$ that in the limit we have $x_1=x_2$ and $f(\beta,x)=\beta_0+(\beta_1+\beta_2)x + \beta_{12}x^2$. That is, $f$ degenerates from a two-dimensional function into a one-dimensional function as the coherence indicated by $\rho$ achieves unity. HDMR uniquely captures this behavior in a variance-preserving manner for general distributions. We illustrate this property in Figure~\ref{corr} with plots of the sensitivity indices as functions of $\rho$ for fixed $\mu$, $\sigma$ and $\beta$. Colored regions indicate increased variance (green) or reduced variance (red) relative to $\rho=0$. For example, Figure~\ref{fig:poly:12} illustrates that the HDMR component function subspace in $X_1\times X_2$ experiences annihilation for $|\rho|\rightarrow 1$.\\

\FloatBarrier

In the next section we make a study of \eqref{poly} using other interpretative diagnostics in machine learning and compare to the results of this section.

\section{Interpretative diagnostics}\label{sec:id} Interpretative diagnostics provide explanation for prediction \citep{guyon}. Many common interpretative diagnostics are dependent upon the choice of $F$  and are said to be class-dependent. For example, variable depth in decision nodes of ensembles of decision trees and node weight pruning in neural networks are commonly employed for supervised machine learning settings to attain measures of variable importance. Other interpretative diagnostics are defined in terms of an objective function $J$. For suitable $J$ finite differences or derivative-based sensitivity analysis are conducted to assess the importance of various subsets of variables. We use the risk function $R$ as the objective function $J$. We compare variable importance and dependence measures using $\textbf{I}^u$ and $\textbf{P}^u$ defined by partial dependence \citep{friedman2001}, derivative-based global sensitivity indices \citep{dgsm1}, and HDMR for the model \eqref{poly}. We illustrate that HDMR preserves explained variance independent of correlation in $\mu$, whereas partial dependence is valid for modest correlation ($\rho<0.3$). 

\subsection{Partial dependence}The partial dependence of $f(x)$ on $x_u$ can be defined a couple ways \citep{friedman2001}. One is given by \begin{equation}\label{pd1}{\color{black}\textbf{P}^u_\text{PD}} f(x) \equiv \textbf{M}^u f(x)=\int_{X_{-u}}f(x)\D\mu_{-u}(x_{-u}) = {\color{black}f^\text{PD}_u(x_u)},\end{equation} which is useful whenever the dependence between $x_u$ and $x_{-u}$ is not too strong. Another formulation of partial dependence is given by \begin{equation}\label{pd2}{\color{black}\textbf{P}^u_\text{PD}} f(x) \equiv \textbf{N}^u f(x)=\int_{X_{-u}} f(x)\D\mu_{-u|u}(x_{-u}) = {\color{black}f^\text{PD}_u(x_u)},\end{equation} which considers the effect of dependencies. Here, the function is averaged with respect to the conditional distribution. If $\tilde{f}(x)=f(x)-f_0$ and the inputs are independent, then partial dependence is related to HDMR, where $\textbf{P}^i_\text{PD}\tilde{f}(x)=\textbf{P}^i_\text{HDMR}\tilde{f}(x)=\tilde{f}_i^\text{HDMR}(x_i)$ for $i\in\mathbb{N}_n$. Because of this relationship, we define the importance functional for partial dependence in a manner similar to HDMR, \[{\color{black}\textbf{I}^i_\text{PD}} f(x) \equiv(S^{a\,\text{PD}}_i, S^{b\,\text{PD}}_i, S_i^\text{PD}).\] 

\subsection{Derivative-based global sensitivity measures} \cite{dgsm1} introduced derivative-based global sensitivity measures. Weighted derivative-based global sensitivity measures (DGSM) are defined as \[{\color{black}\textbf{I}^u_\text{DGSM}}f(x)\equiv  \int_X (D^{\alpha_u}f(x))^2w_u(x_u)\D\mu(x) = {\color{black}S_u^\text{DGSM}},\] where \[D^{\alpha_u}f(x) = \frac{\partial^{|\alpha|} f(x)}{\partial x_1^{\alpha_{u1}}\dotsb\partial x_n^{\alpha_{un}}}\] and $w_u(x_u)$ is a weight function. Putting $\bm{D^\alpha f}=(D^{\alpha_u}f(x): u)$, a normalized DGSM is given as \[{\color{black}\textbf{I}^u_\text{DGSM}}f(x)\equiv\frac{ \int_X (D^{\alpha_u}f(x))^2w_u(x_u)\D\mu(x)}{\norm{\bm{D^\alpha f}}_{2;F}^2} = {\color{black}S_u^\text{DGSM}},\] where \[\norm{\bm{f}}_{2;F}=\left(\sum_{f\in\bm{f}}\norm{f}_{L^2(\Omega,\mathscr{F},\mu)}^2\right)^{1/2}.\] We take $w_u(x_u)=1$. \\

\subsection{Illustration} 
\begin{figure}
\centering
\begingroup
\captionsetup[subfigure]{width=5in,font=normalsize}
\color{black}
\subfloat[Structural, $S^\text{a}_1+S^\text{a}_2$\label{fig:poly:a}]{\includegraphics[width = 5in]{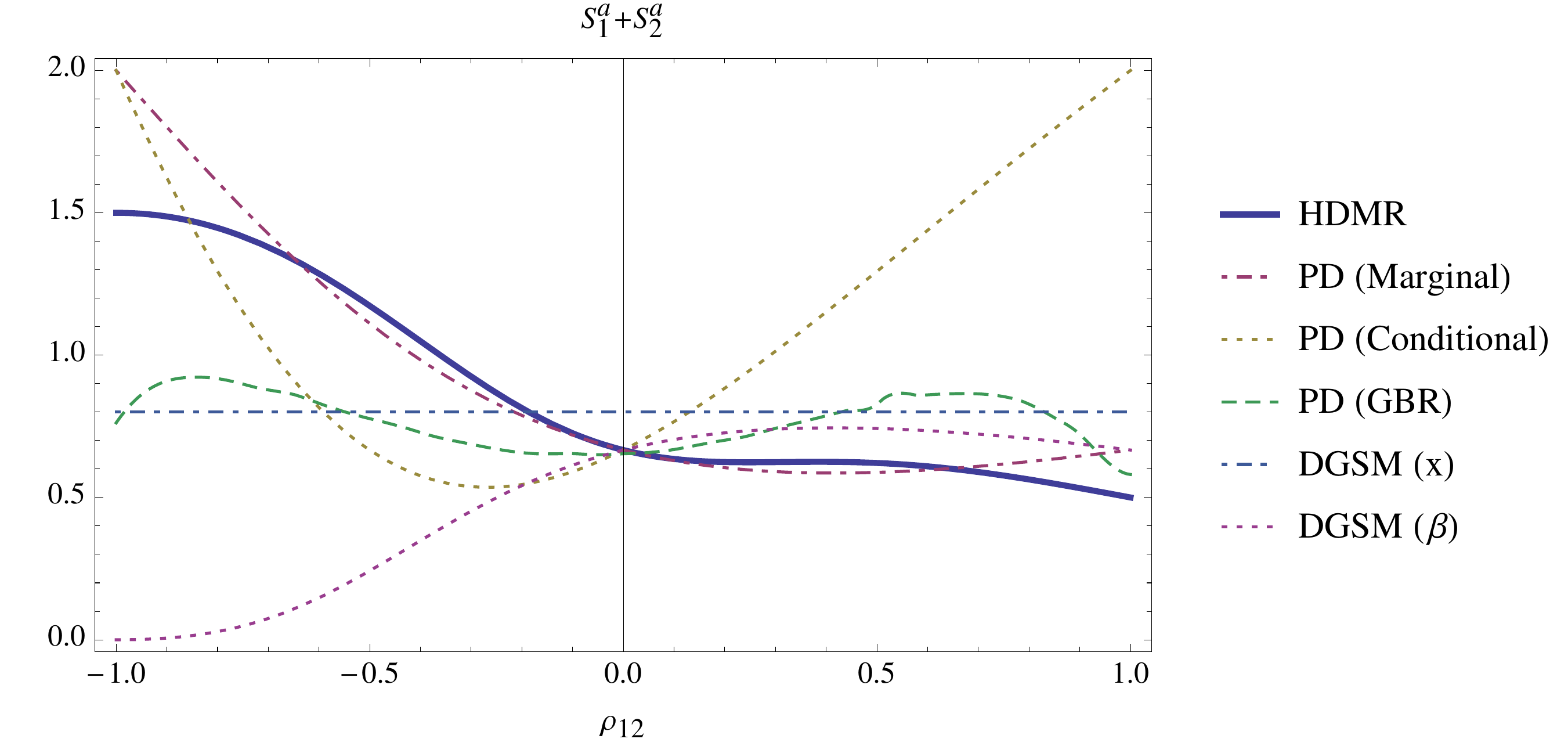}}\\
\subfloat[Correlative, $S^\text{b}_1+S^\text{b}_2$\label{fig:poly:b}]{\includegraphics[width = 5in]{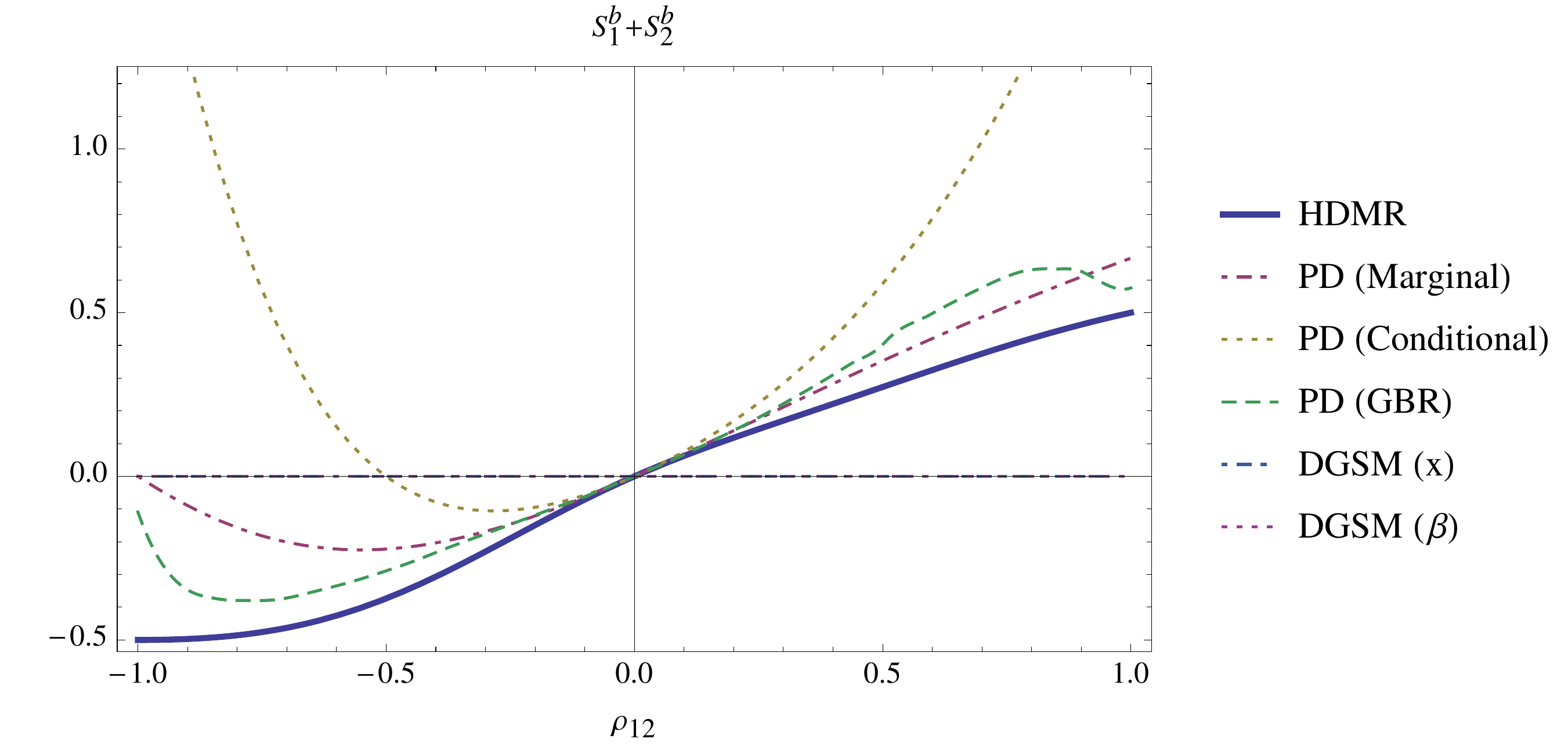}}\\
\subfloat[Overall, $S_1+S_2$\label{fig:poly:a}]{\includegraphics[width = 5in]{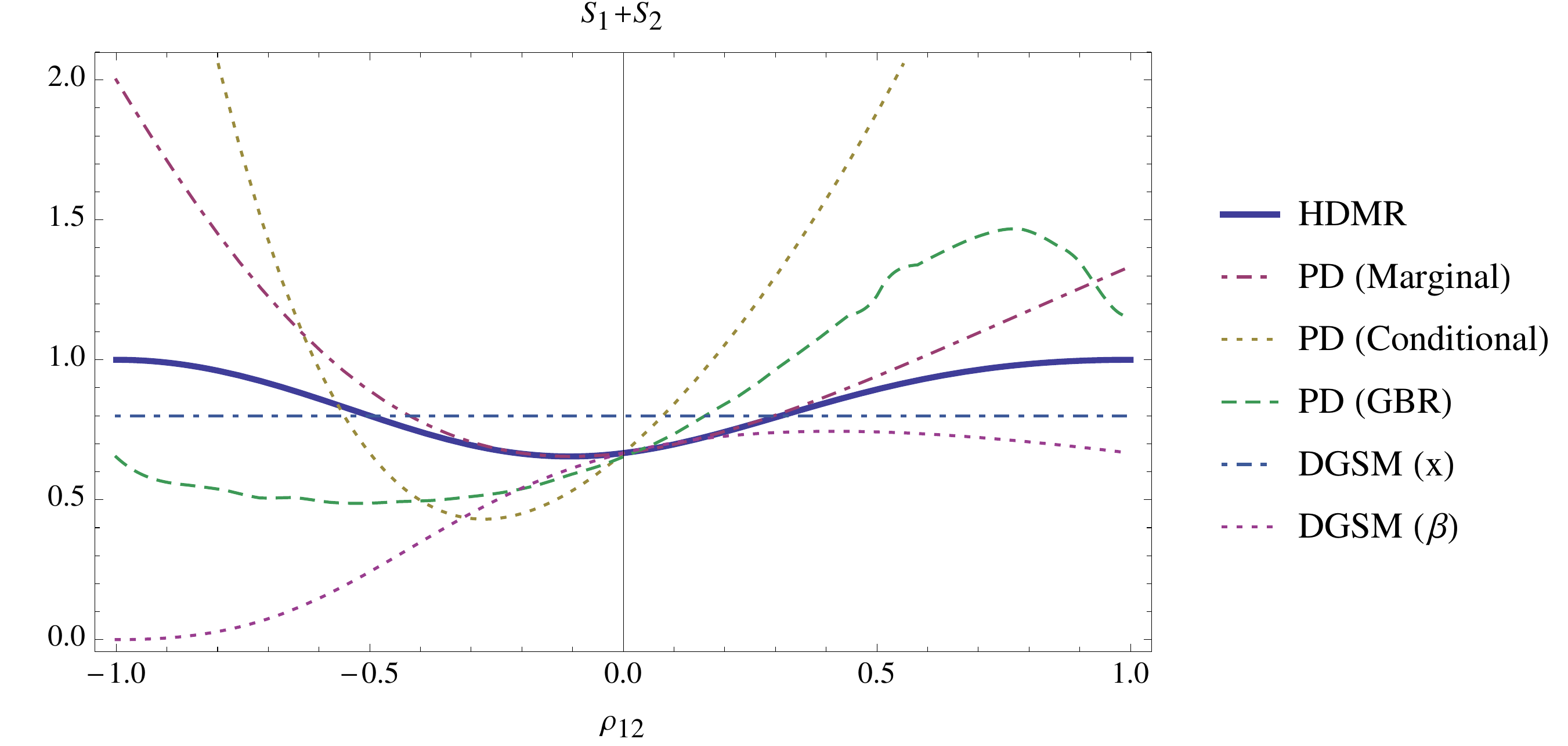}}
\endgroup
{\color{black}\caption{Univariate variable importance by HDMR, PD, and DGSM for $\beta=\bm{1}$, $\mu=\bm{0}$ and $\sigma=\bm{1}$ as a function of $\rho$.}\label{corrpd}}
\end{figure}

\begin{figure}
\centering
\begingroup
\captionsetup[subfigure]{width=5in,font=normalsize}
\color{black}
\subfloat[$\rho = 0$\label{fig:rho0}]{\includegraphics[width = 5in]{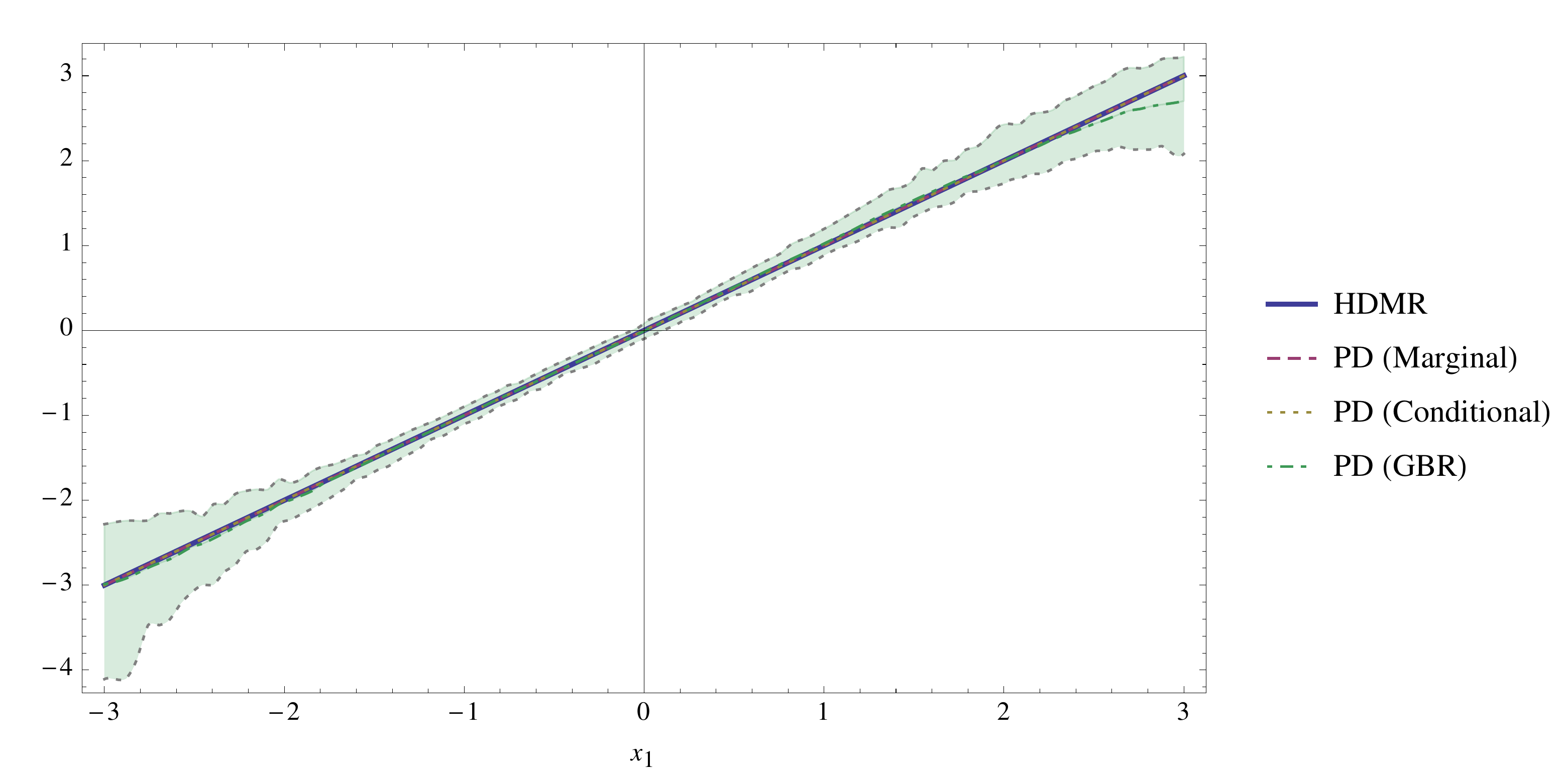}}\\
\subfloat[$\rho = \frac{1}{2}$\label{fig:rho5}]{\includegraphics[width = 5in]{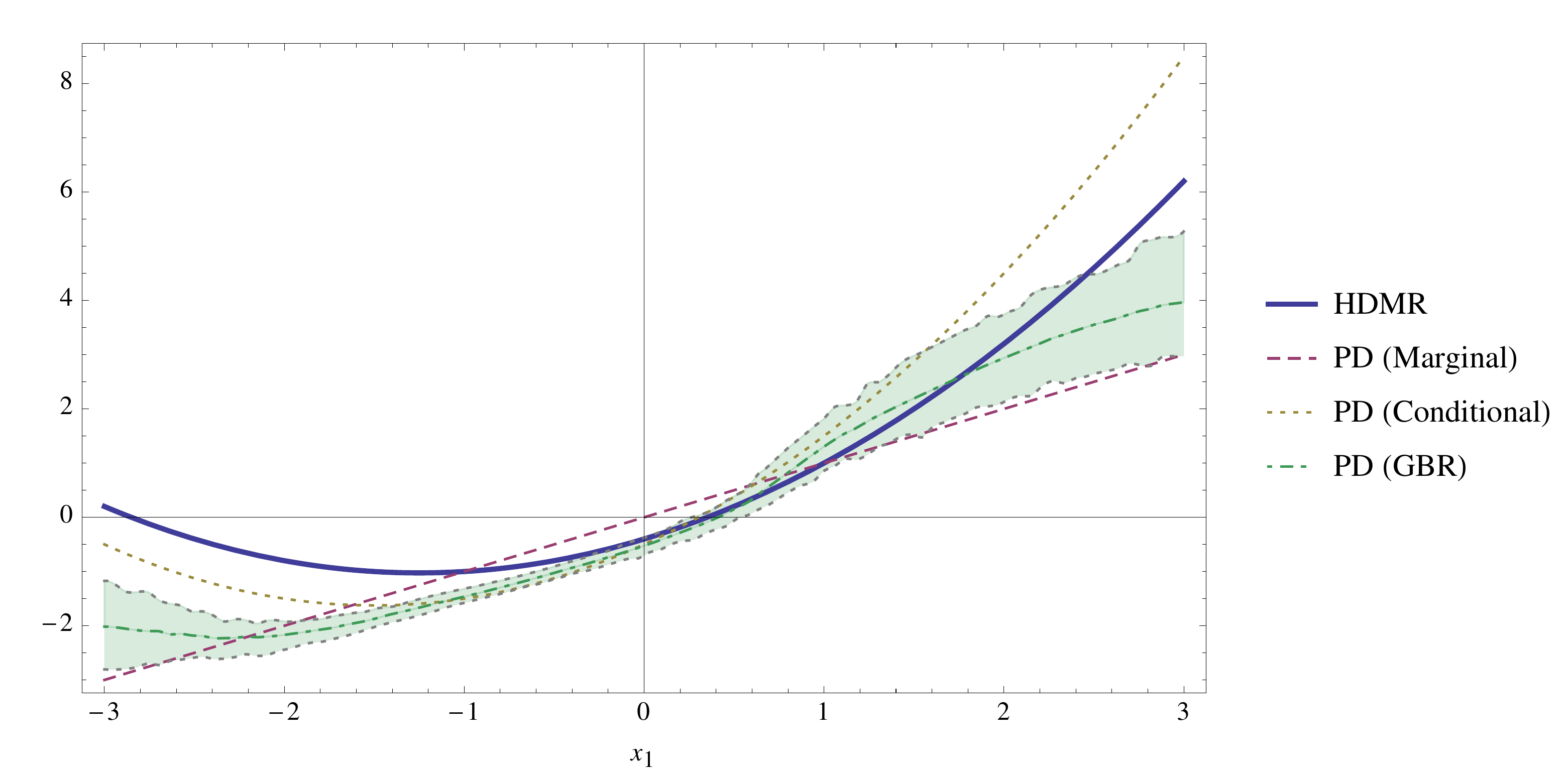}}\\
\subfloat[$\rho = \frac{9}{10}$\label{fig:rho9}]{\includegraphics[width = 5in]{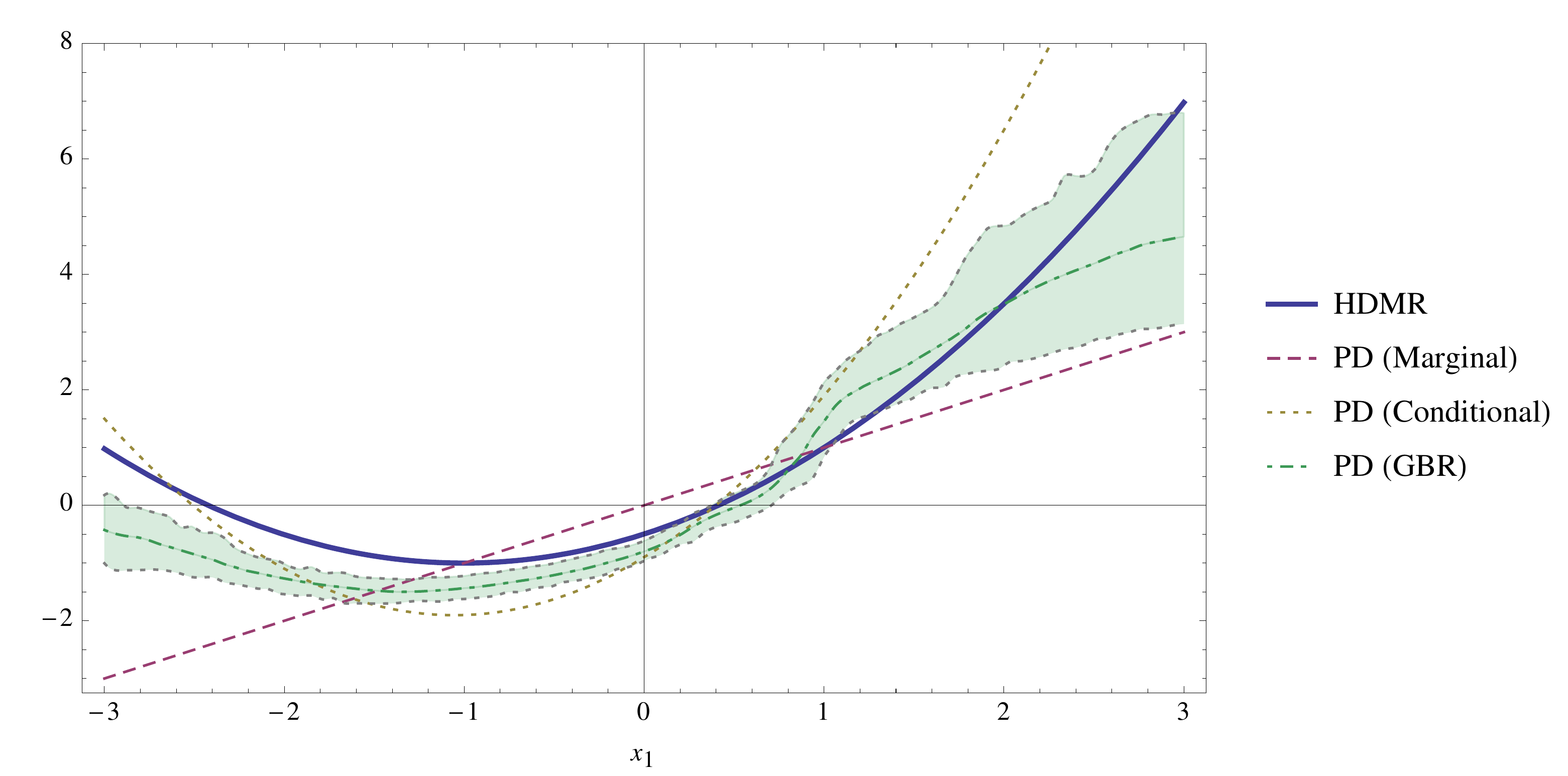}}
\endgroup
{\color{black}\caption{Variable dependence $\tilde{f}_1(x_1)$ for $\rho\in\lbrace0,\frac{1}{2},\frac{9}{10}\rbrace$ for HDMR, PD (Marginal), PD (Conditional), and PD (GBR) ($(\alpha,1-\alpha)$ percentile, $\alpha=0.025$)}\label{depend}}
\end{figure}

We illustrate these interpretative diagnostics for equation~\eqref{poly}, \[f(\beta,x) = \beta_0 + \beta_1x_1 + \beta_2x_2 + \beta_{12}x_1x_2.\] We take $\beta=\bm{1}$, $\mu=\bm{1}$, and $\sigma=\bm{1}$, and examine variable importance as a function of correlation, $\rho$. For DGSM, we also consider $\sigma_f^2(\beta) = \beta_2^2+2 \beta_2 \beta_3 \rho+\beta_3^2+\beta_{12}^2 \left(\rho^2+1\right)$. Whenever $\rho\in\lbrace\text{-}1,1\rbrace$ such that $x_1=x_2$ the function is one-dimensional and all variable importance resides in the univariate importance measures. The component functions are analytically given by 
\begin{align*}
\tilde{f}_1^\text{HDMR}(x_1) &= x_1 + \frac{\rho}{\rho^2+1}(x_1^2 - 1)\\
\tilde{f}_1^\text{PD (marg.)}(x_1) &= x_1\\
\tilde{f}_1^\text{PD (cond.)}(x_1) &= x_1 + \rho(x_1^2+x_1-1).
\end{align*}
Notice that the HDMR component function has non-linear dependence on $\rho$, whereas PD exhibits linear dependence. Variable importance results are exhibited in Table~\ref{tab:vi} and Figure~\ref{corrpd} (note that  the minimum and curvature for the `U'-shaped behavior changes in $\beta$ and for $\beta=\bm{1}$ in Figure~\ref{fig:poly:a} it is not symmetric), and variable dependence results are exhibited in Figure~\ref{depend}. For HDMR and partial dependence, we breakdown the importance measure into structural and correlative terms. We compute importance measures for $\rho\in\lbrace\text{-}1,0,1\rbrace$. All importance measures, with the exception of $\textbf{I}_\text{DGSM}f(x)$, depend upon $\rho$. When $\rho=0$, all coincide except $\textbf{I}_\text{DGSM}f(x)$. Only HDMR preserves variance, $S(\text{-}1)=S(1)=1$. \\

The variance decomposition property of HDMR, i.e. $S(\text{-}1)=S(1)=1$, means that the HDMR expansion correctly captures the contributions of subsets of variables to the variance in the output. \textbf{Observe that partial dependence given by \eqref{pd1} is accurate when correlation is roughly less than 0.3.} As the strength of dependence grows, the univariate partial dependency interpretative diagnostics of \eqref{pd1} become increasingly distorted. The diagnostic given by \eqref{pd2} exhibits far less variance stability in one-dimensional subspaces than that given by \eqref{pd1}. PD is also estimated using a gradient boosting machine (PD (GBR), where GBR-estimated PD is estimated thirty times on a grid for $\rho$, each GBR having 500 estimators of max depth of four and $10^3$ independent random samples. PD (GBR) sensitivity indices are computed from the GBR-estimated PD's using quadrature. GBR estimation of PD \eqref{pd1} is significantly biased as correlation increases. DGSM gives different importance of $f(x)$ than HDMR and partial dependence, even for $\rho=0$, although it coincides for $\sigma_f^2(\beta)$.\\

Figure~\ref{depend} reveals the profile of $\tilde{f}_1(x_1)$ in $\rho$ for HDMR and partial dependence (marginal, conditional, and GBR-estimated). PD \eqref{pd1} does not change in $\rho$, whereas PD \eqref{pd2} does. PD (GBR) significantly deviates from both HDMR and PD \eqref{pd1}. When $\rho=0$, the four coincide.\\

These results illustrate that \textbf{PD-based interpretative diagnostics experience information leakage whenever $\rho>0$}. In a similar manner, GBR-estimated PD exhibits information leakage. It is interesting that GBR-estimated PD diverges from both analytic PD-measures. This may be attributed to biases in the underlying algorthmic implementation, as tree-based ensemble methods such as random forest are known to exhibit biases in variable importance \citep{Strobl2007}.

\begin{table}[h]
\caption{Variable importance measures in $\rho$ for $\beta=\bm{1}$, $\mu=\bm{1}$, and $\sigma=\bm{1}$.}
\label{tab:vi}
\begin{center}
\resizebox*{\textwidth}{!}{
\begin{tabular}{lcccccc}
\toprule
Variable Importance & $S^\text{a}(\rho)$ & $S^\text{b}(\rho)$ & $S(\rho)$ & $S(\text{-}1)$ & $S(0)$ & $S(1)$\\\midrule
$\textbf{I}^1_\text{HDMR}$+$\textbf{I}^2_\text{HDMR}$  & $\frac{2 \left(\rho^4+4 \rho^2+1\right)}{\left(\rho^2+1\right)^2 (\rho (\rho+2)+3)}$ &$\frac{2 \rho \left(2 \rho^3+\left(\rho^2+1\right)^2\right)}{\left(\rho^2+1\right)^2 (\rho (\rho+2)+3)}$&$\frac{\rho-1}{\rho^2+1}+\frac{\rho+5}{\rho (\rho+2)+3}$ & 1 & 2/3 & 1\\
$\textbf{I}^1_\text{PD}$+$\textbf{I}^2_\text{PD}$ (marg.) &  $\frac{2 \left(\rho^2+1\right)}{\rho (\rho+2)+3}$ & $\frac{2 (\rho (\rho+1))}{\rho^2+2 \rho+3}$ & $\frac{2 \left(2 \rho^2+\rho+1\right)}{\rho (\rho+2)+3}$&2 & 2/3 & 4/3\\
$\textbf{I}^1_\text{PD}$+$\textbf{I}^2_\text{PD}$ (cond.) & $\frac{6 \rho^2+4 \rho+2}{\rho^2+2 \rho+3}$ & $\frac{2 \left(\rho \left(2 \rho^3+\rho^2+2 \rho+1\right)\right)}{\rho^2+2 \rho+3}$ & $\frac{2 \left(\rho \left(\rho \left(2 \rho^2+\rho+5\right)+3\right)+1\right)}{\rho (\rho+2)+3}$&4 & 2/3 &4\\
$\textbf{I}^1_\text{PD}$+$\textbf{I}^2_\text{PD}$ (GBR) & - & - & - & 0.65 & 0.65 & 1.14\\
$\textbf{I}^1_\text{DGSM}$+$\textbf{I}^2_\text{DGSM}$  & 4/5 & 0 & 4/5&4/5 & 4/5 & 4/5\\
$\textbf{I}^1_\text{DGSM}$+$\textbf{I}^2_\text{DGSM}$  & $\frac{2 (\rho+1)^2}{\rho \left(\rho^3+4 \rho+4\right)+3}$ & 0 & $\frac{2 (\rho+1)^2}{\rho \left(\rho^3+4 \rho+4\right)+3}$&0 & 2/3 &2/3\\
\bottomrule
\end{tabular}}
\end{center}
\end{table}

\FloatBarrier

\section{Glass boxes from black boxes}\label{sec:gb}  General-purpose black box learning algorithms in machine learning, such as kernel machines or decision tree models, exhibit favorable predictive performances and configuration or tuning costs. Suppose the availability of a black box representation $f\in F$.  We demonstrate how HDMR assesses variable importance and dependence of kernel machines or ensembles of decision trees. In this manner, HDMR is said to be a wrapper method \citep{guyon}. We give HDMR constructions for kernel machines and ensembles of decision trees and provide illustrations for polynomial kernel machines and ensembles of gradient boosting machines. %A cautionary  \emph{ex nihilo nihil fit}: the estimated HDMRs are exact (to numerical precision) representations of the underlying black boxes, and estimation accuracy is dependent on the efficiency of the black boxes in surveying the subspaces of $f\in F$. If the black boxes experience information leakage, then the resultant HDMRs necessarily have approximation error. 

\subsection{Kernel machines}\label{sec:km} HDMR provide interpretative diagnostics for the output of a kernel machine through a decomposition of the inner product of its RKHS. This is similar to the approach used in smoothing spline ANOVA RKHS but such is based on orthogonal projections \citep{ss1,ss2}. We outline the approach below and discuss polynomial kernels. \\

Assume we have a kernel \[K(x,x')=\langle\bm{\phi}(x),\bm{\phi}(x')\rangle_{\ell^2},\] where $\bm{\phi}: X\mapsto\ell^2(\bm{\alpha})$ is a feature map with index set $\bm{\alpha}$ of size $b\in\mathbb{N}$ and $\bm{\phi}(x) = (\phi_\alpha(x): \alpha\in\bm{\alpha})$ is a feature vector. We assume these feature vectors are represented by the Hadamard (entrywise) product of constants and basis elements, $\bm{\phi}(x) = \bm{\beta}\odot\textbf{B}(x)$, where $\textbf{B}(x)=(\text{B}_\alpha(x) : \alpha\in\bm{\alpha})$ are bases and $\bm{\beta}\in\mathbb{R}^{b}$ are coefficients (note that $\textbf{B}(x)$ is known as a Mercer map). The HDMR of the kernel is attained by decomposing the feature vector as \begin{equation}\label{sp}\bm{\phi}(x)=\sum_{\alpha\in\bm{\alpha}}\nu_\alpha\bm{\psi}_\alpha(x)=\sum_{\alpha\in\bm{\alpha}}\nu_\alpha(\bm{\eta}_\alpha\odot\textbf{B}(x)),\end{equation} where $\lbrace\bm{\psi}_\alpha(x)\equiv\bm{\eta}_\alpha\odot\textbf{B}(x)\rbrace$ are component feature vectors and $\lbrace\bm{\eta}_\alpha = (\eta_{\alpha\alpha'} : \alpha'\in\bm{\alpha})\rbrace$ are coefficient vectors attained from $\mu$. These coefficients reflect the hierarchical-orthogonality of the feature vectors and their generation is detailed in the appendices. Putting $\textbf{A}\equiv(\bm{\eta}_\alpha)\in\mathbb{R}^{b\times b}$, we have \begin{equation}\label{beta}\bm{\beta}=\textbf{A}\,\bm{\nu}\end{equation} where $\bm{\nu}=(\nu_\alpha : \alpha\in\bm{\alpha})\in\mathbb{R}^{b}$. Given i) $\bm{\beta}$ from the kernel and ii) $\textbf{A}$ from the measure $\mu$, this system is solved for $\bm{\nu}$, \[\bm{\nu}= \textbf{A}^{-1}\bm{\beta}.\] Using $\lbrace\nu_\alpha\bm{\psi}_\alpha(x)\rbrace$, we form the \textbf{HDMR of $\bm{K}$} by putting \[\phi(x) \equiv \sum_{u\subseteq\mathbb{N}_n}\bm{\Psi}_u(x),\] where the collection $\lbrace\bm{\Psi}_u(x)\rbrace$ is formed as \begin{equation}\label{ss}\left\lbrace\bm{\Psi}_u(x):\forall u\subseteq\mathbb{N}_n,\,\,\,\bm{\alpha}_u\subset\bm{\alpha},\,\,\,\bm{\Psi}_u(x)\equiv\sum_{\alpha\in\bm{\alpha}_u}\nu_\alpha\bm{\psi}_\alpha(x)\right\rbrace.\end{equation} If a subset of component feature vectors is sought, $\lbrace\bm{\Psi}_u(x) : |u|\le T<d\rbrace$, then we have $\textbf{A}\in\mathbb{R}^{a\times b}$, an underdetermined system $a<b$, and a least-norm solution can be efficiently attained using QR decomposition. Given a finite dataset $\textbf{D}$ and a kernel $K$, a kernel machine attains a set coefficients $\lbrace\xi_{x'}\rbrace$ such that
\begin{align*}f(x)&=\sum_{x'\in\textbf{D}}\xi_{x'} K(x,x')\\
&=\sum_{x'\in\textbf{D}}\xi_{x'}\langle\bm{\phi}(x),\bm{\phi}(x')\rangle_{\ell^2}\\
&=\left\langle\sum_{u\subseteq\mathbb{N}_n}\bm{\Psi}_u(x),\sum_{x'\in\textbf{D}}\xi_{x'}\bm{\phi}(x')\right\rangle_{\ell^2}\\
&=\sum_{u\subseteq\mathbb{N}_n}\left\langle\bm{\Psi}_u(x), f\right\rangle_{\ell^2}\\
%&=\bm{\phi}(x)\cdot\left(\sum_{x'\in\textbf{D}}\xi_{x'}\bm{\phi}(x')\right)\\
%&=\left(\sum_{u\subseteq\mathbb{N}_n}\bm{\Psi}_u(x)\right)\cdot f\\
%&=\sum_{u\subseteq\mathbb{N}_n}\left( \bm{\Psi}_u(x)\cdot f\right)\\
&=\sum_{u\subseteq\mathbb{N}_n}f^\text{HDMR}_u(x_u).
\end{align*}

%=\sum_{\alpha\in\bm{\alpha}}\nu_\alpha\bm{\eta}_\alpha
	
\subsubsection*{Polynomial kernels} The (real-valued) polynomial kernel is given by \begin{equation}\label{realpoly}K_\text{poly}(x,x';c,d,\gamma) = (\gamma x\cdot x'+c)^d,\end{equation} where $c,\gamma\ge 0$, $d\in\mathbb{N}$, and $x,x'\in\mathbb{R}^n$. Defining $\bm{\alpha}(n,d)=\lbrace \alpha\in\mathbb{N}_0^{n+1}, |\alpha|= d\rbrace$, \[\textbf{B}(x;n,d) = (x^{\alpha_1\dotsb\alpha_n} : \alpha\in\bm{\alpha}(n,d))\] and \[\bm{\beta}(n,d) = (\beta_\alpha : \alpha\in\bm{\alpha}(n,d)),\] \[\beta_\alpha = \left(\binom{d}{\alpha}c^{\alpha_{0}}\gamma^{d-\alpha_{0}}\right)^{1/2},\] we have (suppressing notation in $n$ and $d$) \[\bm{\phi}(x)=\bm{\beta}\odot\textbf{B}(x).\] Together, 
\begin{align*}
K_\text{poly}(x,x';c,d,\gamma)&=\langle\bm{\phi}(x),\bm{\phi}(x')\rangle_{\ell^2}\\
&=\sum_{\alpha\in\bm{\alpha}} \beta_\alpha^2 x^{\alpha_1\dotsb\alpha_n}x'^{\alpha_1\dotsb\alpha_n}.
\end{align*}  

Defining $\textbf{C}(x;n,d)=(e^{2\pi i\langle\alpha_{1\dotsb n},x\rangle} : \alpha\in\bm{\alpha}(n,d))$ and $\bm{\phi}(x)=\bm{\beta}\odot\textbf{C}(x)$, we write a complex-valued polynomial kernel,  
\begin{align*}
K_\text{poly}(x,x';c,d,\gamma)&=\langle\bm{\phi}(x),\bm{\phi}(x')\rangle_{\ell^2}\\
&=(\gamma e^{2\pi ix}\cdot e^{2\pi ix'}+c)^d\\
&=\sum_{\alpha\in\bm{\alpha}}\beta_\alpha^2 e^{2\pi i\langle\alpha_{1\dotsb n},x\rangle}e^{2\pi i\langle\alpha_{1\dotsb n},x'\rangle}\\
&=\sum_{\alpha\in\bm{\alpha}}\beta_\alpha^2 e^{2\pi i\langle\alpha_{1\dotsb n},x+x'\rangle}.
\end{align*}

\subsubsection*{Illustration: analytic function} 
\begin{figure}[h]
\centering
\includegraphics[scale=0.55]{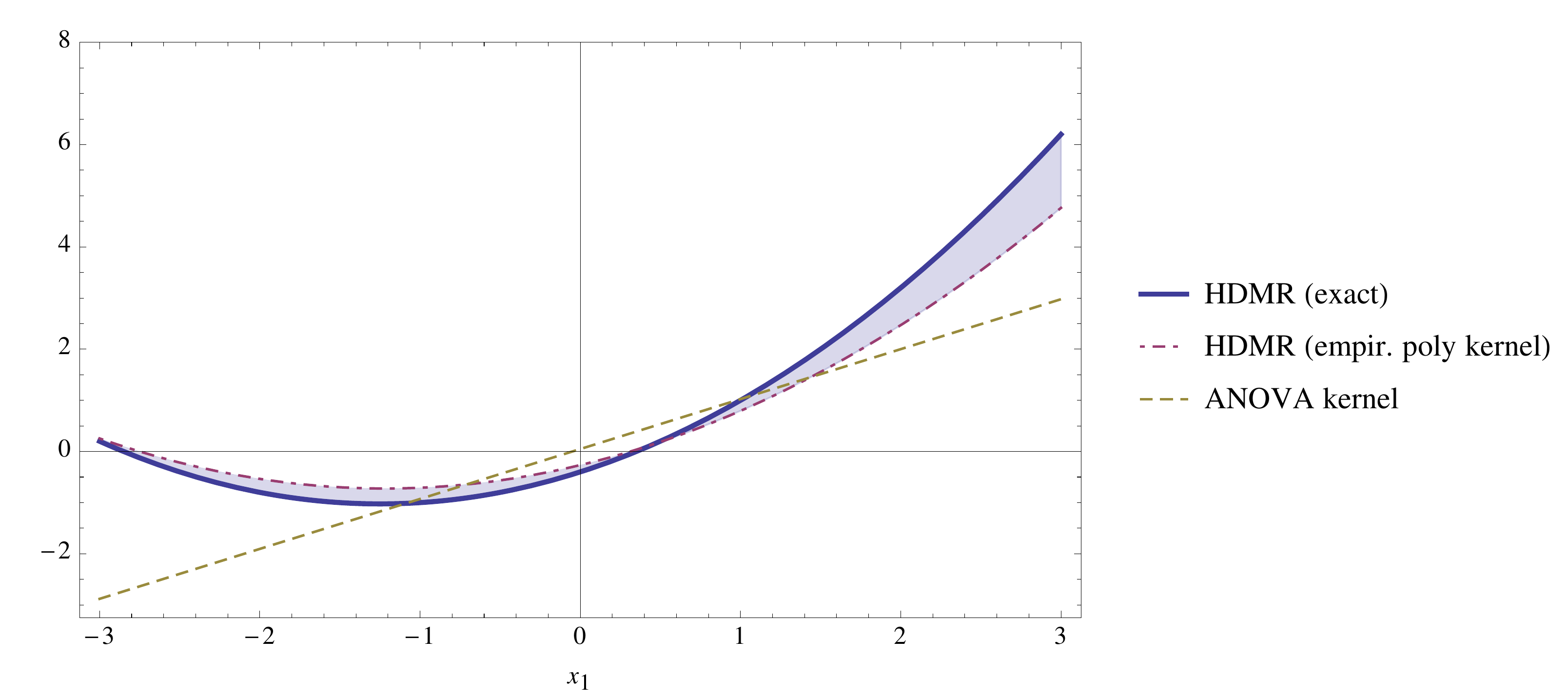}
\caption{Plot of $f_1(x_1)$ for $\rho=\frac{1}{2}$; HDMR (exact) compared to ANOVA-polynomial kernel and empirical-polynomial HDMR}\label{fig:anova}
\end{figure}

We consider the mathematical function \eqref{poly}. In Figure~\ref{fig:anova}, we compare $f_1(x_1)$ as computed by kernel machines---$f^\text{HDMR}_1(x_1)$, attained from the empirical HDMR \eqref{ss} (``empirical-polynomial HDMR'') and $f^\text{ANOVA}_1(x_1)$, attained from the ANOVA representation of the polynomial kernel \eqref{realpoly} (``ANOVA-polynomial kernel''). The ANOVA-kernel of the polynomial kernel is defined as $K_\text{ANOVA}(x,x') = \prod_{i\in\mathbb{N}_n}(1+k_{i}(x_i,x_i'))$ \citep{kernel}, where $\lbrace k_i(x_i,x_i')\rbrace$ are univariate zero-mean polynomial kernels. Both kernel machines use the same empirical measure of $10^3$ data elements with $\rho=\frac{1}{2}$ and are compared to the exact HDMR. As seen in Figure~\ref{fig:anova}, \textbf{the ANOVA-polynomial kernel does not use the correlative information of the input data}, whereas the empirical-polynomial HDMR better approximates the exact underlying HDMR.

\FloatBarrier

\subsection{Decision trees}\label{sec:dt} A decision tree is a function that uses a partition of $X=\cup_{R_i\in R}R_i$, $R=\lbrace R_i\rbrace_{i}$, such that \[g(x) = \sum_{R_i\in R} c_i 1_{R_i}(x).\] The partition $R_u$ is defined using a subset of variables $x_u$ such that a tree is written as \[g_u(x_u) \equiv  \sum_{R_i\in R_u} c_i 1_{R_i}(x_u).\] A collection of tree ensembles, indexed by $\mathcal{I}$, is denoted by \begin{equation}\label{eq:coll}F_\mathcal{I}\equiv\lbrace f^i(x)\rbrace_{i\in\mathcal{I}},\end{equation}  where $f^i(x)$ is a sum of trees across the subspaces $\mathcal{P}_i\subseteq\powerset(\mathbb{N}_n)$ \[f^i(x)\equiv\sum_{u\in\mathcal{P}_i} g^i_{u}(x_{u})=\sum_{u\in\mathcal{P}_i} \sum_{j} g^i_{uj}(x_{u})\] and has depth $d\in\mathbb{N}$ \citep{Breiman2001}. A key property of trained decision tree models is that tree subspaces are highly sparse, $|\mathcal{P}_i|\ll |\powerset(\mathbb{N}_n)|$ for $i\in\mathbb{N}_n$. A linear combination of tree ensembles is defined as \begin{equation}\label{eq:comb}f(x) \equiv \sum_{i\in\mathcal{I}}\beta_i f^i(x).\end{equation} Forming $\bm{F_\mathcal{I}}\equiv(f^i: i\in\mathcal{I}) = (g_u^i: i\in\mathcal{I}, u\in\mathcal{P}_i)$ we take \[\bm{\beta}(\lambda)\equiv(\beta_{ui}(\lambda): i\in\mathcal{I}, u\in\mathcal{P}_i)=(\bm{F_\mathcal{I}}\otimes\bm{F_\mathcal{I}}+\lambda I)^{-1}(\bm{F_\mathcal{I}}\otimes f).\] We index the tree ensemble feature vectors on $\bm{\alpha}$ as
\begin{align*}
\bm{\alpha}&\equiv (\bm{\alpha}_0)\parallel( (i,u)\in\mathcal{I}\times\powerset(\mathbb{N}_n) : i\in\mathcal{I},\,\,\,u\in\mathcal{P}_i)
\end{align*}
where 
\begin{align*}
\bm{\alpha}_0 &= (\lbrace\rbrace,\lbrace\rbrace)
\end{align*}
and define  
\begin{align*}
\bm{\alpha}_{\bullet u} &\equiv \lbrace(i,w)\in\bm{\alpha} : i\in\mathcal{I},\,\,\, w = u\rbrace\\
\bm{\alpha}_{i\bullet} &\equiv \lbrace(j,u)\in\bm{\alpha} : u\in\mathcal{P}_j,\,\,\, j = i\rbrace.
\end{align*}
The tree ensemble feature vectors are given by
\begin{align*}
\bm{\phi}^i(x)&\equiv\bm{\beta}^i(\lambda)\odot\textbf{B}^i(x)\\
\textbf{B}^i(x) &\equiv\left(g^i_{u}(x_{u})1_{\bm{\alpha}_{i\bullet}}(j,u) : (j,u)\in\bm{\alpha}\right)\\
\bm{\beta}^i(\lambda)&\equiv (\beta_i(\lambda)1_{\bm{\alpha}_{i\bullet}}(j,u) : (j,u)\in\bm{\alpha}).
\end{align*} This is written as \[\bm{\phi}(x)=\sum_{i\in\mathcal{I}}(\bm{\beta}^i(\lambda)\odot\textbf{B}^i(x))=\bm{\beta}(\lambda)\odot\textbf{B}(x),\] where \[\textbf{B}(x)=\sum_{i\in\mathcal{I}}\textbf{B}^i(x)\] and \begin{equation}\label{eq:beta}\bm{\beta}(\lambda)=\sum_{i\in\mathcal{I}}\bm{\beta}^i(\lambda).\end{equation} Note that $\bm{\beta}_{0}=0$.
As before, we have the coefficients attained from $\mu$ as $\bm{\eta}_\alpha$ and in organized form, \begin{equation}\label{eq:A}\textbf{A}\equiv (\bm{\eta}_\alpha).\end{equation} Then, weights $\bm{\nu}=(\nu_\alpha : \alpha\in\bm{\alpha})$ are attained that satisfy \begin{equation}\label{eq:beta2}\bm{\beta}(\lambda)=\textbf{A}\,\bm{\nu}\end{equation} as \begin{equation}\label{eq:nu}\bm{\nu}= \textbf{A}^{-1}\bm{\beta}(\lambda).\end{equation} Noting
\begin{equation}\label{eq:gbrhdmr}\bm{\phi}(x)=\sum_{\alpha\in\bm{\alpha}}\nu_\alpha\bm{\psi}_\alpha(x)=\sum_{\alpha\in\bm{\alpha}}\nu_\alpha(\bm{\eta}_\alpha\odot\textbf{B}(x)),\end{equation} we form the \textbf{HDMR of $\bm{F_\mathcal{I}}$} as \begin{equation}\label{ss}\left\lbrace\bm{\Psi}_u(x):\forall u\subseteq\mathbb{N}_n,\,\,\,\bm{\alpha}_u\subset\bm{\alpha},\,\,\,\bm{\Psi}_u(x)\equiv\sum_{\alpha\in\bm{\alpha}_u}\nu_\alpha\bm{\psi}_\alpha(x)\right\rbrace.\end{equation}

We summarize the computations: 

\begin{enumerate}[label=(\roman*)]
\item Equation \eqref{eq:coll}: forming $F_\mathcal{I}$ from the collection of GBR models $\lbrace f^i(x)\rbrace_{i\in\mathcal{I}}$, indexed on $\mathcal{I}$
\item Equations \eqref{eq:comb} and \eqref{eq:beta}: linearly combining the elements of $F_\mathcal{I}$ as $f(x)$ using the coefficients $\bm{\beta}(\lambda)\equiv(\beta_{ui}(\lambda): i\in\mathcal{I}, u\in\mathcal{P}_i)\in\mathbb{R}^\text{b}$, attained from regularized least-squares using with grid-searched $\lambda$ estimated using $K$-fold cross-validation% (assessed using $K$-fold cross-validation)
\item Equation \eqref{eq:A}: attaining $\bm{\eta}_\alpha$ from the measure $\mu$ using a collection of QR-decompositions, each identified to a component function subspace and each QR decomposition having linear cost in the dataset size $\mathcal{O}(N)$ and quadratic cost $\mathcal{O}(|\mathcal{P}(x_u)\cap\mathcal{P}_i|^2)$ (typically small in most HDMR applications) and forming the square matrix $\textbf{A}\in\mathbb{R}^{b\times b}$, where $b$ is total size of the corresponding inner product space $|\cup_i\mathcal{P}_i|$
\item Equations \eqref{eq:beta2} and \eqref{eq:nu}: forming the system $\bm{\beta}=\textbf{A}\cdot\bm{\nu}$, computing the inverse of the square matrix $\textbf{A}$, such as through using Gaussian elimination, and putting $\bm{\nu}\equiv \textbf{A}^{-1}\bm{\beta}$
\item Equations \eqref{eq:gbrhdmr} and \eqref{ss}: forming the HDMR $\bm{F_\mathcal{I}}$.
\end{enumerate}

Projections into the space of decision trees, which are systems of simple functions, produce noisy approximations to continuous functions. This can be addressed through spectral filtering of the component function subspaces, such as projection into smooth subspaces (Fourier). Another consideration is that if the GBR ensemble experience information leakage on the subspaces of $F$ the interpretative diagnostics will be biased (as illustrated in the previous section for non-trivial correlation). The variance-preserving property of HDMR, however, enables introspection of black box learning algorithms for independent and/or correlated variables, to the extent that the black box contains information on the projections of the system. \\%These computations can be formulated for vector-valued functions (see appendices for a brief description of the set-up).\\

We compute the HDMR of two efficient general-purpose black-box supervised machine learning algorithms---the random forest (RF) and the gradient boosting regressor (GBR) machine---for a non-linear mathematical function (Ishigami) with analytic solution and a benchmark dataset (California housing). We demonstrate that RF experiences far more information leakage than GBR and that GBR well approximates HDMR.
%We use ensembles of decision trees through the random forest (RF) and the gradient boosting regressor (GBR) machine are popular, efficient, and general purpose black box algorithms. We demonstrate how an ensemble of GBR models (one or more GBR models) is used to approximate HDMR

\subsubsection*{Illustration 1: analytic test function} In this example we verify that the HDMR of $F_\mathcal{I}$ given by \eqref{ss} well approximates the true HDMR for a test function. We consider the Ishigami function, a non-linear continuous function, which exists in closed-form and is sparse. It is defined as \[ f(x) = \sin x_1 + a \sin^2 x_2 + b\, x_3^4\sin x_1\] with independent uniformly distributed $x=(x_1,x_2,x_3)\in[-\pi,\pi]^3$. Its HDMR is analytic and shown below,
\begin{align*}
f^\text{HDMR}_0 &= \frac{a}{2}\\
f^\text{HDMR}_1(x_1) &= (1+\frac{b\pi^4}{5})\sin x1\\
f^\text{HDMR}_2(x_2) &= -\frac{a}{2}\cos 2x_2\\
f^\text{HDMR}_3(x_3) &= 0\\
f^\text{HDMR}_{12}(x_1,x_2) &= 0\\
f^\text{HDMR}_{13}(x_1,x_3) &= b(x_3^4-\frac{\pi^4}{5})\sin x_1\\
f^\text{HDMR}_{23}(x_2,x_3) &= 0\\
f^\text{HDMR}_{123}(x_1,x_2,x_3) &= 0.
\end{align*}
Its non-trivial sensitivity indices are
\begin{align*}
\mathbb{S}_1 &= \frac{36 \left(\pi ^4 b+5\right)^2}{5 \left(45 \left(a^2+4\right)+20 \pi ^8 b^2+72 \pi ^4 b\right)}\\
\mathbb{S}_2 &= \frac{45 a^2}{45 \left(a^2+4\right)+20 \pi ^8 b^2+72 \pi ^4 b}\\
\mathbb{S}_{13} &= \frac{64 \pi ^8 b^2}{5 \left(45 \left(a^2+4\right)+20 \pi ^8 b^2+72 \pi ^4 b\right)}\\
%T_1 &= \mathbb{S}_1+\mathbb{S}_{13}\\
%T_2 &= \mathbb{S}_2\\
%T_3 &= \mathbb{S}_{13}\\
%R_1 &= \frac{20 \left(\pi ^4 b \left(5 \pi ^4 b+18\right)+45\right)}{225 \left(a^2+4\right)+164 \pi ^8 b^2+360 \pi ^4 b}\\
%R_2 &= \frac{225 a^2}{225 \left(a^2+4\right)+164 \pi ^8 b^2+360 \pi ^4 b}\\
%R_3 &= \frac{64 \pi ^8 b^2}{225 \left(a^2+4\right)+164 \pi ^8 b^2+360 \pi ^4 b}
\end{align*}

%We take $a=7$ and $b=10^{-1}$, giving $\mathbb{S}_1\approx 0.31$, $\mathbb{S}_2\approx 0.44$, and $\mathbb{S}_{23}\approx 0.24$. \\

We utilize an ensemble of GBR's indexed on 
\begin{align*}
\mathcal{I}_i&=\lbrace(\text{depth}_i,\text{subspaces})\rbrace\\
&=\lbrace(k,(1,2,3)): k\in[i]\rbrace.
\end{align*}
 These are linearly combined using regularized least-squares with grid-searched $\lambda$. $J=10^3$ trees are used in each ensemble and $N=5\times 10^3$ data points are sampled from $\text{Uniform}[-\pi,\pi]^3$. 
 
\begin{figure}[h]
\centering
\caption{One-dimensional component functions: GBR approximation, smooth Fourier projection, and analytic}\label{fig:ishigami12}
\begingroup
\captionsetup[subfigure]{width=5in,font=normalsize}
\subfloat[$f^\text{HDMR}_1(x_1)$]{\includegraphics[width = 5in]{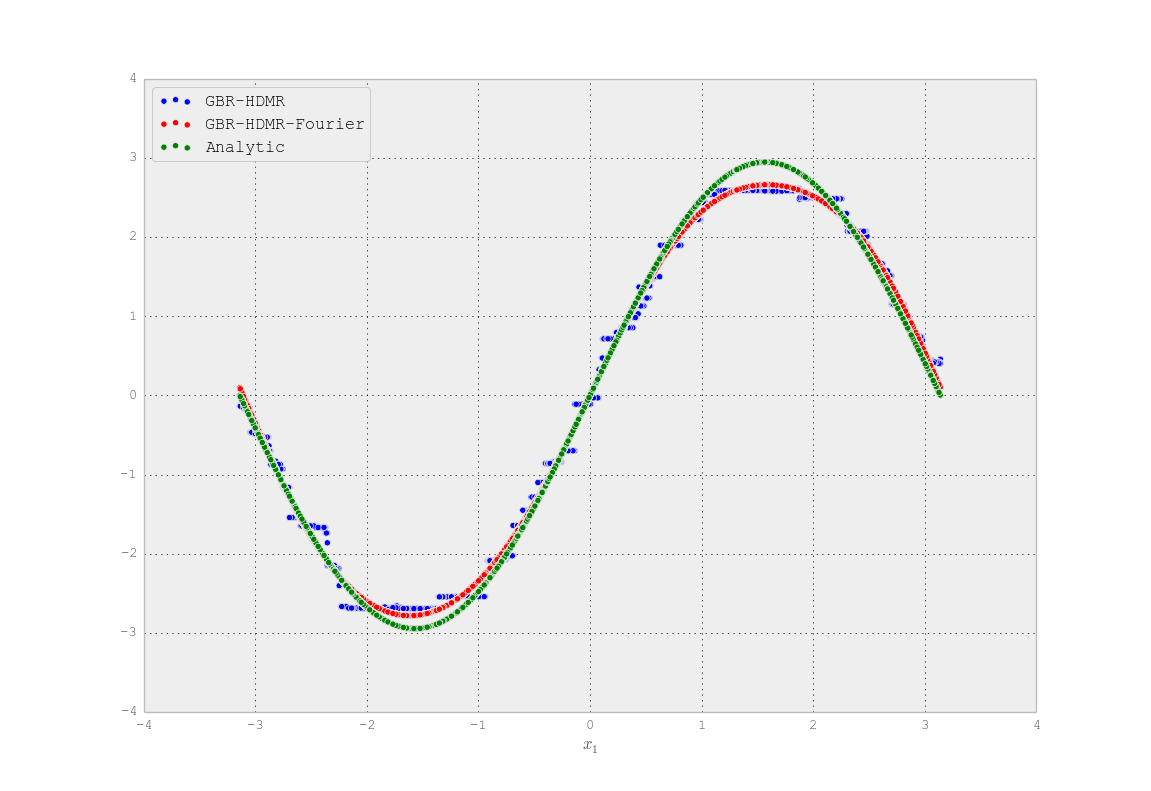}}\\
\subfloat[$f^\text{HDMR}_2(x_2)$]{\includegraphics[width = 5in]{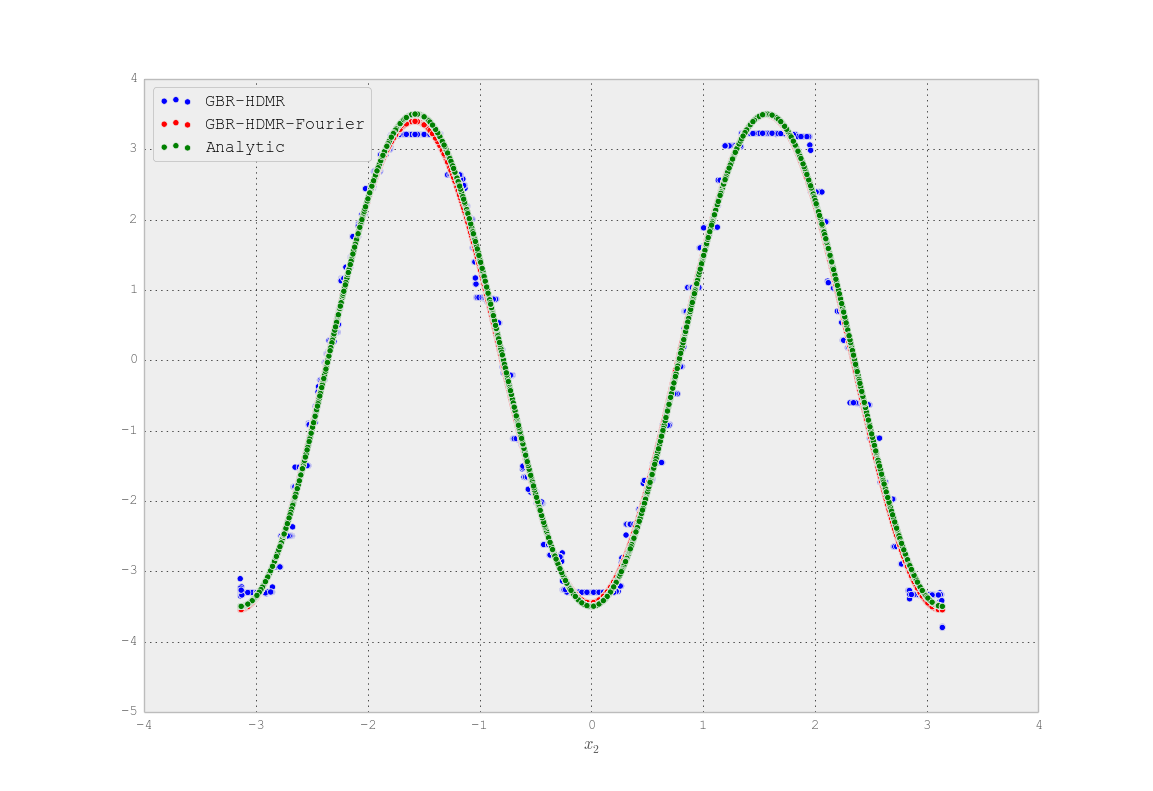}}\\
\endgroup
\end{figure}

\begin{figure}[h]
\centering
\caption{$f^\text{HDMR}_{13}(x_1,x_3)$ by GBR approximation}\label{fig:ishigami3}
\begingroup
\captionsetup[subfigure]{width=5in,font=normalsize}
\color{black}
\subfloat[Decision tree approximation\label{fig:ishigami3dc}]{\includegraphics[width = 5in]{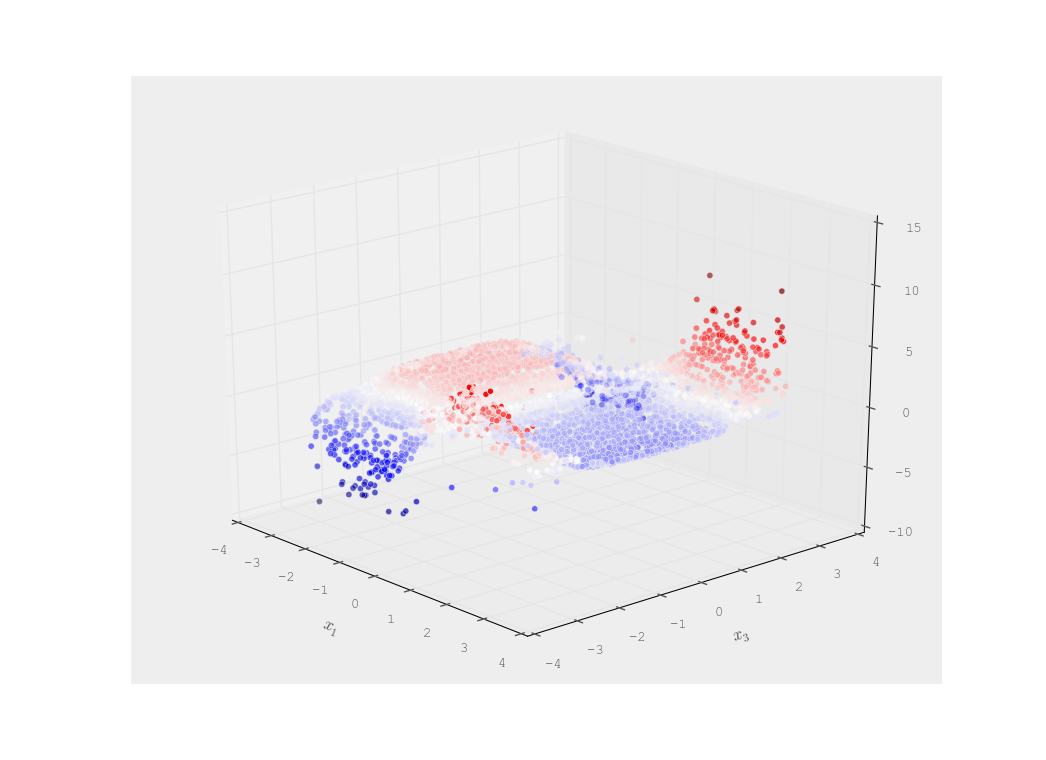}}\\
\subfloat[Analytic]{\includegraphics[width = 5in]{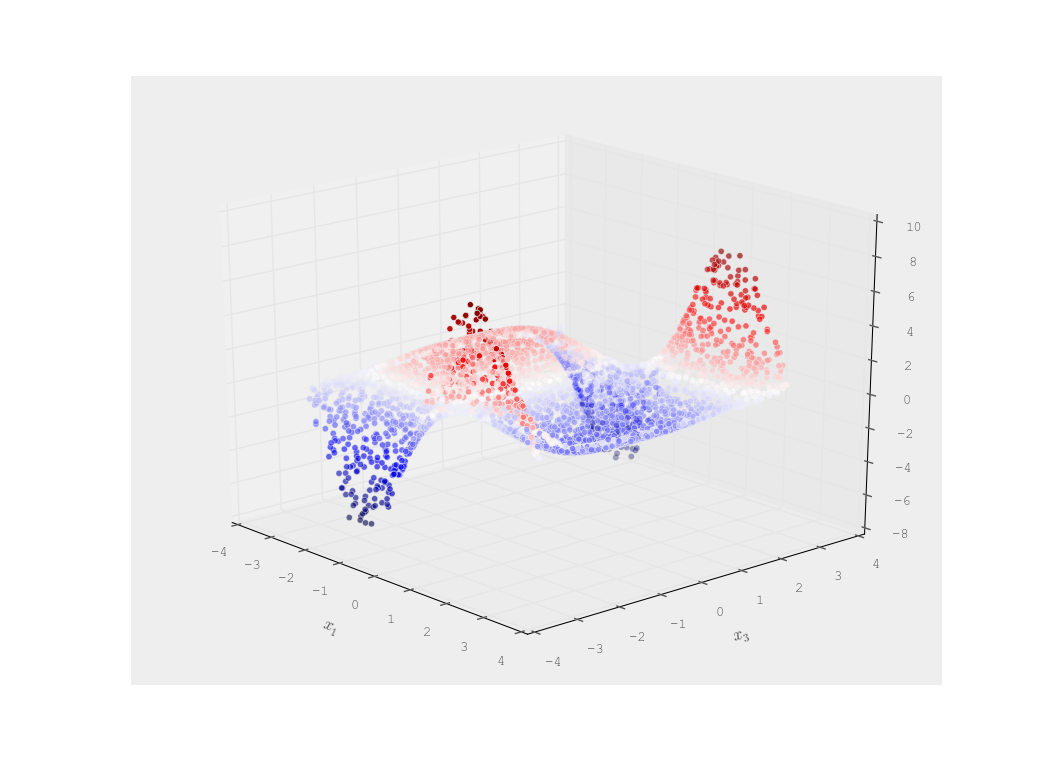}}\\
\endgroup
\end{figure}

In Figures~\ref{fig:ishigami12} and \ref{fig:ishigami3} we estimate $f^\text{HDMR}_1(x_1)$, $f^\text{HDMR}_2(x_2)$, and $f^\text{HDMR}_{13}(x_1,x_3)$ using GBR from $F_{\mathcal{I}_3}$. We also estimate these using RF (located in the appendices; Figures~\ref{fig:ishigami1_rf}, \ref{fig:ishigami2_rf}, and \ref{fig:ishigami3_rf}). We project the one-dimensional component functions into the space of zero-mean band-limited Fourier series (degree four) to attain smooth low-frequency approximations. Sensitivity indices are shown below in Table~\ref{tab:scsa_ishi}. \textbf{The GBR-based decision tree approximations closely approximate the true HDMR component functions, whereas RF performs poorly, experiencing significant information leakage in comparison to GBR.} GBR estimation of the two-dimensional component function $f^\text{HDMR}_{13}(x_1,x_3)$ experiences some information leakage but its general qualitative behavior is satisfactorily reproduced as shown in Figure~\ref{fig:ishigami3dc}. A tree-based measure of variable importance based on the fraction of samples variables contribute through trees is computed as `Tree'  and gives substantially different results than $R$ by HDMR: $x_3$ is the most important variable by fraction of samples but the least important by HDMR. The defect in RF is appears to be insensitive to the number of trees, sample size, or tree depth. 

%\begin{figure}[h]
%\centering
%\includegraphics[scale=0.5]{ishigami1.png}
%\caption{$f^\text{HDMR}_1(x_1)$ by GBR approximation, smooth Fourier projection, and analytic}\label{fig:ishigami1}
%\end{figure}
%
% \begin{figure}[h]
%\centering
%\includegraphics[scale=0.5]{ishigami2.png}
%\caption{$f^\text{HDMR}_2(x_2)$ by GBR approximation, smooth Fourier projection, and analytic}\label{fig:ishigami2}
%\end{figure}

% Random forest was also utilized but performs poorly, where it is unable to accurately learn $f^\text{HDMR}_2(x_2)$. The gradient boosting machine is therefore said to experience far less information leakage than the random forest model. 

\begin{table}[h]
\caption{Structural and correlative sensitivity analysis $\text{P}_\epsilon$ for the Ishigami function ($a=7$, $b=10^{-1}$) using gradient boosting regressor (GBR) machine and mean and standard deviation estimated from 50 random realizations for $\mathcal{I}_1$, $\mathcal{I}_2$, and $\mathcal{I}_3$}
\label{tab:scsa_ishi}
\begin{center}
\resizebox*{\textwidth}{!}{
\begin{tabular}{llllllll}
\toprule
 Subspace   & \mc{1}{Index}  & \mc{1}{ $\mathbb{S}^\text{a}$} &      \mc{1}{$\mathbb{S}^\text{b}$} &     \mc{1}{ $\mathbb{S}$}  &  \mc{1}{ T} & \mc{1}{R} & \mc{1}{Tree}\\\midrule
(1,) & \mc{1}{(Analytic)}& 0.31 & 0.00 &  0.31  & 0.56 & 0.45   &  \\\midrule
& \mc{1}{$\mathcal{I}_1$}& 0.31 $\pm$ 0.00 & 0.00 $\pm$ 0.00 &  0.31 $\pm$ 0.00  & 0.31 $\pm$ 0.00 & 0.40 $\pm$ 0.00   & 0.23 $\pm$ 0.00 \\
& \mc{1}{$\mathcal{I}_2$}& 0.29 $\pm$ 0.00 & 0.01 $\pm$ 0.00 &  0.30 $\pm$ 0.00  & 0.49 $\pm$ 0.01 & 0.42 $\pm$ 0.00   & 0.33 $\pm$ 0.01 \\
& \mc{1}{$\mathcal{I}_3$}& 0.29 $\pm$ 0.00 & 0.01 $\pm$ 0.00 &  0.30 $\pm$ 0.00  & 0.51 $\pm$ 0.01 & 0.42 $\pm$ 0.00   & 0.32 $\pm$ 0.01 \\\midrule
(2,) & \mc{1}{(Analytic)}& 0.44 & 0.00 &  0.44  & 0.44 & 0.36   &  \\\midrule
& \mc{1}{$\mathcal{I}_1$}& 0.46 $\pm$ 0.00 & -0.00 $\pm$ 0.00 &  0.45 $\pm$ 0.00  & 0.45 $\pm$ 0.00 & 0.58 $\pm$ 0.00   & 0.57 $\pm$ 0.00 \\
& \mc{1}{$\mathcal{I}_2$}& 0.44 $\pm$ 0.00 & 0.00 $\pm$ 0.00 &  0.45 $\pm$ 0.00  & 0.45 $\pm$ 0.00 & 0.38 $\pm$ 0.00   & 0.25 $\pm$ 0.01 \\
& \mc{1}{$\mathcal{I}_3$}& 0.44 $\pm$ 0.00 & 0.00 $\pm$ 0.00 &  0.45 $\pm$ 0.00  & 0.46 $\pm$ 0.00 & 0.38 $\pm$ 0.00   & 0.28 $\pm$ 0.01 \\\midrule
(3,) & \mc{1}{(Analytic)}& 0.00 & 0.00 &  0.00  & 0.24 & 0.20   &  \\\midrule
& \mc{1}{$\mathcal{I}_1$}& 0.01 $\pm$ 0.00 & 0.00 $\pm$ 0.00 &  0.02 $\pm$ 0.00  & 0.02 $\pm$ 0.00 & 0.02 $\pm$ 0.00   & 0.21 $\pm$ 0.00 \\
& \mc{1}{$\mathcal{I}_2$}& 0.03 $\pm$ 0.00 & 0.01 $\pm$ 0.00 &  0.04 $\pm$ 0.01  & 0.23 $\pm$ 0.00 & 0.20 $\pm$ 0.00   & 0.42 $\pm$ 0.01 \\
& \mc{1}{$\mathcal{I}_3$}& 0.03 $\pm$ 0.00 & 0.01 $\pm$ 0.00 &  0.04 $\pm$ 0.01  & 0.25 $\pm$ 0.00 & 0.20 $\pm$ 0.00   & 0.40 $\pm$ 0.01 \\\midrule
(1,2) &\mc{1}{(Analytic)}& 0.00 & 0.00 &  0.00  &  &    &  \\
\cmidrule(r){1-5}& \mc{1}{$\mathcal{I}_2$}& 0.00 $\pm$ 0.00 & -0.00 $\pm$ 0.00 &  0.00 $\pm$ 0.00  &  &  & \\
& \mc{1}{$\mathcal{I}_3$}& 0.00 $\pm$ 0.00 & -0.00 $\pm$ 0.00 &  0.00 $\pm$ 0.00  &  &  & \\
\cmidrule(r){1-5}(1,3) &\mc{1}{(Analytic)}& 0.24 & 0.00 &  0.24  &  &    &  \\
\cmidrule(r){1-5}& \mc{1}{$\mathcal{I}_2$}& 0.18 $\pm$ 0.00 & 0.01 $\pm$ 0.00 &  0.19 $\pm$ 0.00  &  &  & \\
& \mc{1}{$\mathcal{I}_3$}& 0.19 $\pm$ 0.00 & 0.01 $\pm$ 0.00 &  0.20 $\pm$ 0.00  &  &  & \\
\cmidrule(r){1-5}(2,3) &\mc{1}{(Analytic)}& 0.00 & 0.00 &  0.00  &  &    &  \\
\cmidrule(r){1-5}& \mc{1}{$\mathcal{I}_2$}& 0.00 $\pm$ 0.00 & 0.00 $\pm$ 0.00 &  0.00 $\pm$ 0.00  &  &  & \\
& \mc{1}{$\mathcal{I}_3$}& 0.00 $\pm$ 0.00 & 0.00 $\pm$ 0.00 &  0.00 $\pm$ 0.00  &  &  & \\
\cmidrule(r){1-5} (1,2,3) & \mc{1}{(Analytic)}& 0.00 & 0.00 &  0.00  &  &    &  \\
\cmidrule(r){1-5}& \mc{1}{$\mathcal{I}_2$}& 0.00 $\pm$ 0.00 & -0.00 $\pm$ 0.00 &  0.00 $\pm$ 0.00  &  &  & \\
& \mc{1}{$\mathcal{I}_3$}& 0.01 $\pm$ 0.00 & 0.00 $\pm$ 0.00 &  0.01 $\pm$ 0.00  &  &  & \\
\bottomrule
\end{tabular}}
\end{center}
\end{table}

\FloatBarrier

\subsubsection*{Illustration 2: California housing dataset} We consider the California housing dataset. This dataset has nine variables, eight predictors and a single response variable `median house value.' We standardize all variables to have zero mean and unit variance and utilize an ensemble of GBR's indexed on \[\mathcal{I}=\lbrace(i,(1,\dotsc,8)): i\in[3]\rbrace,\] each having $5\times10^3$ trees, to estimate the HDMR per \eqref{ss}. 

\FloatBarrier

\paragraph*{Variable importance}  We conduct a structural and correlative sensitivity analysis (SCSA) and show sensitivity indices with $\epsilon\ge 0.01$ below in Table~\ref{tab:cali}. Latitude and longitude each have strong negative correlative contributions to explained variance and participate in an modest-sized interaction. The relative sensitivity indices R (derived from the total T) are compared to the tree-based feature importance measure. HDMR places more emphasis on `MedInc', `Latitude', and `Longitude' than the tree-based indices. 

\begin{table}[h]
\caption{Sensitivity analysis for the California housing dataset, $\text{P}_\epsilon$, with $\epsilon=0.01$}
\label{tab:cali}
\begin{center}
\begin{tabular}{llcccccc}
\toprule
 Subspace   & Variables &    \mc{1}{ $\mathbb{S}^\text{a}$} &      \mc{1}{$\mathbb{S}^\text{b}$} & \mc{1}{ $\mathbb{S}$} & T & R & Tree \\
\midrule
 (1,)        & ('MedInc',)                        & 0.1346 &  0.1228 & 0.2574&0.3151&0.2637&0.1240 \\
 (2,) & ('HouseAge',) & - & - & - & 0.0377 &0.0315&0.0612\\
 (3,)        & ('AveRooms',)                      & 0.0305 &  0.0243 & 0.0548 &0.0966 &0.0808& 0.1322\\
 (4,) & ('AveBedrms',) & - & - & - & 0.0310 &0.0259&0.0839\\
 (5,) & ('Population',) & - & - & - & 0.0145 &0.0121&0.0803\\
  (6,)        & ('AveOccup',)                      & 0.0390 &  0.0168 & 0.0557&0.1310&0.1096&0.1475 \\
 (7,)        & ('Latitude',)                      & 0.7003 & -0.5356 & 0.1646 &0.2880&0.2411&0.1650\\
 (8,)        & ('Longitude',)                     & 0.6585 & -0.5218 & 0.1367 &0.2810&0.2352&0.2058\\
  (2, 6)      & ('HouseAge', 'AveOccup')           & 0.0069 &  0.0081 & 0.0150 &-&-&-\\
 (6, 8)      & ('AveOccup', 'Longitude')          & 0.0066 &  0.0046 & 0.0112 &-&-&-\\
 (7, 8)      & ('Latitude', 'Longitude')          & 0.0563 &  0.0168 & 0.0732 &-&-&-\\\midrule
   &                                    & 1.6374 & -0.8562 & 0.7813 &1.1948&1.00&1.00\\
\bottomrule
\end{tabular}
\end{center}
\end{table}
%
%\begin{table}[h]
%\caption{Total and relative sensitivity indices $\text{T}_\epsilon(f)$ and $\text{R}_\epsilon(f)$ for $\epsilon=0$}
%\label{tab:cali}
%\begin{center}
%\begin{tabular}{llrrr}
%\toprule
%&&\mc{3}{$\text{T}_\epsilon(f)$}\\
%\cmidrule(lr){3-5} Index   & Variable &  \mc{1}{$\mathcal{I}_1$}\ &    \mc{1}{ $\mathcal{I}_2$} &  \mc{1}{$\mathcal{I}_3$} \\
%\midrule
% 1        & MedInc                        & 0.1272 &  0.1231 & 0.2502 \\
% 2        & HouseAge                     & 0.0320 &  0.0255 & 0.0575 \\
% 3        & AveRooms                     & 0.0320 &  0.0255 & 0.0575 \\
% 4 &   & 0.0320 &  0.0255 & 0.0575 \\
% 5 &  & 0.0320 &  0.0255 & 0.0575 \\
% 6        & AveOccup                      & 0.0379 &  0.0172 & 0.0551 \\
% 7        & Latitude                      & 0.8451 & -0.6684 & 0.1767 \\
% 8        & Longitude                     & 0.7865 & -0.6475 & 0.1389 \\\midrule
%   &                                    & 1.8975 & -1.1047 & 0.7928 \\
%\bottomrule
%\end{tabular}
%\end{center}
%\end{table}

%In Figure~\ref{fig:ri_cali} we show the relative importance of subspaces attained a GBR having tree depth two and $5\times 10^3$ trees. %Variable importance conclusions between HDMR and GBR are generally similar, although there are some differences: HDMR reveals `AveOccup' to have much less importance than either longitude or latitude, whereas GBR accords similar importance. GBR shows `HouseAge' to have greater importance than `AveRooms,' whereas `HouseAge' does not appear for HDMR at a 1\% explained-variance cut-off. 

\FloatBarrier

\paragraph*{One-dimensional variable dependence}  

\begin{figure}[h]
\centering
\includegraphics[scale=0.65]{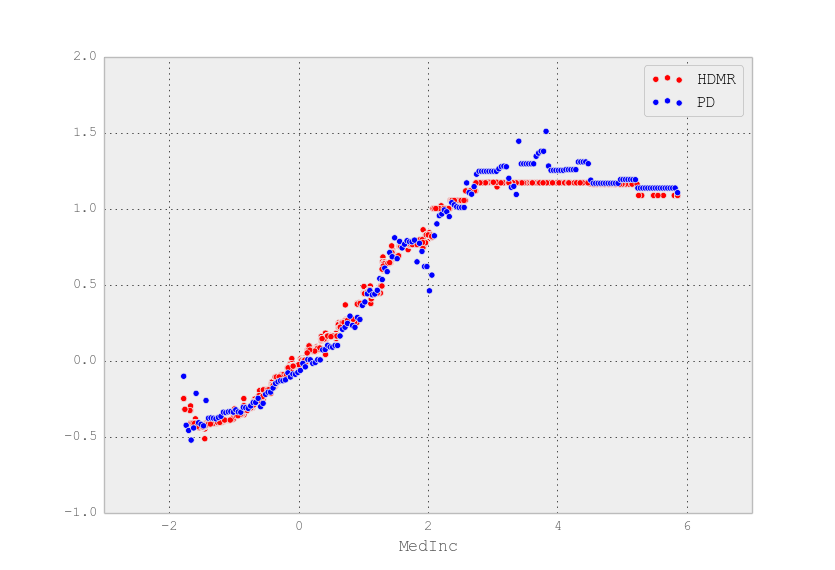}
\caption{HDMR and partial dependence for `MedInc'}\label{fig:medinc}
\end{figure}

Figures~\ref{fig:medinc} and \ref{fig:lat} show HDMR and partial dependence for `MedInc', `Latitude', and 'Longitude.' HDMR and partial dependence are similar for `MedInc,' although HDMR generally is smoother. For 'Latitude' and 'Longitude' they are somewhat different: HDMR profiles are generally vertically shifted to more negative values.   

\FloatBarrier

\begin{figure}[h]
\centering
\includegraphics[scale=0.65]{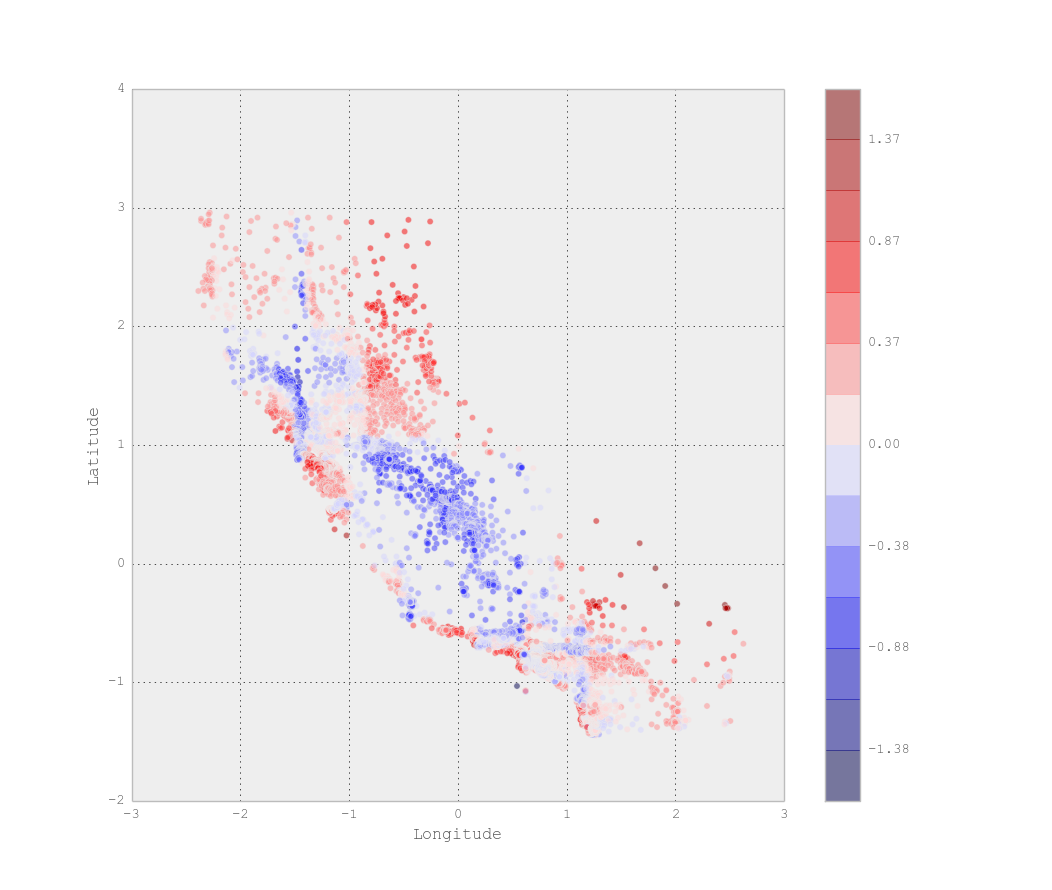}
\caption{HDMR component function in latitude and longitude}\label{fig:ll_hdmr}
\end{figure}

\paragraph*{Two-dimensional variable dependence}
In Figure~\ref{fig:ll_hdmr} we show the HDMR component function of latitude and longitude. In comparison to partial dependence (Figure 10.17 of \cite{esl}), HDMR reveals large positive values in scattered eastern localities and reveals positive contributions in northern-most and eastern-most locations. 

\FloatBarrier

\section{Concluding Remarks}\label{sec:cr} HDMR provides structured information on the variable importance and dependence for square-integrable functions. As such, it provides information necessary and sufficient to interpret the model output in terms of the model input and constitutes a glass box representation. This research highlights key utilities of HDMR to supervised machine learning applications and introduces algorithmic constructions from black box models. One is that HDMR can identify sources of information leakage. This is demonstrated for goodness-of-fit settings for ``big-data'' and popular machine learning interpretative diagnostics and black boxes. Second, HDMR is a useful tool for managing the curse of dimensionality, as it often admits efficient reduced-order representations of dense high-dimensional systems. Third, whenever input variables are correlated, the HDMR component functions are functions of the input distribution parameters including those for correlation. Fourth, HDMR can be applied as a wrapper method for black boxes to provision glass boxes. Collectively, these results suggest HDMR to have broad utility within machine learning. \\

%HDMR utility to theoretical machine learning can be broadened through investigations into information leakage and formulation for other machine learning tasks, such as transductive or unsupervised learning. For example, HDMR characterization of information leakage in estimators is not specific to PGOF but general to a broad class of quadratic forms \citep{rempala3}. High-efficiency estimators can be provisioned for ``big-data'' settings through using HDMR to excise inefficiencies, in a manner similar to the PGOF illustration of this article. HDMR can also be used to identify bias in black box machine learning algorithms for various models and such diagnostics can be used in the continual development and refinement of black boxes, of keen interest to the representation and analysis of complex systems arising from, say, network models. Transductive or unsupervised (auto-encoder) formulations can be developed to extend the HDMR framework to other machine learning tasks. HDMR utility to the practice of machine learning can be broadened by formulating constructions for other black boxes, such as support vector machines (SVM) or ensembles of neural network models, as well as extending the implementations of this article to structured settings involving multiple (conditionally-dependent) outputs, i.e. structured prediction (Li et al 2017).

\section*{Acknowledgements} CB acknowledges support for this project from the National Institute of Dental and Craniofacial Research of the National Institutes of Health (NIH-NIDCR Grant No. K08-DE-021430). HR acknowledges support from the National Science Foundation. We thank Grzegorz A. Rempala for comments helpful to improving the article.

\section*{Computational implementations} PD (GBR) is computationally implemented using the scikit learn (version 0.18) classes \textsl{GradientBoostingRegressor} and \textsl{partial-dependence}. %\emph{Note that the sklearn implementation of partial dependence gives incorrect results for 2D calculations.} This is exhibited in Figure~\ref{fig:pd_2d} for partial dependence in 'Latitude' and 'Longitude' whereby \textsl{partial-dependence} exhibits the marginal density rather than the two-dimensional projection.

\newpage

\addcontentsline{toc}{section}{Appendices}
\renewcommand{\thesubsection}{\Alph{subsection}}

\appendix
\section*{Appendices}

\subsection*{$\chi_n^2$ decomposition} Recall the PGOF statistic \[\chi_n^2= n\sum_{k\in [k_n]}\frac{(\hat{\mu}_n\lbrace k\rbrace-\mu_n\lbrace k\rbrace)^2}{\mu_n\lbrace k\rbrace}\] with empirical probability $\hat{\mu}_n$ defined as \[\hat{\mu}_n\lbrace k\rbrace = n^{-1}\sum_{j\in\mathbb{N}_n} I(X_{nj},k).\] Put $(x_1,\dotsc,x_n)\equiv (X_{n1},\dotsc, X_{nn})$. Observe that the statistic $\chi^2_n$ is a function of the random observation vector $(x_1,\dotsc,x_n)$. In \cite{rempala} $\chi_n^2$ is decomposed into two uncorrelated components, \[\chi_n^2 = n^{-1}(S_n+U_n)-n,\] where
\[S_n = \sum_{i} \mu^{-1}_n\lbrace x_i\rbrace\] 
and
\[U_n = \sum_{\substack{i,j\\i\ne j}}\mu_n^{-1}\lbrace x_i\rbrace I(x_i, x_j)\]
with \[\mathbb{E}\, S_n = n\, k_n,\,\,\,\,\,\mathbb{E}\,U_n=n(n-1),\,\,\,\,\,\mathbb{E}\,\chi_n^2 = k_n - 1.\] This gives \[\Var\, \chi^2_n = n^{-1}\left(\Var\, \mu_n^{-1}\lbrace X_n\rbrace+2(n-1)(k_n-1)\right).\] 
This is a second-order HDMR, \[\chi^2_n(x_1,\dotsc,x_n) = f_0 + \sum_{i} f_i(x_i) + \sum_{i<j} f_{ij}(x_i,x_j),\] where 
\begin{align*}
f_0 &= k_n-1\\
f_i(x_i) &=n^{-1}\left(\mu_n^{-1}\lbrace x_i\rbrace-k_n\right)\\
f_{ij}(x_i,x_j)&=2\, n^{-1}\left(\mu_n^{-1}\lbrace x_i\rbrace\, I(x_i, x_j) -1\right),
\end{align*}
with \[\Var\, f_i = n^{-2}\,\Var\,\mu_n^{-1}\lbrace X_n\rbrace\] and \[\Var\, f_{ij}= 4\,n^{-2}(k_n-1).\] 
Note that if \[(k_nn)^{-1}\Var\, \mu_{\,n}^{-1}\lbrace X_n\rbrace\xrightarrow{n, k_n\rightarrow\infty}0,\] then $(2k_n)^{-1/2}\sum_i f_i(x_i)\rightarrow 0$ in probability. This is satisfied for uniform and power law ($\alpha\in[0,1)$) discrete random variables. Hence, for $k_n\rightarrow\infty$ as $n\rightarrow\infty$, the first-order terms converge to constants and the asymptotic influence of the second-order terms dominates, i.e. the distributional limit of $\chi_n^2$ is determined by the second-order component functions. This is equivalently stated using the sensitivity indices, $\sum_{i<j}\mathbb{S}_{ij}\xrightarrow{n, k_n\rightarrow\infty} 1$. Putting $n / \sqrt{k_n}\rightarrow\lambda$, it turns out that $\chi_n^2\xrightarrow{\lambda=\infty}\text{Gaussian}$, $\chi_n^2\xrightarrow{0<\lambda<\infty}\text{Poisson}$, and $\chi_n^2\xrightarrow{\lambda=0}\text{Degenerate}$. Hence, the distributional limit of $\chi_n^2$ is standard Gaussian for \begin{enumerate*}[label=(\roman*)] \item $n, k_n\rightarrow\infty$ with $\lambda=\infty$ or for \item $n\rightarrow\infty$ and $k_n=k<\infty$ \end{enumerate*}; however, for $n, k_n\rightarrow\infty$ and $\lambda <\infty$ this is not true. Consequently, an improved statistic for $\chi_n^2$ is defined using HDMR, \begin{align*}\chi_{n\text{HDMR}}^2(x_1,\dotsc,x_n)&\equiv f_0+\sum_{i<j}f_{ij}(x_i,x_j) \\&\equiv \chi_n^2(x)-\sum_if_i(x_i).\end{align*} 

%In Figure~\ref{chi}, we exhibit empirical distributions for PGOF, HDMR chi-square, and chi-square for $k_n=2500$ and $n\in\lbrace50,500,5000\rbrace$, having corresponding $\lambda\in\lbrace1,10,100\rbrace$, for the power law distribution with $\alpha=1/2$. We show statistics for the estimators in Table~\ref{tab:est}. For $\lambda=100$, Figure~\ref{fig:est5000} shows both $\chi_n^2$ and $\chi_{n\text{HDMR}}^2$ are asymptotically $\chi^2$. For $\lambda = 10$, Figure~\ref{fig:est500} shows that $\chi_n^2$ is not asymptotically $\chi^2$ but $\chi_{n\text{HDMR}}^2$ is. For $\lambda = 1$, Figure~\ref{fig:est50} shows that neither $\chi_n^2$ nor $\chi^2_{n\text{HDMR}}$ is asymptotically $\chi^2$. In particular, $\chi^2_{n\text{HDMR}}$ is expressed in terms of a Poisson law (see appendices for a precise characterization), although observe that the empirical distribution is truncated due to undersampling of the power law tails. \textbf{These results show that the PGOF statistic $\chi_n^2$ is inefficient and experiences \emph{information leakage} in comparison to the HDMR $\chi_{n\text{HDMR}}^2$}. For example, for $\lambda = 1$ we observe PGOF to have a relative efficiency of $\sim10\%$ to that of HDMR.

A precise characterization of the asymptotic behavior of $\chi_n^2$ is stated below. Note that this result also holds for power-law distributions. 
\begin{theorem}[Limit theorem for $\chi_n^2$ for the uniform distribution] Assume $\mu_n\lbrace x\rbrace = k_n^{-1}$ for $x\in K_n$ and $n=1,2,\dotsc$, as well as \[n/\sqrt{k_n}\rightarrow\lambda.\] Then, 
\[
\frac{\chi_n^2-k_n}{\sqrt{2 k_n}} \xrightarrow{d}
\begin{cases}
0 & \text{when }\lambda = 0\\
\frac{\sqrt{2}}{\lambda} Z - \frac{\lambda}{\sqrt{2}},\,\,\, Z\sim\text{Poisson}\left(\frac{\lambda^2}{2}\right) & \text{when }\lambda\in(0,\infty)\\
N\sim\text{Gaussian}(0,1)& \text{when } \lambda = \infty.
\end{cases}
\]
\end{theorem}

\subsection*{Simple product function} For the function $f(x)=\prod_{i=1}^n x_i$, the partition of variance is given by \[\color{black}\Var\,(f) = \sigma_f^2 = \sum_i \sigma_i^2 + \sum_{i<j}\sigma_{ij}^2 + \dotsc + \sigma_{1\dotsc n}^2,\]
where
\begin{align*}
\sigma_u^2 &= \int_{X_u}f^2_u(x_u)\D\nu_u(x_u)\\
\nu_u(x_u)&=\prod_{i\in u}\nu_i(x_i)
\end{align*}
and
\begin{align*}
\textcolor{black}{\sigma_f^2} &\textcolor{black}{\,\,=\mu^{2n}\left(\left(1+\rho^2\right)^n-1\right)}\\%^3^{-n} (a-b)^{-n} \left(a^3-b^3\right)^n-4^{-n} (a+b)^{2 n}\\
\sigma_i^2 &=\mu^{2n}\rho^2\\%\frac{1}{3} 4^{-n} (a-b)^2 (a+b)^{2 n-2}\\
\sigma_{ij}^2 &=\mu^{2n}\rho^4\\%\frac{1}{9} 2^{-2 n-1} (a-b)^4 (a+b)^{2 n-4}\\
\vdots&\\
\textcolor{black}{\sigma_{i_1\dotsb i_k}^2} &\textcolor{black}{\,\,=\mu^{2n}\rho^{2k}}.%\frac{1}{9} 2^{-2 n-1} (a-b)^4 (a+b)^{2 n-4}\\
\end{align*}

\begin{figure}[h]
\centering
\includegraphics[scale=0.55]{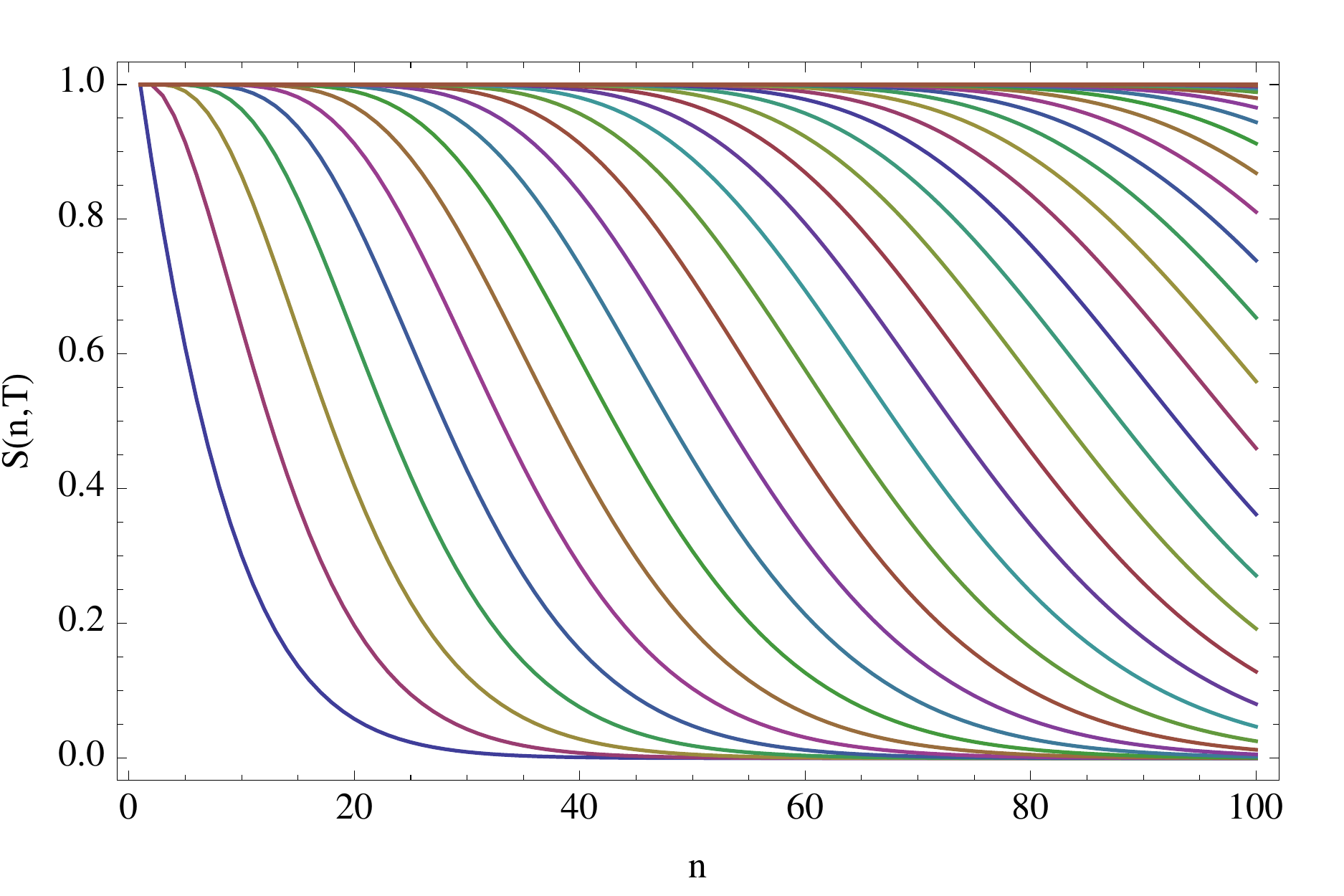}
\caption{$\mathbb{S}(n,T)=\sum_{k=1}^Tp\lbrace k\rbrace$ for $(n,T)\in\lbrace1,\dotsc,100\rbrace^2$ and $\rho=\frac{1}{2}$}
\label{fig:polyn}
\end{figure}

Figure~\ref{fig:polyn} shows the percent of explained variance of $f^T(x_1,\dotsc,x_n)$ in $(n,T)$ for $\rho=\frac{1}{2}$. The curves correspond to different values of $T$, where $(1,\dotsc,n)$ is left-to-right. As exhibited, increasing $T$ increases the explained variance.

%${\color{black}S(n,T)=\sum_{k=1}^Tp\lbrace k\rbrace}$ for $(n,T)\in\lbrace1,\dotsc,100\rbrace^2$, \textcolor{black}{$\rho=\frac{1}{2}$}

\begin{figure}[h]
\centering
\includegraphics[scale=0.55]{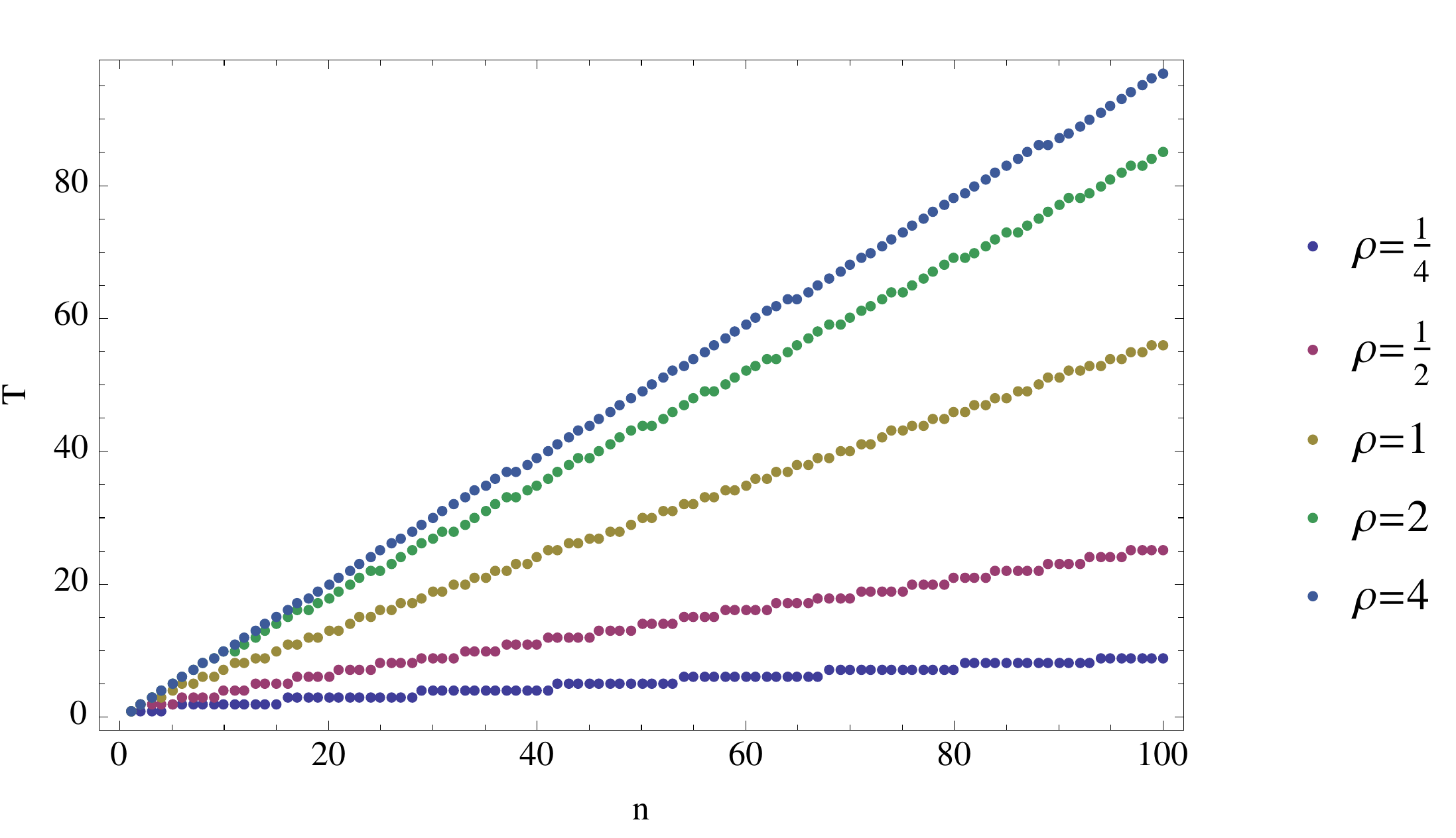}
\caption{$\min T$ such that $S(n,T)\ge 0.9$ for $(n,\rho)\in\lbrace1,\dotsc,100\rbrace\times\lbrace\frac{1}{4},\frac{1}{2},1,2,4\rbrace$}
\label{fig:polyn2}
\end{figure}

Figure~\ref{fig:polyn2} reveals that the necessary value of $T$ to attain a given percent of explained variance depends upon $\rho$.

\FloatBarrier

\subsection*{Polynomials as linear combinations of monomials}

Defining the monomial \[x_u = \prod_{i\in u}x_i\] with $iid$ $x$, having common $\mu$ and $\sigma^2$, we consider a linear combination of monomials, \[e_T(\beta,x) = \sum_{\substack{u\\|u|\le T}}\beta_u x_u.\]  The variance of $e_T$ is \[\Var\,(e_T) = \mathbb{E}[(\sum_{\substack{u\\|u|\le T}}\beta_u(x_u - \mu^{|u|}))^2] = \sum_{\substack{u\\|u|\le T}}\beta_u^2\Var\,(x_u) + 2\sum_{\substack{u<v\\|u|,|v|\le T}}\beta_u\beta_v\Cov(x_u,x_v),\] where, putting $\rho=\sigma/\mu$, \[\Var\,(x_u) = \mu^{2|u|}((1+\rho^2)^{|u|}-1).\] \[\Cov(x_u,x_v) = \mu^{|u|+|v|}((1+\rho^2)^{|u\cap v|} - 1).\] %so that \[\sum_{j}\norm{\beta_{|u|=j}}_2^2\mu^{2j}((1+\rho^2)^{j}-1) + \sum_{j,k,l}\binom{n}{j}\binom{n}{k}\mu^{j+k}((1+\rho^2)^{l} - 1)\] 

$e_T(\beta,x)$ is factorized as
%\begin{align*}
%e_T(x) &= x_i\left(1+\sum_{\substack{j\\j\ne i}} x_j + \dotsc + \sum_{\substack{j_1<\dotsc<j_{T-1}\\j_1,\dotsc,j_{T-1}\ne i}}x_{j_1\dotsc j_{T-1}}\right)\\
%& + \left(1 + \sum_{\substack{j\\j\ne i}} x_{j}  + \dotsc + \sum_{\substack{j_1<\dotsc<j_{T}\\j_1,\dotsc,j_{T}\ne i}}x_{j_1\dotsc j_{T}}\right)\\
%&= x_{ij}\left(1 + \sum_{\substack{k\\k\ne i,j}} x_k + \dotsc + \sum_{\substack{k_1<\dotsc<k_{T-2}\\k_1,\dotsc,k_{T-2}\ne i,j}}x_{k_1\dotsc k_{T-2}}\right)\\
%& + (x_i+x_j)\left(1 + \sum_{\substack{k\\k\ne i,j}} x_k + \dotsc + \sum_{\substack{k_1<\dotsc<k_{T-1}\\k_1,\dotsc,k_{T-1}\ne i,j}}x_{k_1\dotsc k_{T-1}}\right)\\
%& + \left(1 + \sum_{\substack{k\\k\ne i,j}} x_{j} + \dotsc + \sum_{\substack{k_1<\dotsc<k_{T}\\k_1,\dotsc,k_{T}\ne i,j}}x_{k_1\dotsc k_{T}}\right).
%\end{align*}
%
%\[a_{v} = \sum_{\substack{u\\|u|\le T\\|u\cap v|=0}}\beta_{u} x_u,\]
\begin{align*}
e_T(x) &= x_i\left(\sum_{\substack{u\\|u|\le T-1\\|i\cap u|=0}}\beta_{iu} x_u\right) + \left(\sum_{\substack{u\\|u|\le T\\|i\cap u|=0}}\beta_{u} x_u\right)\\
&= x_{ij}\left(\sum_{\substack{u\\|u|\le T-2\\|ij\cap u|=0}}\beta_{iju} x_u\right) + x_i\left(\sum_{\substack{u\\|u|\le T-1\\|ij\cap u|=0}}\beta_{iu} x_u\right) + x_j\left(\sum_{\substack{u\\|u|\le T-1\\|ij\cap u|=0}}\beta_{ju} x_u\right) + \left(\sum_{\substack{u\\|u|\le T\\|ij\cap u|=0}}\beta_{u} x_u\right),
\end{align*}
%we define \[x_{[u,t]} = (x_{i_1\dotsb i_t} :  i_1<\dotsb<i_{t},\,\,\, |i_1\dotsb i_t \cap u|=0)\] and write  
%%and defining \[a_{ij} = \sum_{k\le T-j}\binom{n-i}{k}\mu^k,\]
%
%\begin{align*}
%e_T(x) &= x_i\left(\beta_i+\sum_{t=1}^{T-1}\beta_{i[i,t]} x_{[i,t]}\right) + \left(\beta_0 + \sum_{t=1}^{T}\beta_{\Var\,nothing[i,t]} x_{[i,t]}\right)\\
%%& + \left(1 + \sum_{\substack{j\\j\ne i}} x_{j}  + \dotsc + \sum_{\substack{i_1<\dotsc<i_{T}\\i_1,\dotsc,i_{T}\ne i}}x_{i_1\dotsc i_{T}}\right)\\
%%&= x_{ij}\left(\beta_{ij}+\sum_{t=1}^{T-2}\beta_{ij[ij,t]} x_{[ij,t]}\right) + x_i\left(\beta_0 + \sum_{t=1}^{T}\beta_{\Var\,nothing[i,t]} x_{[i,t]}\right)
%\end{align*}
For $\beta=\bm{1}$, we call $e_T(x)$ an elementary symmetric polynomial and can compute its conditional expectations as
\begin{align*}
\mathbb{E}[e_T] &= a_{00}\\
\mathbb{E}[e_T|\mathscr{F}_i] &= a_{11}x_i + a_{10}\\ 
\mathbb{E}[e_T|\mathscr{F}_{ij}] &= a_{22}x_{ij} + a_{21}(x_i+x_j) + a_{20}.\\ 
\vdots&\\
\mathbb{E}[e_T|\mathscr{F}_{u}] &=\sum_{w\subseteq u}a_{|u||w|}x_w,
\end{align*}
where  \[a_{rs} = \sum_{\substack{u\\|u|\le T-s\\|r\cap u|=0}}\beta_{ru}\mu^{|u|} = \sum_{k\le T-s}\binom{n-r}{k}\mu^k.\] 

%For $\beta\in\mathbb{R}^{b}$, we have\\

The component functions of $e_T(x)$ are defined recursively,

\begin{align*}
f_0 &= \mathbb{E}[e_T]\\
f_i(x_i) &= \mathbb{E}[e_T|\mathscr{F}_i]  - f_0\\
f_{ij}(x_i,x_j) &= \mathbb{E}[e_T|\mathscr{F}_{ij}] - f_i(x_i) - f_j(x_j) - f_0\\
\vdots&\\
f_u(x_u) &= \mathbb{E}[e_T|\mathscr{F}_u] - \sum_{w\subset u} f_w(x_w).
\end{align*}

For $T=n$, we have 
 \[\sigma_f^2 = (\mu +1)^{2 n} \left(\left(\tau^2+1\right)^n-1\right)\]
 \[f_u(x_u) = (1+\mu)^{n-|u|}\prod_{i\in u}(x_{i}-\mu)\]
 \[\sigma^2_u = \tau^{2|u|} (\mu +1)^{2n}\] 
 \[q\lbrace k\rbrace = \sum_{i_1<\dotsb<i_k} \mathbb{S}_{i_1\dotsc i_k} = \frac{\binom{n}{k} \tau^{2k}}{\left(\tau^2+1\right)^n-1}\] where \[\tau = \frac{\sigma}{\mu+1}= \frac{\rho}{1+\frac{1}{\mu}}.\] When $\tau<1$, $q\lbrace k\rbrace>p\lbrace k\rbrace$ for $k\ll n$. For the uniform distribution on the unit interval, we have $\tau=\frac{1}{3 \sqrt{3}}\approx 0.19$. Furthermore, \[\lim_{\substack{\mu,\sigma\rightarrow\infty\\\sigma/\mu=\rho}}\tau=\rho.\]
 
% Hence, the improvement factor is $1+\frac{1}{\mu}$
 
  \begin{figure}[h]
\centering
\includegraphics[scale=0.55]{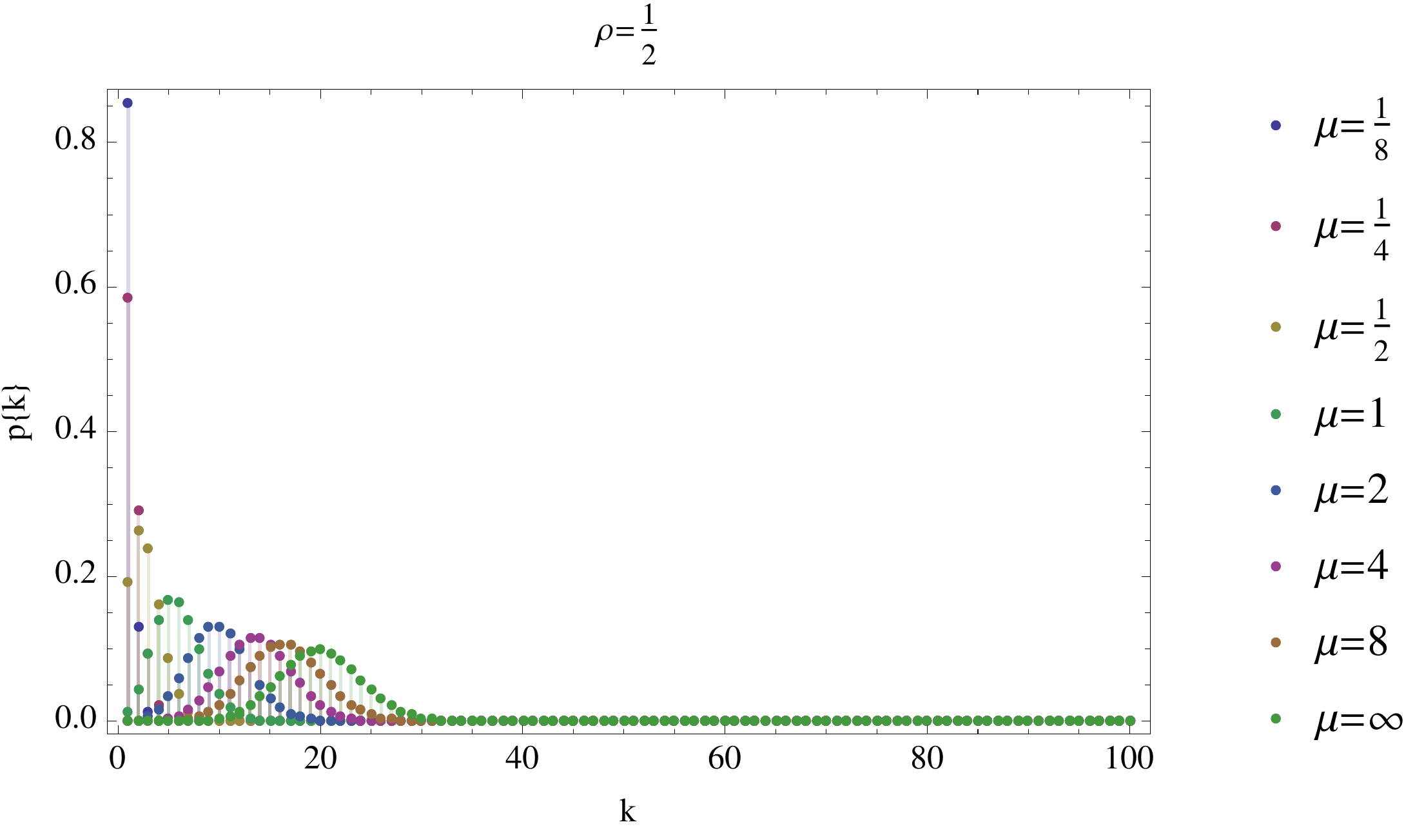}
\caption{$q\lbrace k\rbrace=\sum_{i_1<\dotsc<i_k}\mathbb{S}_{i_1\dotsc i_k}$ for $\rho=\frac{1}{2}$}\label{fig:linearprobrho12}
\end{figure}

 \begin{figure}[h]
\centering
\includegraphics[scale=0.55]{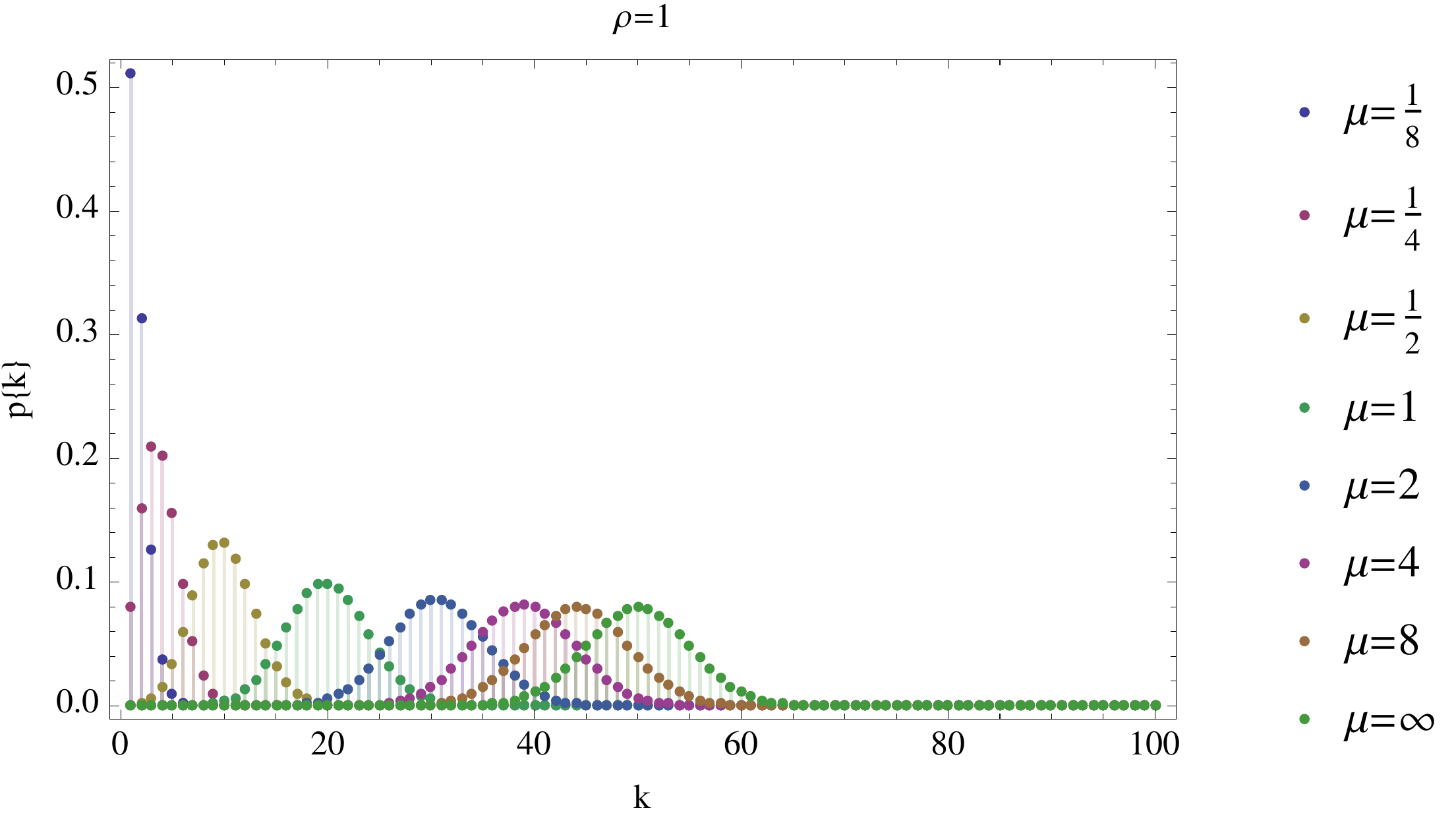}
\caption{$q\lbrace k\rbrace=\sum_{i_1<\dotsc<i_k}\mathbb{S}_{i_1\dotsc i_k}$ for $\rho=1$}\label{fig:linearprobrho1}
\end{figure}

 \begin{figure}[h]
\centering
\includegraphics[scale=0.55]{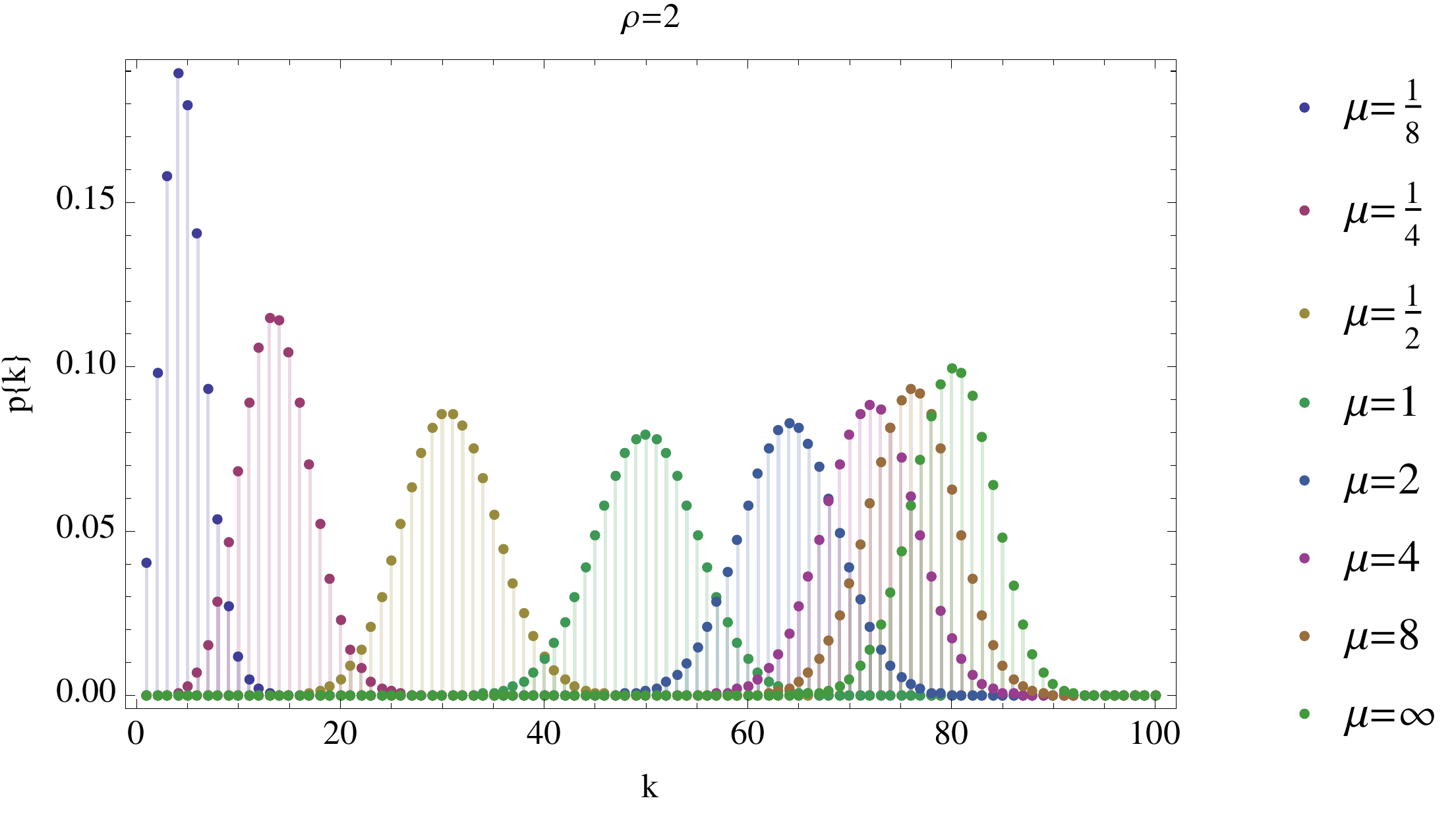}
\caption{$q\lbrace k\rbrace=\sum_{i_1<\dotsc<i_k}\mathbb{S}_{i_1\dotsc i_k}$ for $\rho=2$}\label{fig:linearprobrho2}
\end{figure}

\FloatBarrier

%These results are attained using the following.

\subsection*{Feature vector construction for hierarchically-orthogonal subspaces}
%Suppose we have the polynomial feature vector $V^d =(x_u^\alpha: |\alpha|\le d, \alpha\in\mathbb{N}_0^{|u|})$ with $|V^d| = K$ of $f(x)$. We form $\Phi_d(x) \equiv  \concat_{\substack{u\\|u|\le d}}Q_u^d(\norm{\alpha}_0=|u|)$ from orthonormal bases $\lbrace Q^d_u\rbrace$ attained from the Gram-Schmidt process, where $|\Phi_d(x)|=B$ and $\concat$ is the column-wise concatenation operator. This gives is \[\Phi_d(x)\gamma = f(x),\] with coefficients \[\gamma = \left(\Phi_d\otimes\Phi_d\right)^{-1}(\Phi_d\otimes f).\] These outer products can be calculated exactly, as they are computations of moments of $\mu$. The component functions are given by \[f_u(x_u) = \gamma_u\cdot Q_u(\norm{\alpha}_0 = |u|),\] which can be computed as \[(f_u(x_u) : u) = (\Phi_d(x)\odot\gamma)\cdot I\] where $I\in\lbrace0,1\rbrace^{B\times C}$ is a tall matrix with $C=1+\sum_{i\in[d]}\binom{n}{i}$ columns, each column $u$ containing ones at rows corresponding to $Q_u(\norm{\alpha}_0=|u|)$. We have \[\text{Span}\bigcup_{\substack{u\\|u|\le d}}Q_u^d(\norm{\alpha}_0=|u|) = \mathcal{V}_0\oplus\sum_i\mathcal{V}_i\oplus\sum_{i_1<i_2}\mathcal{V}_{i_1i_2}\oplus\dotsc\oplus\mathcal{V}_{1\dotsc d} = \Upsilon_u^d. \] 
Suppose for the measure space $(X_u,\mathscr{X}_u,\mu_u)$ of subspace $u\subseteq\lbrace1,\dotsc,n\rbrace$, where $\mu_u(x_u)=\int_{X^{-u}}\mu(x_u,\D x_{-u})$, we have a collection of basis vectors indexed on $\powerset(u)$ and having dimensions $b=(b_v\in\mathbb{N}: v\in\powerset(u))$. This is denoted by \[\textbf{B}\equiv\lbrace\text{B}_v\equiv(\text{B}_{v1},\dotsc,\text{B}_{vb_v}): v\in\powerset(u)\rbrace.\] We can construct a non-orthogonal basis vector $\Phi_u$ having dimension $b_u$ whose elements are hierarchically-orthogonal with respect to $\mu$ (and $\mu_u$). To do this, we order $\powerset(u)$ in size and use the Gram-Schmidt process to generate $\textbf{Q}=(\text{Q}_v\equiv(\text{Q}_{v1},\dotsc,\text{Q}_{vb_v}): v\in\powerset(u))$ from $\textbf{B}$. If the bases are given on an empirical measure as column vectors of data $\textbf{B}\in\mathbb{R}^{N\times\Sum(b)}$, then QR decomposition is conducted on $\textbf{B}$, returning $\textbf{Q}$. Then we put $\Phi_u=\text{Q}_u$. Empirically, this is $\Phi_u\in\mathbb{R}^{N\times b_u}$. Repeating this process for all subspaces of interest $\bm{\alpha}$ we form the feature vector \[\Phi=\Phi_{\varnothing}\parallel\Phi_{\lbrace1\rbrace}\parallel\dotsb\parallel\Phi_{\lbrace n\rbrace}\parallel\Phi_{\lbrace1,2\rbrace}\parallel\dotsb=\bigparallel_{\alpha\in\bm{\alpha}}\Phi_\alpha.\] Putting \[\Phi\gamma = f,\] the coefficients are given as \[\gamma = (\gamma_\alpha: \alpha\in\bm{\alpha})=\left(\Phi\otimes\Phi\right)^{-1}(\Phi\otimes f).\] These outer products can be calculated exactly, as they are computations of moments of $\mu$. The component functions are given by \[f_u(x_u) = \langle\gamma_u, \Phi_u(x_u)\rangle_{\ell^2}.\] %which can be computed as \[(f_u(x_u) : u) = (\Phi_d(x)*\gamma)\cdot I\] where $I\in\lbrace0,1\rbrace^{B\times C}$ is a tall matrix with $C=1+\sum_{i\in[d]}\binom{n}{i}$ columns, each column $u$ containing ones at rows corresponding to $Q_u(\norm{\alpha}_0=|u|)$.\\

%Suppose we have the feature-vectors \[((\phi_{v1},\dotsc,\phi_{vb_v}): v\subseteq u, \phi_v\in)\] on $X_u$ indexed on $(\bm{\alpha}_v: v\subseteq u)$, denoted by $\phi_u =(\phi_\alpha(x_u): \alpha\in\bm{\alpha})$ with $|V_u| = K$. 

%Putting \[\text{proj}_{u_j}(v_k) = \frac{\langle u_j, v_k\rangle}{\langle u_j, u_j\rangle}u_j = \frac{s_{kj}}{t_j} u_j = c_{kj} u_j,\;\;\; j<k,\] where 
%\begin{equation} 
%c_{kj} =  \frac{s_{kj}}{t_j},
%\end{equation}
%Gram-Schmidt orthonormalization involves computing \[Q_u^d = \left(e_k : e_k = \frac{u_k}{\norm{u_k}_2},\,\,\,u_k \gets v_k - \sum_{\substack{j\\j<k}}c_{kj}u_j,\,\,\, v_k\in V_u^d\right),\] 

%which uses the moments of $\mu_X$. Ordering $\bm{\alpha}$ by 

\subsection*{Correlated expansion}

A non-orthogonal HDMR basis $\lbrace \Phi_0,\Phi_1,\Phi_2,\Phi_{12}\rbrace$ is formed as 

\resizebox*{\textwidth}{!}{
\begin{minipage}{\linewidth}
\begin{align*}
\footnotesize
\Phi_0 &= (1)\\
\Phi_i &= \left(\frac{x_i-\mu_i}{\sigma_i},\frac{(x_i-\mu_i)^2}{\sigma_i^2}-\sqrt{2}\right), \,\,\, i=1,2\\
\Phi_{12}&= \left(\frac{-\rho \sigma_2^2 \left(\left(\rho^2-1\right) \sigma_1^2+(x_1-\mu_1)^2\right)+\left(\rho^2+1\right) \sigma_1 \sigma_2 (x_1-\mu_1) (x_2-\mu_2)+\rho \sigma_1^2 \left(-(x_2-\mu_2)^2\right)}{\left(\rho^2+1\right) \sqrt{\rho^2+\frac{4}{\rho^2+1}-3} \sigma_1^2 \sigma_2^2}\right).
\end{align*}
\end{minipage}}

The coefficients are given by
\begin{align*}
\gamma_0 &= \left(\beta_{0}+\beta_{1} \mu_1+\beta_{2} \mu_2+\beta_{12} \mu_1 \mu_2+\beta_{12} \rho \sigma_1 \sigma_2\right)\\
\gamma_1 &= \left(\sigma_1 (\beta_{1}+\beta_{12} \mu_2),\frac{\sqrt{2} \beta_{12} \rho \sigma_1 \sigma_2}{\rho^2+1}\right) \\
\gamma_2 &= \left(\sigma_2 (\beta_{2}+\beta_{12} \mu_1),\frac{\sqrt{2} \beta_{12} \rho \sigma_1 \sigma_2}{\rho^2+1}\right) \\
\gamma_{12} &= \left(\beta_{12} \sqrt{\rho^2+\frac{4}{\rho^2+1}-3} \sigma_1 \sigma_2\right)
\end{align*}

\subsection*{Analytic test function}

 \begin{figure}[h]
\centering
\includegraphics[scale=0.5]{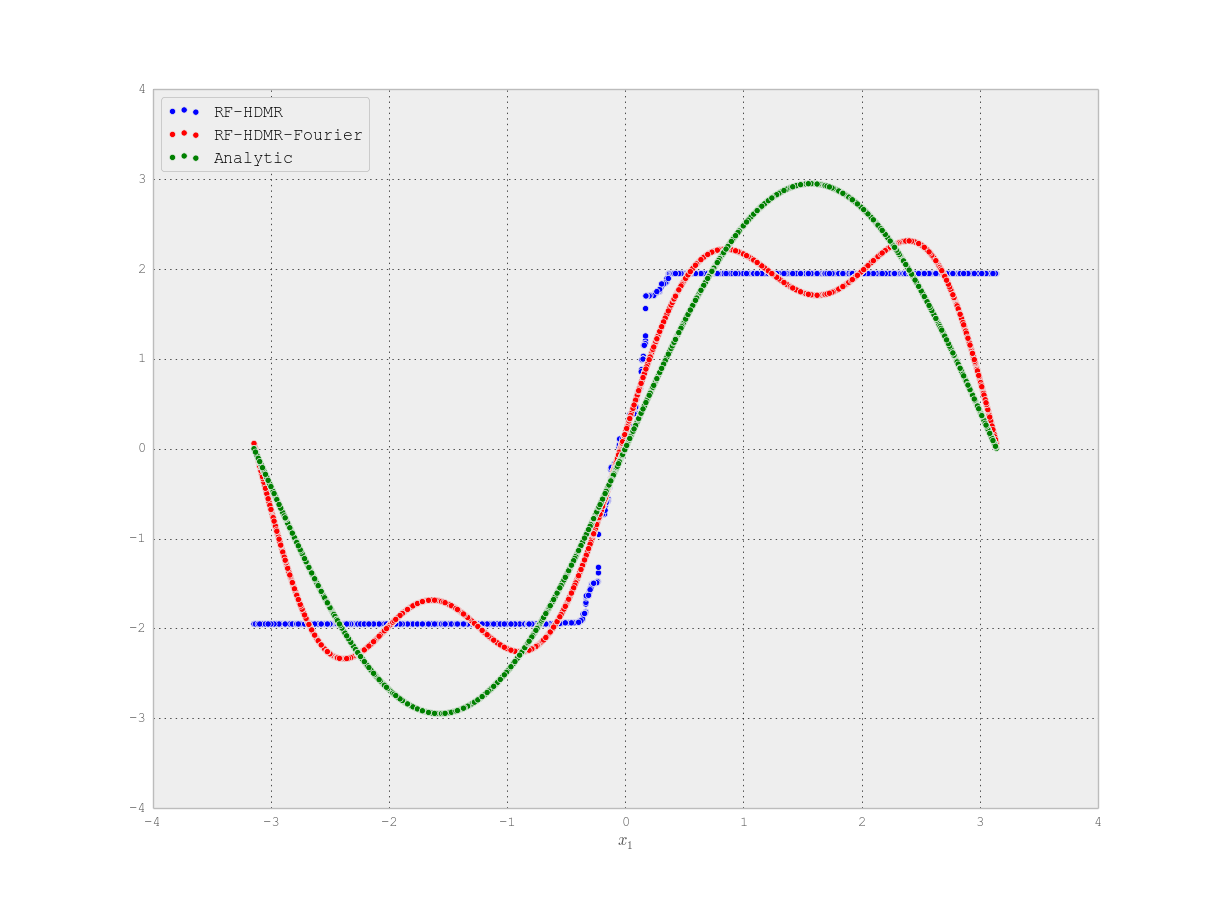}
\caption{$f^\text{HDMR}_1(x_1)$ by RF approximation, smooth Fourier projection, and analytic}\label{fig:ishigami1_rf}
\end{figure}

 \begin{figure}[h]
\centering
\includegraphics[scale=0.5]{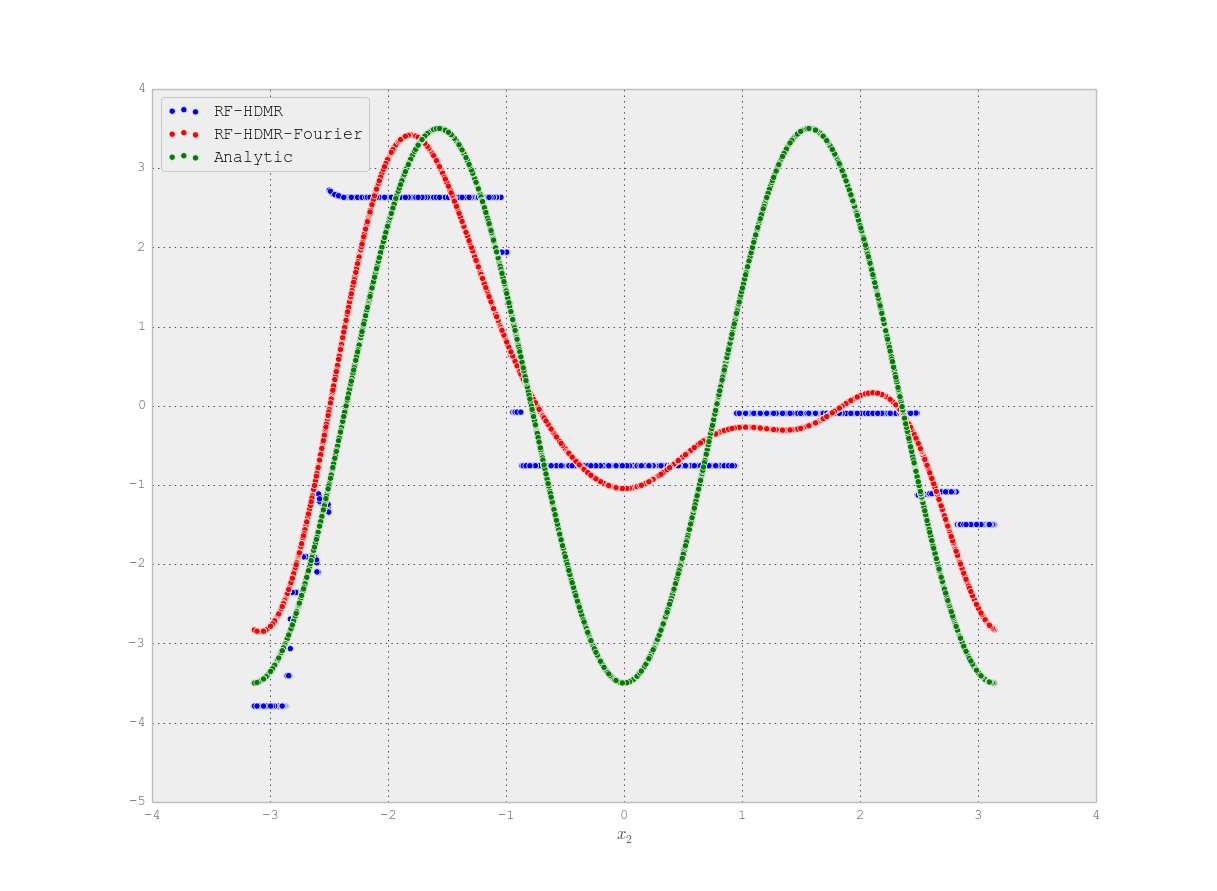}
\caption{$f^\text{HDMR}_2(x_2)$ by RF approximation, smooth Fourier projection, and analytic}\label{fig:ishigami2_rf}
\end{figure}

 \begin{figure}[h]
\centering
\caption{$f^\text{HDMR}_{13}(x_1,x_3)$ by RF approximation}\label{fig:ishigami3_rf}
\begingroup
\captionsetup[subfigure]{width=5in,font=normalsize}
\color{black}
\subfloat[Decision tree approximation]{\includegraphics[width = 5in]{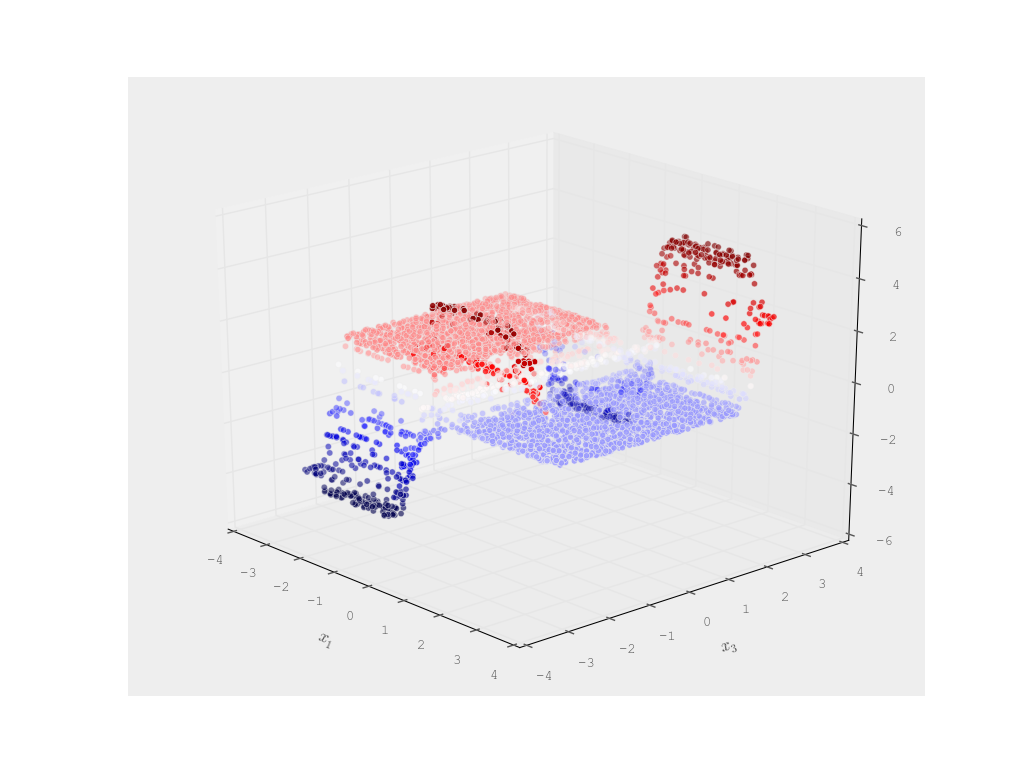}}\\
\subfloat[Analytic]{\includegraphics[width = 5in]{ishigami3_analytic.png}}\\
\endgroup
\end{figure}

\FloatBarrier

\subsection*{California housing dataset} 

%\begin{figure}[h]
%\centering
%\includegraphics[scale=0.6]{gbr_ri.png}
%\caption{Relative importance of subspaces in California housing dataset from GBR}\label{fig:ri_cali}
%\end{figure}

\begin{figure}[h]
\centering
\includegraphics[scale=0.5]{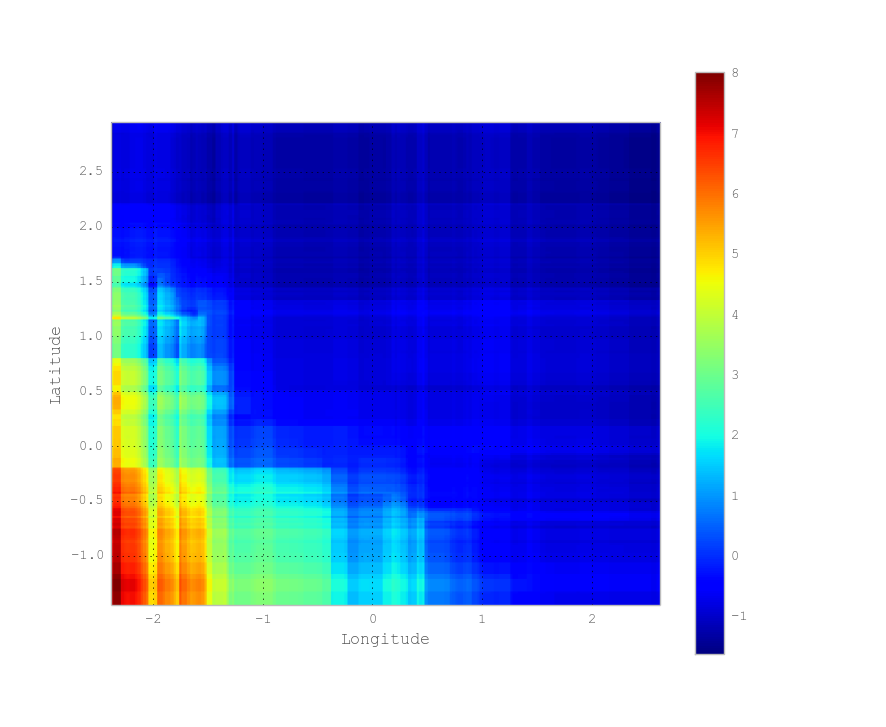}
\caption{Partial dependence in latitude and longitude as computed by sklearn classes \textsl{GradientBoostingRegressor} and \textsl{partial-dependence}}\label{fig:pd_2d}
\end{figure}

\begin{figure}[h]
\centering
\includegraphics[scale=0.65]{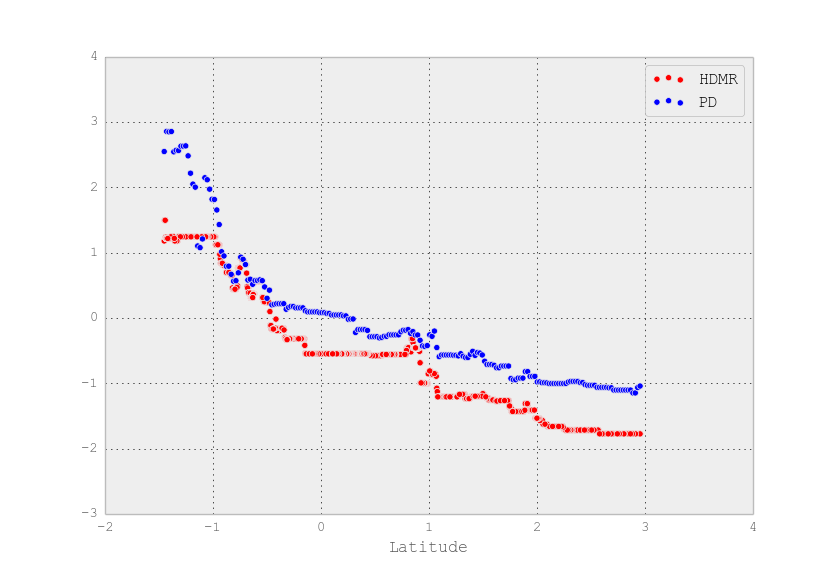}
\caption{HDMR and partial dependence for `Latitude'}\label{fig:lat}
\end{figure}

\begin{figure}[h]
\centering
\includegraphics[scale=0.65]{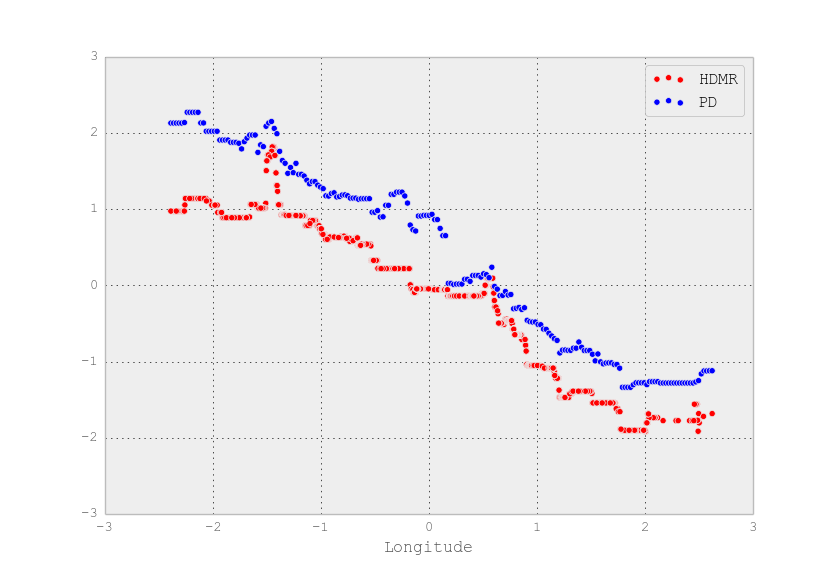}
\caption{HDMR and partial dependence for `Longitude'}\label{fig:long}
\end{figure}

%\begin{figure}[h]
%\centering
%\includegraphics[scale=0.5]{pd_ll_cali.tiff}
%\caption{Partial dependence in latitude and longitude \citep{esl} (Figure reprinted with kind permission of Springer-Verlag)}\label{fig:pd_ll_cali}
%\end{figure}
\FloatBarrier

\clearpage
\vskip 0.2in
\bibliographystyle{plainnat}

\end{document}